\documentclass{jfm}

\usepackage{graphicx}
\usepackage{natbib}
\usepackage{color}
\usepackage{epsfig}
\usepackage{relsize}

\usepackage{amsmath}    % need for subequations
\usepackage{verbatim}   % useful for program listings
\usepackage{subfigure}  % use for side-by-side figures
\usepackage{hyperref}
\usepackage{tabularx}

\usepackage[normalem]{ulem}

\usepackage[dvipsnames]{xcolor}

\ifCUPmtlplainloaded \else
  \checkfont{eurm10}
  \iffontfound
    \IfFileExists{upmath.sty}
      {\typeout{^^JFound AMS Euler Roman fonts on the system,
                   using the 'upmath' package.^^J}%
       \usepackage{upmath}}
      {\typeout{^^JFound AMS Euler Roman fonts on the system, but you
                   dont seem to have the}%
       \typeout{'upmath' package installed. JFM.cls can take advantage
                 of these fonts,^^Jif you use 'upmath' package.^^J}%
       \providecommand\upi{\pi}%
      }
  \else
    \providecommand\upi{\pi}%
  \fi
\fi

\ifCUPmtlplainloaded \else
  \checkfont{msam10}
  \iffontfound
    \IfFileExists{amssymb.sty}
      {\typeout{^^JFound AMS Symbol fonts on the system, using the
                'amssymb' package.^^J}%
       \usepackage{amssymb}%
         \let\leq=\leqslant
         
      }{}
  \fi
\fi

\ifCUPmtlplainloaded \else
  \IfFileExists{amsbsy.sty}
    {\typeout{^^JFound the 'amsbsy' package on the system, using it.^^J}%
     \usepackage{amsbsy}}
    {\providecommand\boldsymbol[1]{\mbox{\boldmath $##1$}}}
\fi

\newsavebox{\astrutbox}
\sbox{\astrutbox}{\rule[-5pt]{0pt}{20pt}}

\newcommand\etal{\mbox{\textit{et al.}}}

\newcommand{\zh}{\boldsymbol}

\title[Deformation and dewetting of liquid films under gas jets]{Deformation and dewetting of liquid films under gas jets}

\author[Chinasa J. Ojiako \etal]%
{C\ls H\ls I\ls N\ls A\ls S\ls A\ns J.\ns O\ls J\ls I\ls A\ls K\ls O$^{1,2}$, \ns
R\ls A\ls D\ls U\ns C\ls I\ls M\ls P\ls E\ls A\ls N\ls U$^{3,4,5}$, \ns
H.\ns C.\ns H\ls E\ls M\ls A\ls K\ls A\ns B\ls A\ls N\ls D\ls U\ls L\ls A\ls S\ls E\ls N\ls A$^{2}$, \ns
R\ls O\ls G\ls E\ls R\ns S\ls M\ls I\ls T\ls H$^{1}$ \ns
 \and D\ls M\ls I\ls T\ls R\ls I\ns T\ls S\ls E\ls L\ls U\ls I\ls K\ls O$^{1}$ \ns
}

\affiliation{$^1$ Department of Mathematical Sciences, Loughborough University,\\ Loughborough, LE11 3TU, UK,\\
$^2$ Department of Chemical Engineering, Loughborough University,\\ Loughborough, LE11 3TU, UK,\\
$^3$ Mathematics Institute, Zeeman Building, University of Warwick, Coventry, CV4 7AL, UK,\\
$^4$ Mathematical Institute, Andrew Wiles Building, Radcliffe Observatory Quarter,\\ Woodstock Road, Oxford, OX2 6GG, UK,\\
$^5$ Department of Mathematics, Imperial College London, London, SW7 2AZ, UK.}

\pubyear{}
\volume{}
\pagerange{}
\date{?; revised ?; accepted ?. - To be entered by editorial office}

\DeclareFontFamily{OT1}{pzc}{}
\DeclareFontShape{OT1}{pzc}{m}{it}{<-> s * [1.10] pzcmi7t}{}
\DeclareMathAlphabet{\mathpzc}{OT1}{pzc}{m}{it}

\usepackage{amsmath}

\DeclareFontFamily{OT1}{pzc}{}
\DeclareFontShape{OT1}{pzc}{m}{it}{<-> s * [1.10] pzcmi7t}{}
\DeclareMathAlphabet{\mathpzc}{OT1}{pzc}{m}{it}

\begin{document}

\maketitle

\begin{abstract}
We study the deformation and dewetting of liquid films under impinging gas jets using experimental, analytical and numerical techniques. We first derive a reduced-order model (a thin-film equation) based on the long-wave assumption and on appropriate decoupling of the gas problem from that for the liquid. The model not only provides insight into relevant flow regimes, but is also used in conjunction with experimental data to guide more computationally prohibitive direct numerical simulations (DNS) of the full governing equations. A unique feature of our modelling solution is the use of an efficient iterative procedure in order to update the interfacial deformation based on stresses 
originating from computational data.We show that both gas normal and tangential stresses are equally important for achieving accurate predictions. The interplay between these techniques allows us to study previously unreported flow features. These include finite-size effects of the host geometry, with consequences for flow and vortex formation inside the liquid, as well as the specific individual contributions from the non-trivial gas flow components on interfacial deformation. Dewetting phenomena are found to depend on either a dominant gas flow or contact line motion, with the observed behaviour (including healing effects) being explained using a bifurcation diagram of steady-state solutions in the absence of the gas flow.

\end{abstract}

\begin{keywords}
%Authors should not enter keywords on the manuscript, as these must be chosen by the author during the online submission process and will then be added during the typesetting process 
%(see http://journals.cambridge.org/data/\linebreak[3]relatedlink/jfm-\linebreak[3]keywords.pdf for the full list)
\end{keywords}
\vspace{-0.5cm}
\section*{Introduction}

The process of impingement of a gas jet onto a liquid layer is important in numerous industrial applications. For example, it 
is used in steel production in the basic oxygen furnace process \cite[e.g.][]{turkdogan1996fundamentals,Hwang2012}, in coating applications 
in the gas-jet wiping process \cite[e.g.][]{Thornton1976,Lacanette_etal_2006} and in immersion lithography to remove water from a 
photoresist coated wafer \cite[e.g.][]{Berendsen_etal_2012,BERENDSEN2013505}. A closely related process is that 
of impingement of a gas plasma jet (instead of simply a gas jet) onto a layer of a liquid which appears, for example, 
in the arc welding process \cite[e.g.][]{Berghmans_1972}, in medical applications such as wound healing and skin treatment \cite[e.g.][]{Tian_2014,C7CP07593F} 
and in environmental applications such as water treatment and disinfection \cite[e.g.][]{Foster_2017}.

The impact of gas jets onto layers of liquids has been previously studied mainly for the case when the layer of the liquid is relatively thick.  A gas jet impinging onto a liquid layer exerts normal and tangential stresses on its surface, which result in its deformation creating a cavity 
and flow inside the liquid. Most of the previous research was focused on analysing the shape of the cavity and its stability. An early experimental study was performed by \cite{banks1963experimental}, who identified three regimes, namely, a steady cavity, an oscillating cavity and splashing. They focused on the analysis of steady cavities and suggested scaling approaches to establish a relation between the impact of the jet and the depth of the cavity. \cite{turkdogan1966fluid} carried out the \cite{banks1963experimental} experiments with liquids of different densities but focused on the effects of the gas nozzle diameter and the nozzle distance from liquid surface. \cite{cheslak1969cavities} performed an analysis similar to  \cite{banks1963experimental} and concluded that the occurrence of splashing or a smooth cavity depends on the jet velocity, while the viscosity of the liquid and surface tension are less important. \cite{molloy1970impinging} studied not only the effect of the gas jet on the cavity, but also the effect of the liquid properties. Previous analytical work investigating the shapes of steady cavities has been mainly based on a conformal mapping approach, in which the flow in the liquid is neglected and the system is assumed to be two dimensional, although both of the assumptions are clearly not valid in practice. The first analytical work using a conformal mapping method was done by \cite{olmstead1964depression}, who studied the cavity shape at relatively small gas velocities in the case of small cavity. \cite{vanden1981deformation} used a similar approach but solved the problem using a different numerical procedure, which allowed analysis of the system for larger gas velocities. A more recent analytical work based on a conformal mapping approach was done by \cite{he2010deformation}, who analysed the cavity shape without requiring it to be small. 
\cite{Mordasov2016} employed the balance equations for forces at the gas--liquid interface and not the balance equation for pressure as was used in most previous studies and obtained good agreement with experiments. Despite previous analytical approaches, detailed understanding of the cavity instability mechanisms is still missing. More recent work on gas jets impinging onto liquids has been mainly focused on experimental and DNS investigations \cite[see e.g.][]{nguyen2006computational,SOLORZANOLOPEZ20114991,Liu_etal_2015,munoz2012numerical,ADIB2018510}

There has been less investigation for the case when the layer of the liquid is relatively thin.  In such a case, if the gas jet flow is sufficiently strong, the film ruptures and dewetting is initiated. Gas-jet induced dewetting of thin liquid films was first considered by \cite{Berendsen_etal_2012,BERENDSEN2013505} both experimentally and using modelling,  
via a reduced-order thin-film equation. In \cite{Berendsen_etal_2012}, the authors focused on analysing the liquid film rupture times and the influence of surfactants and found good agreement between and experimental and modelling results. In \cite{BERENDSEN2013505}, the authors additionally analysed the effect of the movement of the gas jet. 

In the present study, we expand on a comprehensive theoretical and experimental investigation of the deformation and dewetting of (thin and moderately thin) liquid films in a cylindrical beaker under the influence of an impinging gas jet that is generated by maintaining a constant gas flow rate from a stationary cylindrical tube.  Our goal is two-fold. On the one hand, we aim to provide an improved theoretical characterisation of the interfacial deformation process 
that lies at the centre of the physical systems, both in terms of balance of forces and interaction with the surrounding flow fields in the liquid and gas phases.
On the other hand, we wish to examine challenging features related to dewetting dynamics that merit further understanding and mathematical description (here 
to be performed using a dynamical systems approach) as a prototypical case for more complex real-world scenarios. 
To obtain initial insight into relevant flow regimes and timescales of the system, we use a systematically derived thin-film equation that in the axisymmetric case coincides with the model of \cite{Berendsen_etal_2012}. The equation is obtained under the long-wave assumption,  that has been extensively used in the literature \cite[see e.g.][]{Thiele_2007,Craster_Matar_2009}, and by decoupling the problem for the gas from that for the liquid, under the so-called quasi-static assumption. This involves modelling the gas--liquid interface as a solid wall for the gas problem, which is valid when the typical velocity in the gas is much larger than that in the liquid, see e.g. \cite{tuck1975air} and also more recent work by  \cite{JFM2011} and \cite{JFM2015}. 
  Under such an assumption, the gas effects enter through the normal and tangential stresses exerted by the gas on the gas--liquid interface. In some previous studies, approximate expressions for these stresses were used that were typically constructed on the basis of general unbounded domains. Often, the gas influence was modelled via only the imposed gas normal stress, typically of a Gaussian form, ignoring the gas tangential stress \cite[see e.g.][]{Kriegsmann_etal_1998,Lunz_Howel_2018}, or via the imposed gas tangential stress, ignoring the gas normal stress \cite[see e.g.][]{Sullivan_etal_2008,Davis_etal_2010}. In our study, we incorporate the gas effects into the thin-film equation using detailed DNS for the gas phase. This allows us to obtain accurate functional expressions 
 for the gas normal and tangential stresses, and thus close the liquid problem and develop an accurate ``one-sided" model. We also use an iterative procedure for the computation of the gas normal and tangential stresses exerted onto the gas--liquid interface which 
 has two notable advantages: 1.~it produces significantly more accurate results by taking into account a realistically computed gas flow rather 
than an approximated prescribed formula and 2.~it allows the study of non-trivial geometrical settings (here finite-size effects) since the functional 
form of the stresses can now incorporate detailed nonlinear features which are otherwise difficult to predict.

The thin-film equation is built on the basis of experimental insight and developed in tandem with the more computationally expensive 
DNS for the full coupled system of the governing equations for our setup. 
We make use of two different packages, the CFD package in COMSOL \cite[see e.g.][]{pryor2011multiphysics} with a moving mesh interface 
and the volume-of-fluid $\mathpzc{Gerris}$ package \cite[see e.g.][]{popinet2}. These two  numerical methodologies offer distinct advantages and features improving 
our understanding of the system. DNS are used to estimate the range of validity of the reduced-order model and allow us, on the one hand, to access regimes 
that would be inaccessible with the reduced-order model and, on the other hand, to analyse flow characteristics that would be very difficult to image in the experiments 
reliably (e.g. the velocity field). In addition to studying dewetting, we also analyse the post-dewetting dynamics, when the flow of the gas is switched off. An insight into the expected behaviours for various parameter values is provided by a bifurcation diagram of steady state solutions for the system in the absence of 
the gas flow.

The manuscript is organised as follows: In \S~1 we describe our experimental setup. In~\S~2 we present the governing equations for the system and derive a reduced order model. Next, \S~3 explains our computational framework. In \S~4 we present and discuss the results. Finally, in \S~5 we give our conclusions. 

\section{Experimental setup}

\begin{figure}
\centering
\includegraphics[scale=0.75]{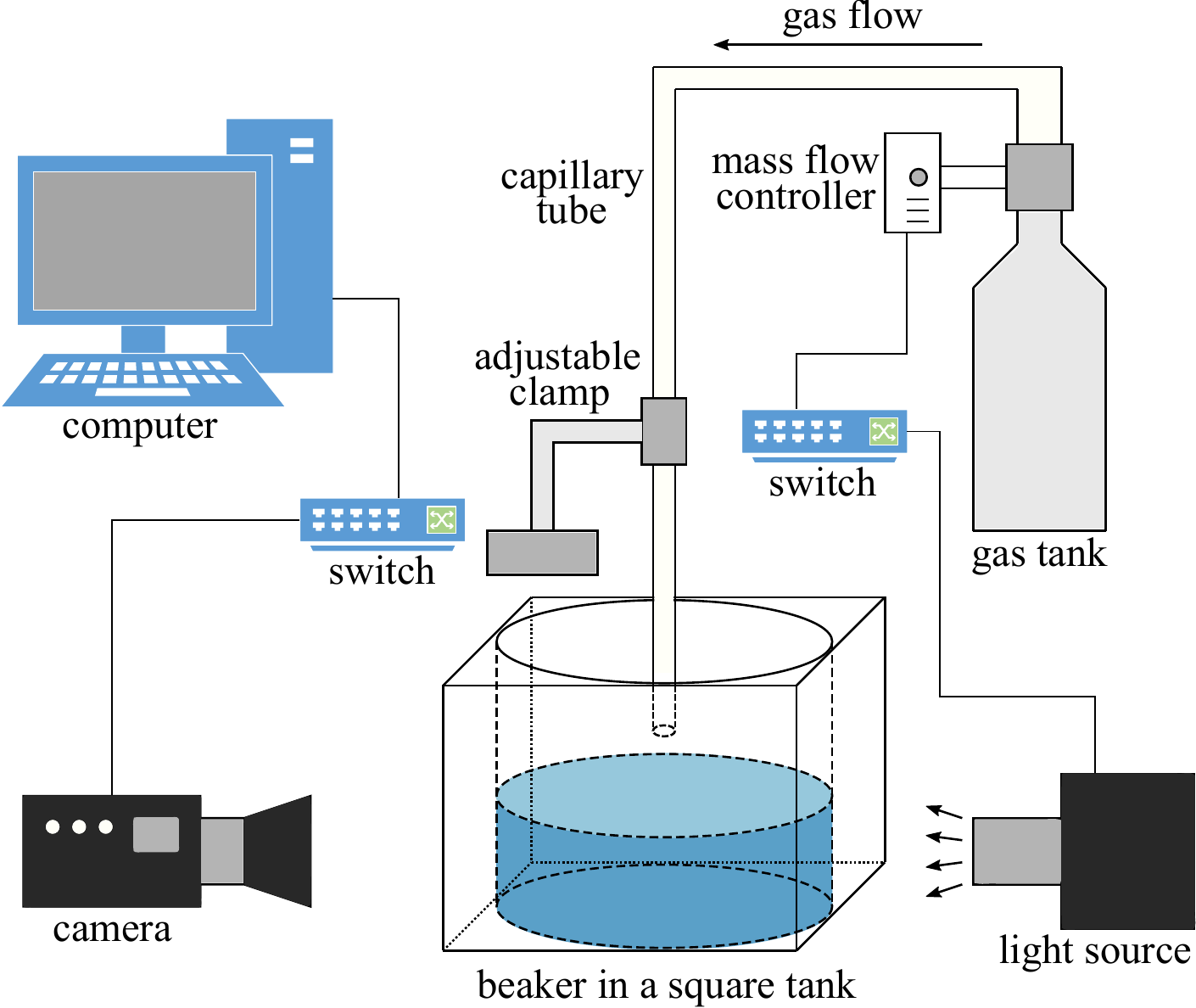}
\caption{Schematic representation of the experimental setup.}
\label{pre.setup1}

\vspace{1.3cm}
\centering
\includegraphics[width=0.85\textwidth]{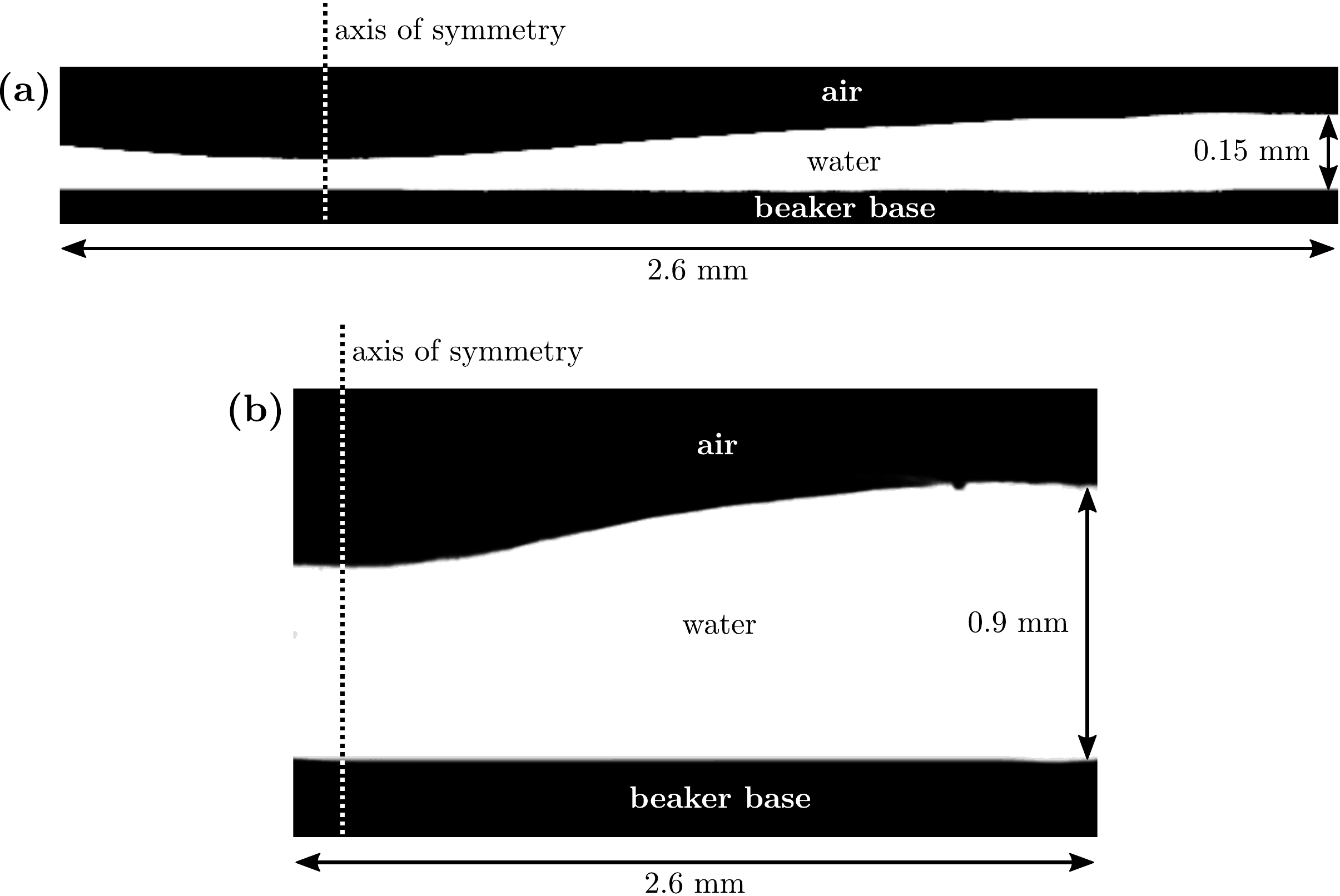}
\caption{Experimental results for (a) an air jet of flow rate $0.15\,\mathrm{slpm}$ impinging onto a film of water of thickness $0.2\,\mathrm{mm}$ and (b) an air jet of flow rate $0.5\,\mathrm{slpm}$ impinging onto a film of water of thickness $1\,\mathrm{mm}$. 
The images show regions of width $2.6\,\mathrm{mm}$.
}
\label{fig:experimental_example}
\end{figure}

A schematic representation of the experimental setup is shown in figure~\ref{pre.setup1}. We consider a layer of a liquid in a transparent cylindrical beaker  and we study the deformation of the surface of the liquid under the influence of an impinging gas jet. The beaker is $6\,\mathrm{cm}$ in height and $3\,\mathrm{cm}$ in diameter and it is placed in a transparent square tank to minimise the distortion of the image. The liquid used in the experiments is water at room temperature. The beaker is made of an acrylic polymer with the static contact angle of $30^\circ$ for water, see Appendix~\ref{appendix:contact_angle}. The gas jet is generated by maintaining a gas flow at a constant rate from a stationary cylindrical tube (nozzle) with its axis coinciding with the axis of the beaker. The inner diameter of the nozzle is $1.6\,\mathrm{mm}$. The nozzle is connected to a compressed gas tank, and the flow rate is controlled by a mass flow controller (MKS, PR4000B). The gas used in most of the experiments is air at room temperature. The nozzle is fixed with a clamp which can be moved to adjust the distance from the nozzle to the surface of the liquid. We typically consider a distance of $5\,\mathrm{mm}$. The position can be read from a movable calibrated scale.

A high-speed camera (Photron Fastcam, M2.1)  with the resolution $512\times512$ pixels and $2000$~fps coupled with a long-distance lens (Infinity, KC) is placed on one side of the square tank in order to record images of the deformed liquid layer.  The camera is connected to a computer to enable  gathering of the data for analysis. A light source (Kern Dual Fiber Unit LED) is placed on the opposite side of the square tank to provide illumination for the images. The camera is fixed on an adjustable $x$-$y$-$z$ stage allowing us to modify the camera's position properly and capture images 
in the beaker at different places. In particular, the initial position of the camera is adjusted by placing a graticule at the centre of an empty beaker and moving the camera along the stage until the image of the graticule comes into focus. This also allows measuring the size of the interrogation window. An example of an image of the graticule is included in the supplementary material. The recorded images were analysed with the software package ImageJ \cite[e.g.][]{ImageJ}. Examples of processed recorded images (with increased contrast) are shown in figure~\ref{fig:experimental_example} for a relatively thin water film (of undisturbed thickness $0.2\,\mathrm{mm}$) in panel (a) and a relatively thick water film (of undisturbed thickness $1\,\mathrm{mm}$) in panel (b). The corresponding raw images are included in the supplementary material. In panels (a) and (b), the water films were deformed by air jets of  flow rates $0.15$ and $0.5\,\mathrm{slpm}$  (standard litres per minute), respectively.  Given that the typical size of the interrogation window is $2.6\,\mathrm{mm}\,\times\,2.6\,\mathrm{mm}$ and the camera resolution is $512\times 512$ pixels, we can conclude that the error of the measurements is $O(10)$~$\mu$m.  A deformation of the gas-liquid interface can be clearly seen in both cases. The gas-liquid interface is constructed by curve fitting in the software package Matlab. 

\section{Mathematical model}

\subsection{Problem statement and full governing equations}
\label{sect:problem_statement}

\begin{figure}
\centering
\includegraphics[scale=0.6]{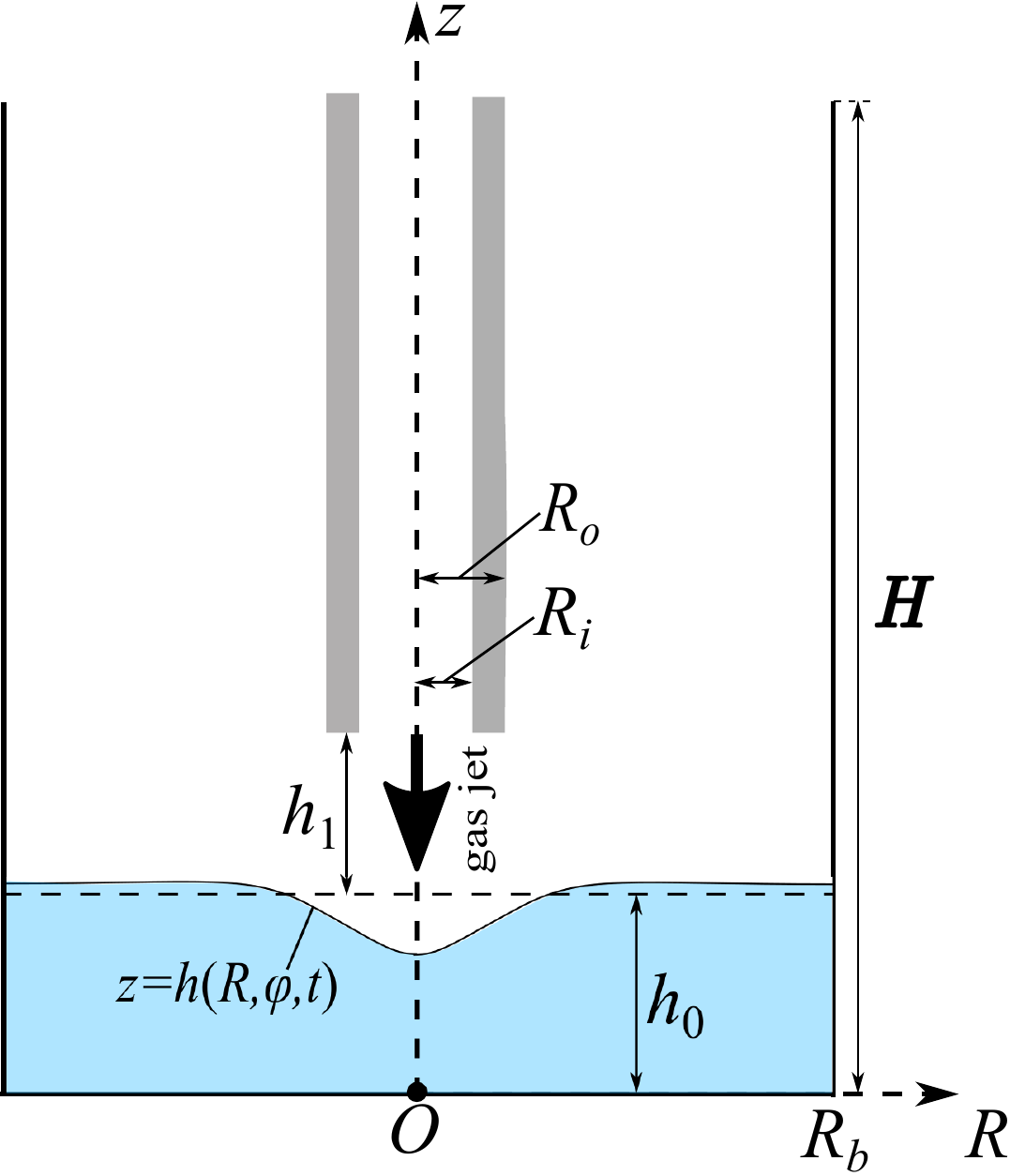}
\caption{Schematic representation of a gas jet impinging on\textcolor{blue}{to} the surface \textcolor{blue}{of} a liquid in a cylindrical beaker.}
\label{schematics1}
\end{figure}

A schematic representation of the model system is shown in figure~\ref{schematics1}. We denote the radius of the beaker by $R_b$ and its height by $H$. 
The thickness of the undisturbed liquid layer is denoted by $h_0$. 
The inner and outer radii of the nozzle are denoted by $R_i$ and $R_o$, respectively, and the distance between the nozzle and the undisturbed 
gas--liquid interface is denoted by $h_1$. We introduce cylindrical polar coordinates $(R,\varphi,z)$ with the $z$-axis pointing upwards along the axis of the 
beaker in the direction opposite to gravity $\boldsymbol{g}$, and with the bottom of the beaker coinciding with the $z=0$ plane. 
The deformed gas--liquid interface is denoted by $\Sigma$ and is given by the equation ${f(R,\varphi,z,t)=0}$. In the simplest case, the interface is given by the graph of a function, $z=h(R,\varphi,t)$, and hence $f=h-z$. We assume that the liquid and the gas are of the same constant temperature. 

As the gas jet strikes the liquid surface at the centre, the surface of the liquid deforms and a cavity appears. Motion in the liquid, in the form of eddies is also generated. The eddies affect the mass transfer and mixing in the liquid. In order to describe such a system, the Navier--Stokes and continuity equations are used and the corresponding boundary conditions must be satisfied. 

As the typical velocity in the liquid will be assumed to be small compared to the speed of sound in the liquid, it is appropriate to model the liquid problem by the incompressible Navier--Stokes equations:
\begin{equation}
\rho_l\frac{\mathrm{D}\zh{u}}{\mathrm{D} t}=\nabla\cdot\zh{\sigma}_l +\rho_l \zh{g},\qquad \nabla\cdot\zh{u}=0,
\label{NSleq}
\end{equation}
where $\rho_l$ is the liquid density, which is  assumed to be constant, $\zh{u}$ is the velocity field in the liquid, $D/Dt={\partial }/{\partial t}+\zh{u}\cdot\nabla$ is the usual material derivative, and 
$\zh{\sigma}_l$ is the viscous stress tensor for the liquid given by 
\begin{equation}
\zh{\sigma}_l=-p_l\zh{\delta}+2\mu_l\zh{S}_l,
\label{stress1}
\end{equation}
where $p_l$ is the pressure in the liquid, $\mu_l$ is the viscosity of the liquid, which is assumed to be constant, $\zh{\delta}$ is the identity tensor and $\zh{S}_l=\frac{1}{2}\bigl[(\nabla\zh{u})^T +\nabla\zh{u}\bigr]$ is the strain-rate tensor.

 In the present study, we consider gas velocities up to $50\,\mathrm{m}/\mathrm{s}$. For such gas velocities, compressibility effects may also be neglected in the gas phase. For such gas velocities it is still appropriate to assume that the gas flow is laminar. The incompressible Navier--Stokes equations in the gas then take the form
\begin{equation}
\rho_g\frac{\mathrm{D}\zh{v}}{\mathrm{D} t}=\nabla\cdot\zh{\sigma}_g +\rho_g \zh{g},\qquad 
\nabla\cdot\zh{v}=0,
\label{NSleq}
\end{equation}
where $\rho_g$ is the gas density, $\zh{v}$ is the gas velocity and $\zh{\sigma}_g$ is the viscous stress tensor for the gas given by 
\begin{equation}
\zh{\sigma}_g=-p_g\zh{\delta}+2\mu_g\zh{S}_g, 
\label{stress2}
\end{equation}
where $p_g$ is the gas pressure, $\mu_g$ is the viscosity of the gas, which is assumed to be constant, and $\zh{S}_g=\frac{1}{2}\bigl[(\nabla\zh{v})^T +\nabla\zh{v}\bigr]$ is the strain-rate tensor in the gas.

We impose no-slip and no-penetration conditions both for the liquid and for the gas at the solid boundaries ($\zh{u}=\zh{0}$ and $\zh{v}=\zh{0}$) except at the beaker side wall, where no-penetration still applies, but instead of no-slip, we impose the Navier slip condition to allow for the motion of the contact line \cite[see, e.g.,][]{sibley2012slip}.  So for the liquid when $R=R_b$  we have
\begin{equation}
\zh{u}\cdot\hat{\zh{t}}_w=\frac{\beta_{NS}^l}{\mu_l} \hat{\zh{n}}_w\cdot \zh{\sigma}_l\cdot\hat{\zh{t}}_w,
\label{eqslip_l}
\end{equation}
where $\hat{\zh{n}}_w$ is a unit normal vector to the side wall pointing into the beaker (i.e. $\hat{\zh{n}}_w=-\hat{\zh{R}}$, where $\hat{\zh{R}}$ is a unit vector pointing in the $R$ direction), $\hat{\zh{t}}_w$ is a unit tangent vector to the wall, $\beta_{NS}^l$ is the slip coefficient for the liquid.  
For the gas when $R=R_b$ we have
\begin{equation}
\zh{v}\cdot\hat{\zh{t}}_w=\frac{\beta_{NS}^g}{\mu_g} \hat{\zh{n}}_w\cdot \zh{\sigma}_g\cdot\hat{\zh{t}}_w,
\label{eqslip_g}
\end{equation}
where $\beta_{NS}^g$ is the slip coefficient for the gas.

Note that in (\ref{eqslip_l}) and (\ref{eqslip_g}), there are, in general, two independent tangent directions to the wall, i.e. each of the equations results in two scalar equations, by taking $\hat{\zh{t}}_w=\hat{\zh{k}}$ and $\hat{\zh{t}}_w=\hat{\zh{\varphi}}$, where $\hat{\zh{k}}$ and $\hat{\zh{\varphi}}$ are unit vectors pointing in the $z$
and $\varphi$ directions. However, we will later assume axisymmetry, i.e. no dependence on $\varphi$, and then it will be sufficient to only consider $\hat{\zh{t}}_w=\hat{\zh{k}}$.

In addition, we impose a fixed contact angle condition at the contact line, so that when the interface is given by $z=h(R,\varphi,t)$, we have
\begin{equation}
\frac{\partial h}{\partial R}=\tan(90^\circ-\theta_c)= \cot \theta_c,
\end{equation} 
when $R=R_b$ and $z=h(R_b,\varphi,t)$, where $\theta_c$ is the angle the liquid makes with the wall at the contact line.

In a subset of our numerical simulations presented below, at the bottom of the beaker we also impose the Navier slip condition instead of the no-slip condition. This allows us to study dewetting induced by the gas jet using DNS where topological transitions of the gas--liquid interface are allowed. We do this with the volume-of-fluid package $\mathpzc{Gerris}$ \cite[e.g.][]{popinet2}, as will be explained below. The contact angle at the bottom of the beaker will be denoted by $\theta_{eq}$. We also performed DNS in the CFD finite-element package COMSOL \cite[e.g.][]{pryor2011multiphysics}, and our implementation allows for mesh movements so that the mesh deformations follow the gas--liquid interface motion. Such an implementation allows us to analyse the deformations of the interface very accurately but forbids topological transitions. 

At the gas inlet, when $z=h_0+h_1$ and $0\leq R\leq R_i$, we impose the fully developed laminar Poiseuille velocity profile: 
\begin{equation}
\zh{v}=-v_\mathrm{max} \bigg(1-\frac{R^2}{R_i^2}\bigg)\hat{\zh{k}},
\label{eqn_inlet}
\end{equation}
where $v_\mathrm{max}=2q_g/\upi R_i^2$, with $q_g$ denoting the imposed gas flow rate. 

At the gas outlet, when $z=H$ and $R_o<R<R_b$, we impose normal flow, and prescribe normal stress, i.e. we require $\hat{\zh{k}}\cdot\zh{\sigma}_g\cdot\hat{\zh{k}}=-p_a$, where $p_a$ is the atmospheric pressure.

Finally, we discuss conditions that must be satisfied at the gas--liquid interface $\Sigma$. First, we have the kinematic condition
\begin{equation}
\frac{\mathrm{D}f}{\mathrm{D}t}=0,
\end{equation} 
where we remind that $f$ is a function such that $\Sigma$ is given by the equation ${f(R,\varphi,z,t)=0}$.
Continuity of velocity must also be satisfied at the interface, $\zh{u}=\zh{v}$, and we must have dynamic balance of stress at $\Sigma$:
\begin{equation}
\hat{\zh{n}}\cdot\zh{\sigma}_l-\hat{\zh{n}}\cdot\zh{\sigma}_g=\gamma\kappa\hat{\zh{n}}. 
\label{eq:stress_balance1}
\end{equation} 
Here, $\hat{\zh{n}}$ is the unit normal vector to the interface pointing into the liquid. The term on the right-hand side is due to the Laplace pressure, where $\gamma$ is the gas--liquid surface tension coefficient (which is assumed to be constant) and $\kappa=\nabla\cdot\hat{\zh{n}}$ is twice the mean curvature of the interface $\Sigma$. 

Note that to study dewetting induced by the gas jet in a numerical formulation where topological transitions are not allowed, we also include a Derjaguin (or  disjoining) pressure in the stress balance condition. This approach is applicable when $\Sigma$ is a graph of a function, $z=h(R,\varphi,t)$. The stress balance condition then becomes
\begin{equation}
\hat{\zh{n}}\cdot\zh{\sigma}_l-\hat{\zh{n}}\cdot\zh{\sigma}_g=\gamma\kappa\hat{\zh{n}}+\Pi(h)\hat{\zh{n}}. 
\label{eq:stress_balance1}
\end{equation}
The disjoining pressure represents an effective interaction between the gas--liquid interface and the liquid--substrate interface. It can be written as $\Pi(h)=-\mathrm{d} V(h)/dh$, where $V(h)$ is the so-called binding potential \cite[e.g.][]{de2013capillarity}. The disjoining pressure is assumed to be of the form \cite[e.g.][]{PISMEN200211,Galvagno_etal_2014}
\begin{equation}
\Pi(h)=-\frac{A}{h^3}+\frac{B}{h^6},
\label{eq:disj_p1}
\end{equation}
where the first term results from the long-range attractive forces (with $A$ representing the Hamaker constant) and the second term results from the short-range repulsive forces. The second term prevents the liquid film from breaking down, and instead of this occurring we obtain a very thin precursor film. In practice, where the film thickness is equal to the thickness of the precursor film, we may assume that a dry spot has appeared. At equilibrium, the thickness of the precursor film corresponds to the minimum of the binding potential $V(h)$ and is equal to 
\begin{equation}
h_{eq}=(B/A)^{1/3}, 
\label{eq:h_eq}
\end{equation}
with the contact angle at the apparent contact line is given by \cite[e.g.][]{Rauscher_Dietrich_2008,Hughes_etal_2015}
\begin{equation}
\theta_{eq}=\cos^{-1}\biggl(1+\frac{V(h_{eq})}{\gamma}\biggr).
\label{eq:theta_eq}
\end{equation}
Note that given the precursor thickness, $h_{eq}$, and the equilibrium contact angle, $\theta_{eq}$, the constants $A$ and $B$ can be recovered using relations (\ref{eq:h_eq}) and (\ref{eq:theta_eq}) given above.

%%%%%%%%%%%%%%%%%%%%%%%%%%%%%%%%%%%%%%%%%

\subsection{Thin-film model}

Solving the full system of governing equations is a computationally expensive task. We therefore aim to simplify the problem by deriving an accurate reduced-order model. Such a model not only provides insight into the fundamental features of the system in an efficient way but also serves as  a mechanism to guide the more computationally prohibitive DNS tools towards suitable regimes with a much more informed view of an otherwise vast parameter space. The first step is to decouple the problem for the gas from that for the liquid, which is possible when the typical velocity in the liquid is much smaller than that in the gas. Then, for the gas problem it is appropriate to neglect the motion of the liquid and to use the quasi-static assumption, i.e.  it is appropriate to model the interface as a rigid wall and solve the gas problem independently, see e.g. \cite{tuck1975air} who states that such an assumption is appropriate when the typical liquid velocity is less than approximately $4\%$ of the typical gas velocity, which is always the case in our study  \cite[see also][for more recent studies, where the quasi-static assumption was used in the modelling of a liquid film sheared by a turbulent gas]{JFM2011,JFM2015}.

The solution of the gas problem can then be used to obtain the stress  exerted by the gas onto the gas--liquid interface, which can then be fed into the normal and tangential stress balance conditions at the interface.  We  denote such a stress by $\zh{s}_g$ so that 
\begin{equation}
\zh{s}_g=-\zh{n}\cdot\zh{\sigma}_g.
\end{equation}
For the analysis in this section, we first consider the general non-axisymmetric case, and for convenience we use Cartesian coordinates $(x,y,z)$ (so that the $z$ direction remains as before. 

We non-dimensionalise the equations using $h_0$ as the length scale, $U_0$ as the velocity scale (to be specified later), $h_0/U_0$ as the time scale and $\mu_l U_0/h_0$ as the scale for pressure and the gas stress, so from now on all the variables will be assumed to be dimensionless. We thus introduce dimensionless variables via the following mappings:
\begin{eqnarray}
&\displaystyle (x,y,z,h)\mapsto h_0 (x,y,z,h),    \qquad  (u,v,w)\mapsto U_0 (u,v,w), &\\
&\displaystyle  t\mapsto \frac{h_0}{U_0} t,\qquad (p_l,\zh{s}_g)\mapsto  \frac{\mu_l U_0}{h_0} (p_l,\zh{s}_g),&
\end{eqnarray}
where we denote by $u$, $v$ and $w$ the $x$, $y$ and $z$ components of the velocity, respectively.

The incompressible Navier--Stokes and continuity equations in the liquid become
\begin{eqnarray}
Re(u_t+uu_x+vu_y+wu_z)&=&-p_{lx}+u_{xx}+u_{yy}+u_{zz},\\
Re(v_t+uv_x+vv_y+wv_z)&=&-p_{ly}+v_{xx}+v_{yy}+v_{zz},\\
Re(w_t+uw_x+vw_y+ww_z)&=&-p_{lz}+w_{xx}+w_{yy}+w_{zz}-G,\\
u_x+v_y+w_z&=&0,
\end{eqnarray}
where $Re$ and $G$ are the Reynolds and the gravity numbers, respectively, given by
\begin{equation}
Re=\frac{\rho_l U_0 h_0}{\mu_l},\qquad G=\frac{\rho_l g h_0^2}{\mu_l U_0}.
\end{equation}  
The no-slip and no-penetration conditions at the bottom of the beaker become
\begin{equation}
u=0,\quad v=0,\quad w=0\quad\text{at}\quad z=0.
\end{equation}
The kinematic condition at the interface, $z=h(x,y,t)$, takes the form
\begin{equation}
w=h_t+uh_x+vh_y.
\end{equation}
The normal stress balance condition takes the form:
\begin{multline}
-p_l+\frac{2}{1+h_x^2+h_y^2}\big[h_x^2u_x+h_y^2v_y+w_z+h_xh_y(u_y+v_x)-h_x(u_z+w_x)-h_y(v_z+w_y)\big]\\
=\frac{1}{Ca} \frac{(1+h_x^2)h_{yy}-2h_xh_yh_{xy}+(1+h_y^2)h_{xx}}{(1+h_x^2+h_y^2)}-N_s+\overline{\Pi}(h),
\end{multline}
at $z=h(x,y,t)$, where  $Ca$ is the Capillary number given by
\begin{equation}
Ca=\frac{\mu_lU_0}{\gamma},
\end{equation}
$N_s$ is the dimensionless normal stress exerted by the gas on the gas--liquid interface
\begin{equation}
N_s=\zh{s}_g\cdot\hat{\zh{n}},
\end{equation}
which under the quasi-static assumption can be assumed to be a functional of the interface shape, $N_s=N_s[h]$, and finally, $\overline{\Pi}(h)$ is the dimensionless disjoining pressure, given by
\begin{equation}
\overline{\Pi}(h)=-\frac{\overline{A}}{h^3}+\frac{\overline{B}}{h^6},
\end{equation}
where $\overline{A}=A/\mu_lU_0h_0^2$ and $\overline{B}=B/\mu_lU_0h_0^5$.

Taking $\hat{\zh{t}}=\hat{\zh{t}}_1={(1, 0, h_x)}/{\sqrt{1+h_x^2}}$ and $\hat{\zh{t}}=\hat{\zh{t}}_2={(0, 1, h_y)}/{\sqrt{1+h_y^2}}$ in the tangential stress balance condition, we obtain 
\begin{equation}
\frac{2h_x(u_x-w_z)+(h_x^2-1)(u_z+w_x)+h_y(u_y+v_x)\\+h_xh_y(v_z+w_y)}{[(1+h_x^2+h_y^2)(1+h_x^2)]^{1/2}}=-T_{s1},
\end{equation}
\begin{equation}
\frac{2h_y\big(v_y-w_z\big)+\big(h_y^2-1\big)\big(v_z+w_y\big)+h_x\big(u_y+v_x\big)\\+h_xh_y\big(u_z+w_x\big)}{[\big(1+h_x^2+h_y^2\big)\big(1+h_x^2\big)]^{1/2}}=-T_{s2}
\end{equation}
at $z=h(x,y,t)$, where $T_{si}$, $i=1,2$, are the $\hat{\zh{t}}_1$ and $\hat{\zh{t}}_2$ components of the tangential stress exerted by the gas on the gas--liquid interface, $T_{si}=\zh{s}_g\cdot\hat{\zh{t}_i}$,  
which under the quasi-static assumption can be assumed to be functionals of the interface shape, $T_{si}=T_{si}[h]$. The  tangential stress exerted by the gas on the gas--liquid interface is then expressed as 
\begin{equation}
\zh{T}_{s}=T_{s1}\,\hat{\zh{t}}_1+T_{s2}\,\hat{\zh{t}}_2.
\end{equation}

Next, we utilise the so-called thin-film or long-wave approximation, namely, we assume that the undisturbed film thickness, $h_0$, is much smaller than the characteristic horizontal length scale $\ell$ over which variations in the film thickness occur, and we introduce the so-called thin-film parameter $\epsilon=h_0/\ell\ll1$. We now use the following additional rescalings of variables that are standard for the thin-film approximation:
\begin{equation}
x=\frac{1}{\epsilon}\xi, \quad y=\frac{1}{\epsilon}\eta, \quad t=\frac{1}{\epsilon}\tau, \quad w={\epsilon}W,\quad p_l=\frac{1}{\epsilon}P_l.
\end{equation}
To derive the thin-film equation, we consider the asymptotic limit $\epsilon\rightarrow 0$. Then, to keep capillary effects at leading order, we assume that $Ca$ is asymptotically bounded above and below by $\epsilon^3$. To neglect inertia at leading order, we assume that $Re\ll 1/\epsilon$. We also assume that $G=O(1/\epsilon)$, so that gravitational effects may enter at leading order. For the disjoining pressure, it is appropriate to assume that $\overline{\Pi}=O(1/\epsilon)$.  
In addition, for the dimensionless gas stress, we need to assume that $N_s=O(1/\epsilon)$ and $T_{si}=O(1)$, $i=1,2$. We then introduce the following rescaled parameters:
\begin{equation}
\widetilde{C}a=\frac{Ca}{\epsilon^3}, \qquad \widetilde{G}=\epsilon G,
\end{equation}
so that $\widetilde{C}a$ is asymptotically bounded above and below by non-zero constants and $\widetilde{G}=O(1)$, and the following rescaled gas normal stress:
\begin{equation}
\widetilde{N}_s=\epsilon N_s,
\end{equation}
so that $\widetilde{N}_s=O(1)$, and the rescaled disjoining pressure:
\begin{equation}
\widetilde{\Pi}=\epsilon \overline{\Pi},
\end{equation}
so that $\widetilde{\Pi}=O(1)$.

The problem at leading order becomes:
\begin{eqnarray}
u_{zz}=P_{l\xi},\qquad v_{zz}=P_{l\eta},\qquad P_{lz}=-\widetilde{G}, \qquad u_\xi+v_\eta+W_z=0,
\end{eqnarray}
with the no-slip and no-penetration conditions
\begin{equation}
u=v=W=0
\end{equation}
at $z=0$ and the tangential and normal stress balances
\begin{equation}
u_z=T_{s1}, \qquad v_z=T_{s2}, \qquad P_l=\widetilde{N}_s - \widetilde{\Pi}(h) - \frac{1}{\widetilde{C}a}(h_{\xi\xi}+h_{\eta\eta})
\end{equation}
at $z=h(\xi,\eta,\tau)$. We also have the kinematic condition, which can be rewritten as 
\begin{equation}
h_\tau+\nabla\cdot\zh{q}=0,
\label{eq:kinematic_flux}
\end{equation}
where $\nabla=(\partial/\partial \xi, \partial/\partial \eta)$ and
\begin{equation}
\zh{q}=\bigg(\int_0^hu\,\mathrm{d}z,\int_0^hv\,\mathrm{d}z\bigg) 
\label{eq:flux}
\end{equation}
is the flux vector parallel to the plane $z=0$.

Then we find that the pressure at leading order is given by 
\begin{equation}
P_l=-\widetilde{G}(z-h)+\widetilde{N}_s-\widetilde{\Pi}(h)-\frac{1}{\widetilde{C}a}\nabla^2 h,
\label{eq:leang_Pl}
\end{equation}
and the velocity components at leading order are given by
\begin{eqnarray}
&& (u, v)=\bigg(\frac{z^2}{2}-hz\bigg)\nabla P_l+z\,\zh{T}_s,\\
&& w=-\bigg(\frac{z^3}{6}-\frac{z^2h}{2}\bigg)\nabla^2 P_l+\frac{z^2}{2}\nabla P_l\cdot\nabla h -\frac{z^2}{2}\nabla\cdot \zh{T}_s.
\end{eqnarray}
Substituting the leading-order expressions for $u$ and $v$ into the expression for the flux vector (\ref{eq:flux}), we find that at leading order
\begin{equation}
\zh{q} =-\frac{h^3}{3}\nabla P_l+\frac{h^2}{2}\zh{T}_s.
\label{eq:flux2}
\end{equation}
Substituting this expression into the kinematic condition (\ref{eq:kinematic_flux}) gives the following evolution equation for the film thickness, the so-called thin-film equation:
\begin{equation}
h_\tau+\nabla\cdot\bigg(-\frac{h^3}{3}\nabla P_l+\frac{h^2}{2}\zh{T}_s\bigg)=0.
\label{eq:th_eq1}
\end{equation}
Scaling back to the dimensionless variables $x$, $y$ and $t$, we obtain
\begin{equation}
h_t+\nabla\cdot\bigg(-\frac{h^3}{3}\nabla p_l+\frac{h^2}{2}\zh{T}_s\bigg)=0,
\label{eq:th_eq2}
\end{equation}
where the dimensionless leading-order pressure is given by 
\begin{equation}
p_l=-G(z-h)+N_s-\overline{\Pi}(h)-\frac{1}{Ca}\nabla^2 h
\label{eq:leang_pl_1}
\end{equation}
and $\nabla=(\partial/\partial x, \partial/\partial y)$. 

For convenience, it is possible to eliminate one of the dimensionless parameters by, for example, multiplying the thin-film equation by $Ca$ and appropriately rescaling time. This is equivalent to choosing $U_0=\gamma/\mu_l$, for which the time scale becomes $\mu_l h_0/\gamma$ and the scale for pressure and the gas stress becomes $\gamma/h_0$. Then the pressure in the thin-film equation takes the form
\begin{equation}
p_l=-Bo(z-h)+N_s-\overline{\Pi}(h)-\nabla^2 h,
\label{eq:leang_pl}
\end{equation}
where $Bo$ is the Bond number given by
\begin{equation}
Bo=G\,Ca=\frac{\rho_lgh_0^2}{\gamma},
\label{eq:Bond_number}
\end{equation}
and the dimensionless coefficients in the disjoining pressure are $\overline{A}=A/\gamma h_0^2$ and $\overline{B}=B/\gamma h_0^5$.

For the validity of the thin-film equation in terms of dimensionless parameters that are independent of $U_0$, we must have $Bo=O(\epsilon^2)$ and $La\ll1/\epsilon^4$, where $La$ is the Laplace number given by
\begin{equation}
La=\frac{Re}{Ca}=\frac{\gamma \rho_l h_0}{\mu_l^2}.
\end{equation}
The latter condition is needed for inertia to be negligible. We must additionally have $N_s=O(\epsilon^2)$ and $T_{si}=O(\epsilon^3)$, $i=1,2$. The validity of the thin-film equation for the experimental parameter values that we have used is discussed \S~\ref{sect:results}.

Finally, going back to cylindrical polar coordinates $(R,\varphi,z)$ and assuming axisymmetry, we obtain the following equation:
\begin{equation}
h_t+\frac{1}{R}\bigg[-\frac{R h^3}{3}p_{lR}+\frac{Rh^2}{2}T_s\bigg]_R=0,
\label{eq:th_eq_axisymm}
\end{equation}
where $T_s$ denotes the gas tangential stress in the $R$-direction and pressure is given by
\begin{equation}
p_l=-Bo(z-h)+N_s-\overline{\Pi}(h)-\frac{1}{R}(Rh_R)_R.
\end{equation}
Note that here $R$ is assumed to be non-dimensionalised using $h_0$ as the length scale. To solve the thin-film equation (\ref{eq:th_eq_axisymm}) numerically, we also need to impose appropriate boundary conditions. Conditions at $R=0$ follow from the symmetry assumption:
\begin{equation}
h_R=h_{RRR}=0\qquad\text{at}\qquad R=0.
\end{equation}
At the side wall, we will assume for simplicity that the contact angle is $90^\circ$, so that
\begin{equation}
h_R=0\qquad\text{at}\qquad R=\overline{R}_b,
\end{equation}
where $\overline{R}_b=R_b/h_0$, and we will impose zero flux in the $R$ direction, so that
\begin{equation}
q\equiv-\frac{h^3}{3}p_{lR}+\frac{h^2}{2}T_s=0\qquad\text{at}\qquad R=\overline{R}_b.
\end{equation}

For analysing flow patterns in the liquid film, it is also useful to give the $R$ and $z$ velocity components in cylindrical polar coordinates:
\begin{eqnarray}
&& u^R=p_{lR}\bigg(\frac{z^2}{2}-hz\bigg)+T_s\, z,\\
&& w=-\frac{1}{R}(R\,p_{lR})_R\bigg(\frac{z^3}{6}-\frac{h\,z^2}{2}\bigg)+\frac{p_{lR}\,h_R\,z^2}{2} -\frac{1}{2R}(R\,T_s)_R\, z^2.
\end{eqnarray}
 
Note that the main model equation~(\ref{eq:th_eq_axisymm}) provides a highly efficient route towards studying mechanistic aspects of the flow and generating an understanding of the underlying physics, thus forming a valuable part of our methodology toolkit.

To close the thin-film model, we need to specify stress contributions $N_s$ and $T_s$. There are different possible approaches for incorporating these. 
As mentioned in the introduction, often approximate  
expressions for these stresses were used that were typically constructed on the basis of general unbounded domains. For example, \cite{Kriegsmann_etal_1998} and \cite{Lunz_Howel_2018} assumed a Gaussian form for the normal stress and completely ignored the shear stress. We choose to integrate the gas effects into the the thin-film equation more carefully by means of exploiting detailed knowledge of the flow field in the gas 
extracted from DNS. In the first instance and for comparison purposes, we suggest computationally informed functional expressions for $N_s$ and $T_s$, thus developing an accurate ``one-sided" model. 
However, more generally, we also introduce and utilise an iterative numerical procedure for computing $N_s$ and $T_s$, which provides an accurate update mechanism that is computationally inexpensive and powerful in the context of predictive modelling, especially in the context of finite-size effects generated by the presence of the lateral walls of the beaker.

\section{Computational framework}

Complementing the experimental and analytical investigations we also considered two distinct numerical platforms to simulate the
 unsimplified and fully coupled Navier--Stokes and continuity equations in both liquid and gas phases within the 
target physical system.
The two packages (described in more detail in the paragraphs to follow) offer distinct advantages and features that aid our understanding of the flow characteristics. 
They act not only to bridge the gap between the previous approaches, but also to access regimes that would be inaccessible with a reduced-order model approach 
on the one hand, as well as easily allow the inspection of quantities in the flow that would be very difficult to image reliably on the other.

First, we implemented the setup in the commercial software platform COMSOL Multiphysics 5.3a. We used the CFD module which is a standard tool to simulating systems that involve complex fluid flow models.  A two-dimensional axisymmetric geometry was built using the parameters from the experiments.  COMSOL uses an unstructured mesh finite-element approach, which is highly suitable for tracking details near specific target regions of the domain. However, for the present problem, we found it challenging and computationally highly expensive to accurately describe the evolution of the gas--liquid interface and topological transitions, occurring e.g. in the dewetting process, using the built-in level-set and phase-field methods. Specific difficulties were encountered in conserving volume. Addressing this would require a prohibitively large number of degrees of freedom even with a powerful machine. We thus utilised a more computationally efficient moving-mesh approach in which the gas-liquid interface is modelled as a sharp surface separating the two phases and the mesh deformations follow the deformations of the interface. However, such an implementation is not directly suitable for describing topological transitions such as in the dewetting process 
(as this functionality is not available in COMSOL altogether for the moving-mesh formulation). Thus, as discussed in \S~\ref{sect:problem_statement}, we included the disjoining pressure into the normal stress balance condition to study dewetting which prevents the liquid film from breaking down so that a dry spot is modelled with a very thin precursor film. 
We should note that this approach still has limitations in modelling dewetting. 
For example, it is not suitable for describing dewetting on hydrophobic surfaces for which the contact angle is greater than $90^\circ$.

To overcome the limitations of our COMSOL implementation with respect to topological transitions, 
we also implemented the setup in the open-source package $\mathpzc{Gerris}$ \cite[e.g.][]{popinet2}. Well-known in the interfacial flow community for more than a decade, its strengths lie in the adaptive mesh refinement and parallelisation capabilities that make it an ideal testbed for multi-scale flow problems. The transparent structure of the code allows for careful validation of any in-house implemented 
extensions, as have been employed here. For example, one particular region of interest in the flow is the near-wall region where dewetting can be  
considered without the need to introduce a precursor film. The interface-capturing techniques, coupled with well-established contact line models \cite[e.g.][]{Afkhami} 
and control over any imposed Navier-slip-type conditions, provide an added perspective to the overall investigation.
The chosen refinement strategy concentrates on adequately addressing the sensitive regions near the gas nozzle and the walls in contact with the liquid,
while adaptive refinement is used to steer degrees of freedom towards any changes in interfacial position, as well as changes in components 
of the velocity field and vorticity
in order to accurately capture non-trivial flow regions in both the liquid and the gas.

The runs in the sections to follow have been executed in parallel on local computing facilities,
typically amounting to $O(10^3)$ CPU hours in $\mathpzc{Gerris}$, depending on flow parameters. While the chosen adaptive strategy restricts 
the number of grid nodes to $O(10^5)$, a gain of two orders of magnitude over a uniform grid with the same minimum cell size, the delicate 
interplay between the gas--liquid coupling requires special measures from a linear algebra and stability viewpoint, leading to 
relatively small time steps in  order to ensure for mesh-independent results.  The Courant--Friedrichs--Lewy (CFL)-limited procedure, 
alongside criteria based on surface tension effects (the timestep needs to be smaller than the period of 
the shortest capillary wave) and viscosity lead to a timestep $\Delta t = O(10^{-4})\,\mathrm{s}$ during the entire evolution of the flow. 
By contrast, we found that the computations in COMSOL 
were significantly faster for a similar number of degrees of freedom (we typically used $O(10^5)$ triangular elements), with a 
typical computation for the fully coupled problem taking approximately $O(10^2)$ CPU hours. 
 This is due to a relaxation of the timestep tolerance in COMSOL that allows it to increase to $\Delta t = O(10^{-1})\,\mathrm{s}$, 
thus requiring significantly fewer iterations of the underlying large scale solver. While resulting in a welcome speedup, it is unclear 
whether this less conservative strategy adopted by the commercial software could in principle cope with unexpected changes in the flow conditions.
Ultimately, even taking the runtime advantages into account, we have already noted above some 
of the restrictions on the study of dewetting phenomena, which is one of the central topics of the present investigation. Thus the functionalities 
of each of the two packages become complementary and contribute towards a versatile computational framework for our setup.

We have validated both implementations extensively by systematically decreasing the cell size or increasing 
the number of mesh elements until the convergence in the numerical results was achieved.  Quantitative measurements (norm-based estimates of both converged and dynamic features of the flow) have underpinned 
the verifications of such grid-size studies and have guided us in designing appropriate temporal and spatial adaptivity strategies. 
Comparisons to analytical and experimental data have been used as reference when possible, while the 
most expensive (and accurate) calculations in regimes outside the reach of the other techniques have been used otherwise.
We have thus established a robust and mesh-independent
solution strategy for the numerical results presented in the following section.

\section{Results}
\label{sect:results}

Throughout this section, we consider the following geometrical parameters in both the mathematical model and in the numerical simulations: the diameter of 
the beaker is $3\,\mathrm{cm}$ and its height is $6\,\mathrm{cm}$; the inner diameter of the gas nozzle is $1.6\,\mathrm{mm}$ and its distance from the undisturbed 
liquid surface is $5\,\mathrm{mm}$, as in the experiments. Also, the liquid is water and the the gas is air at room temperature. 
The density and viscosity of water at room temperature are $\rho_l=1000\,\mathrm{kg}\,\mathrm{m}^{-3}$ and $\mu_l=8.9\times10^{-4}\,\mathrm{Pa}\,\mathrm{s}$, respectively. 
For the density and viscosity of air, 
we use $\rho_g=1.22\,\mathrm{kg}\,\mathrm{m}^{-3}$ and $\mu_g=1.81\times 10^{-5}\,\mathrm{Pa}\,\mathrm{s}$. 
The surface tension coefficient for the air-water interface is set to $\gamma=72\times10^{-3}\,\mathrm{N}\,\mathrm{m}^{-1}$.
Regarding the results obtained with the thin-film equation (\ref{eq:th_eq_axisymm}), for the asymptotic validity of the equation $Bo\ll 1$ is required and also $La\ll 1/Bo^2$. It can be verified that for a water film $La$  becomes smaller than $1/Bo^2$ if $h_0<0.226\,\mathrm{mm}$ (and then $Bo<0.007$). Thus, strictly speaking for the validity of the thin-film model the thickness of the film must be less than $\approx 0.23\,\mathrm{mm}$. As regards gas flow rates suitable for the validity of the thin-film equation, we note that for a film of thickness $0.226\,\mathrm{mm}$ and an air jet flowing at the rate $0.2\,\mathrm{slpm}$, the maximum values of the normal and tangential stresses non-dimensionalised with $\gamma/h_0$ turn out to be $0.007$ and $0.0008$, respectively, which is appropriate for  
the validity of the model. For an air jet flowing at the rate $0.4\,\mathrm{slpm}$, the maximum values of the dimensionless normal and tangential stresses  
turn out to be $0.028$ and $0.0025$, respectively, which is also  
close to the region of the validity of the model. However, it will be shown below that the thin-film equation turns out to produce good agreement with DNS and experiments also for film thicknesses significantly larger than $0.226\,\mathrm{mm}$ (of $O(1)\,\mathrm{mm}$) and for gas flow rates significantly higher than $0.4\,\mathrm{slpm}$ (of $O(1)\,\mathrm{slpm}$). For even thicker water films and higher gas flow rates, inertial effects become important and the derived thin-film equation is not valid. Such parameter regimes can be accessed with the developed DNS framework. 

\subsection{Decoupled gas problem}
\label{sect:gas_only}

\begin{figure}
\centering
\includegraphics[height=4.85cm]{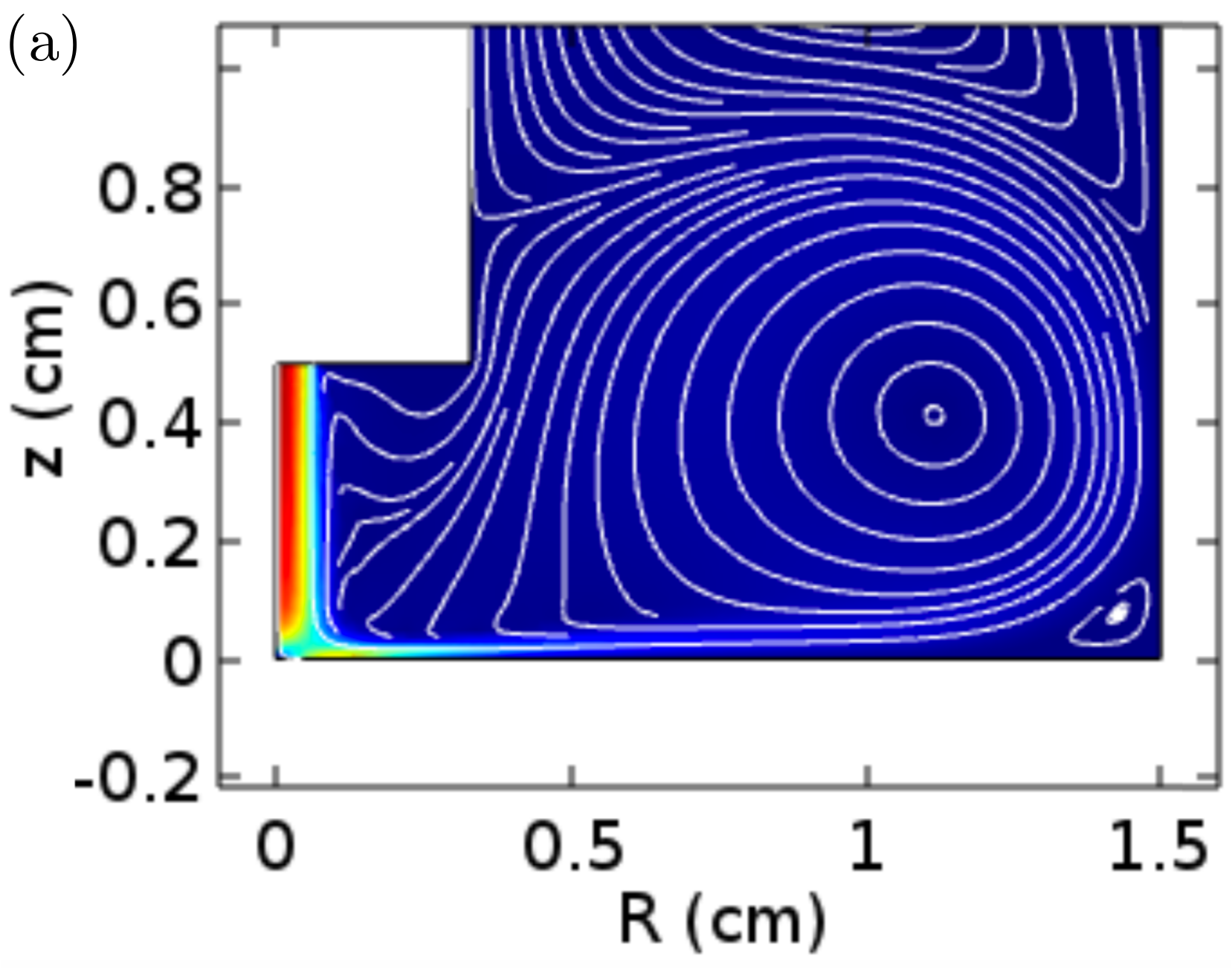}\,\,\,
\includegraphics[height=4.85cm]{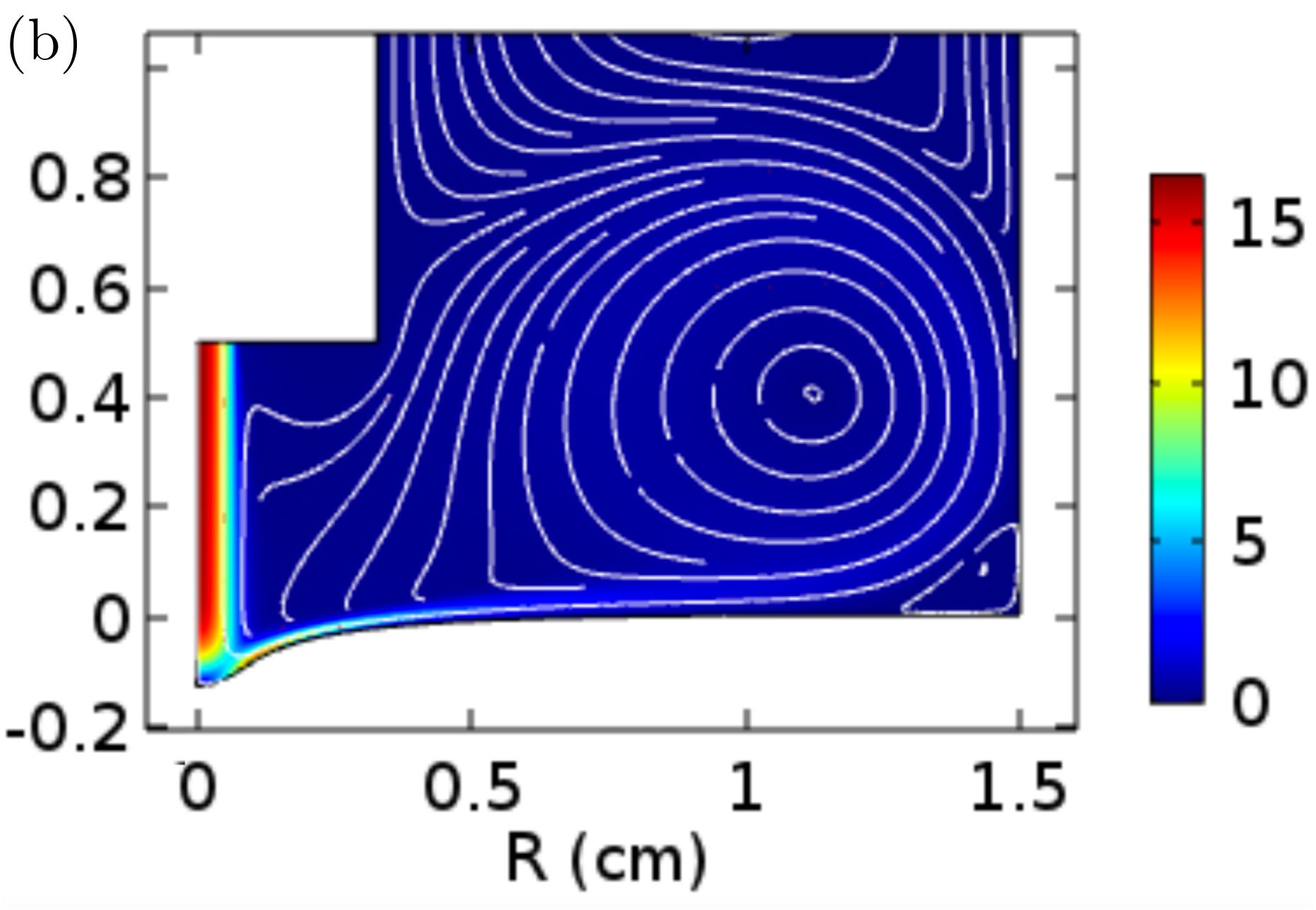}
\caption{Numerical solutions at steady state of the decoupled gas problem for the gas flow rate $q_g=1\,\mathrm{slpm}$ obtained using COMSOL. 
The colour scheme illustrates the gas speed in metres per second as indicated in the colour bars and 
the white thin lines show streamlines. (a) Gas jet impinging onto the interface modelled as a flat wall. (b) Gas jet impinging onto the interface modelled as a deformed wall, with the shape of the deformation obtained by solving the problem for the liquid film using the thin-film equation (\ref{eq:th_eq_axisymm}) with the gas stresses obtained from the solution shown in panel (a), when the interface is modelled as a flat wall.}
\label{fig:gas_only1}
\end{figure}

First, we discuss the decoupled gas problem and explain how the gas normal and tangential stresses exerted on the interface are computed using an iterative procedure.  
Under the quasi-static assumption, we model the liquid surface as a solid wall obtaining a problem for the gas only, and initially we assume that the interface is flat. The gas flow rapidly develops into a steady state. The resulting gas flow pattern at steady state obtained using COMSOL is shown in figure~\ref{fig:gas_only1}(a) for the gas flow rate $q_g = 1\, \mathrm{slpm}$, which corresponds to the maximum gas speed of approximately $16\,\mathrm{m}\,\mathrm{s}^{-1}$. In this section, we assume that the undisturbed interface is located at $z=0$. We note that $\mathpzc{Gerris}$ simulations agree with the COMSOL results. 
The colour scheme indicates the gas speed in metres per second and the thin white lines show streamlines. We can observe that the gas jet impinges onto the lower wall, and then the gas flows in the direction parallel to the wall radially outwards with its speed decreasing as the radial distance increases. We can also observe that a relatively large recirculation zone (eddy) is generated in the gas, and there is also a small, relatively slow eddy in the bottom-right corner. This solution of the gas problem for the interface modelled as a flat solid wall is used to compute the normal and tangential stresses exerted by the gas jet on the interface at different radial locations. Next, we use these stresses in the thin-film equation~(\ref{eq:th_eq_axisymm}) to solve the problem for the liquid film and therefore obtain the deformation of the interface resulting from these stresses. The thin-film equation is solved numerically in Matlab using finite-difference approximations for the spatial derivatives and Matlab's {\it ode15s} solver for stepping in time. For this gas flow rate, the solution evolves into a steady state within a few seconds. An example of a deformed interface computed in this way is shown in figure~\ref{fig:gas_only1}(b). A liquid film of thickness $5\,\mathrm{mm}$ was used for illustrative purposes (although we note that the thin-film equation is not expected to be valid for such a thickness). Next, we use this deformed interface as the lower boundary for the gas domain and again assume that it is a solid wall and recompute the solution of the decoupled gas problem. It can be seen in  figure~\ref{fig:gas_only1}(b) that the computed solution is qualitatively similar to the one in  figure~\ref{fig:gas_only1}(a). We extract the gas stresses from this solutions and use these updated stresses to solve the thin-film equation again to obtain an updated steady-stated interface shape. This procedure is repeated until a converged  steady-state interface shape is achieved. 
Typically, we find that for the relatively thin films considered in the present study 2--3 iterations are sufficient 
to obtain a converged profile to within graphical accuracy. It is expected that for thicker liquid films and stronger interfacial deformations more iterations might be needed. Other fluid systems with properties further away from the present modelling assumptions are also anticipated to give rise to a more challenging convergence procedure.

\begin{figure}
\centering
\includegraphics[scale=0.28]{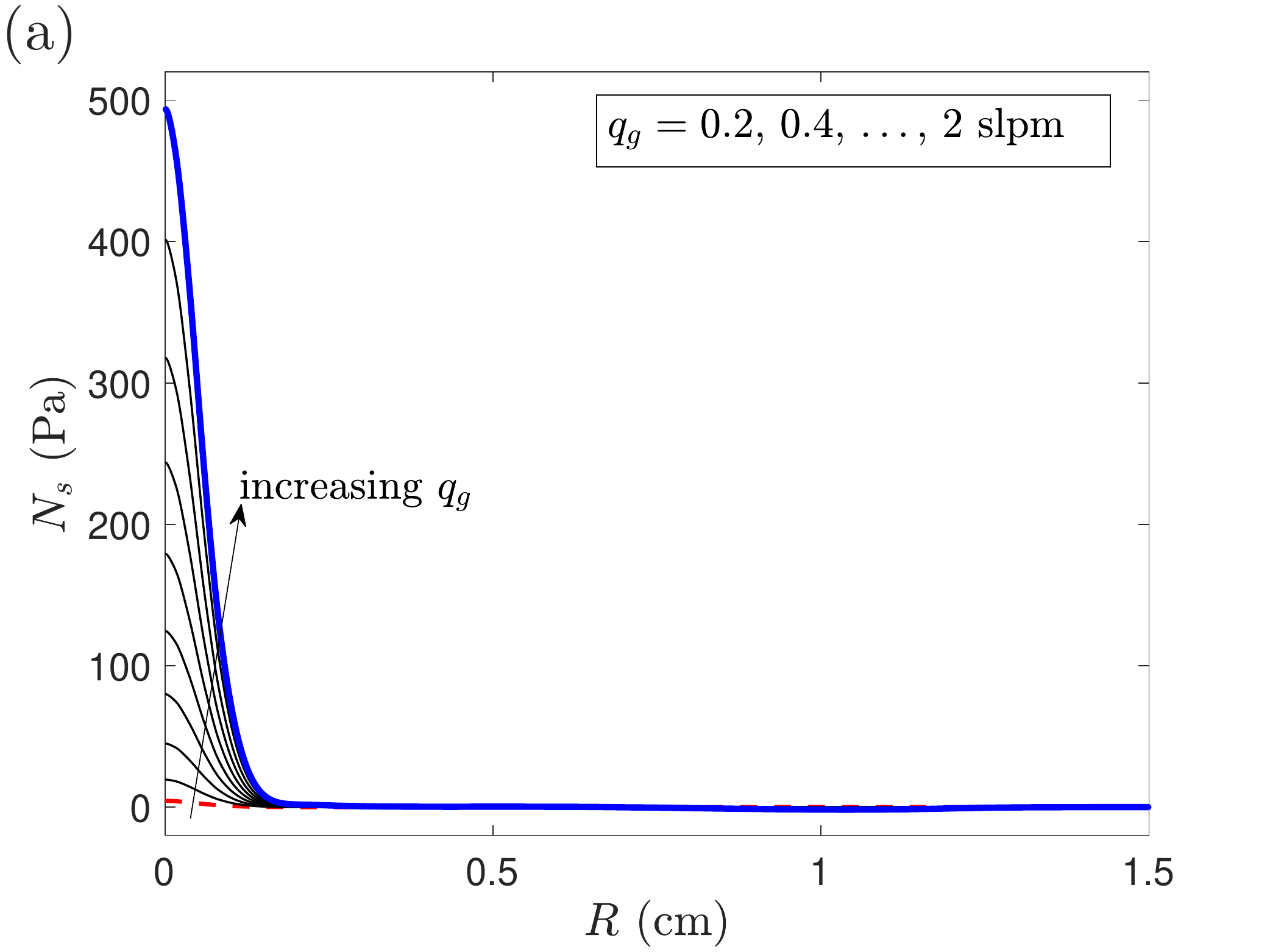}
\hspace{0.01\textwidth}
\includegraphics[scale=0.28]{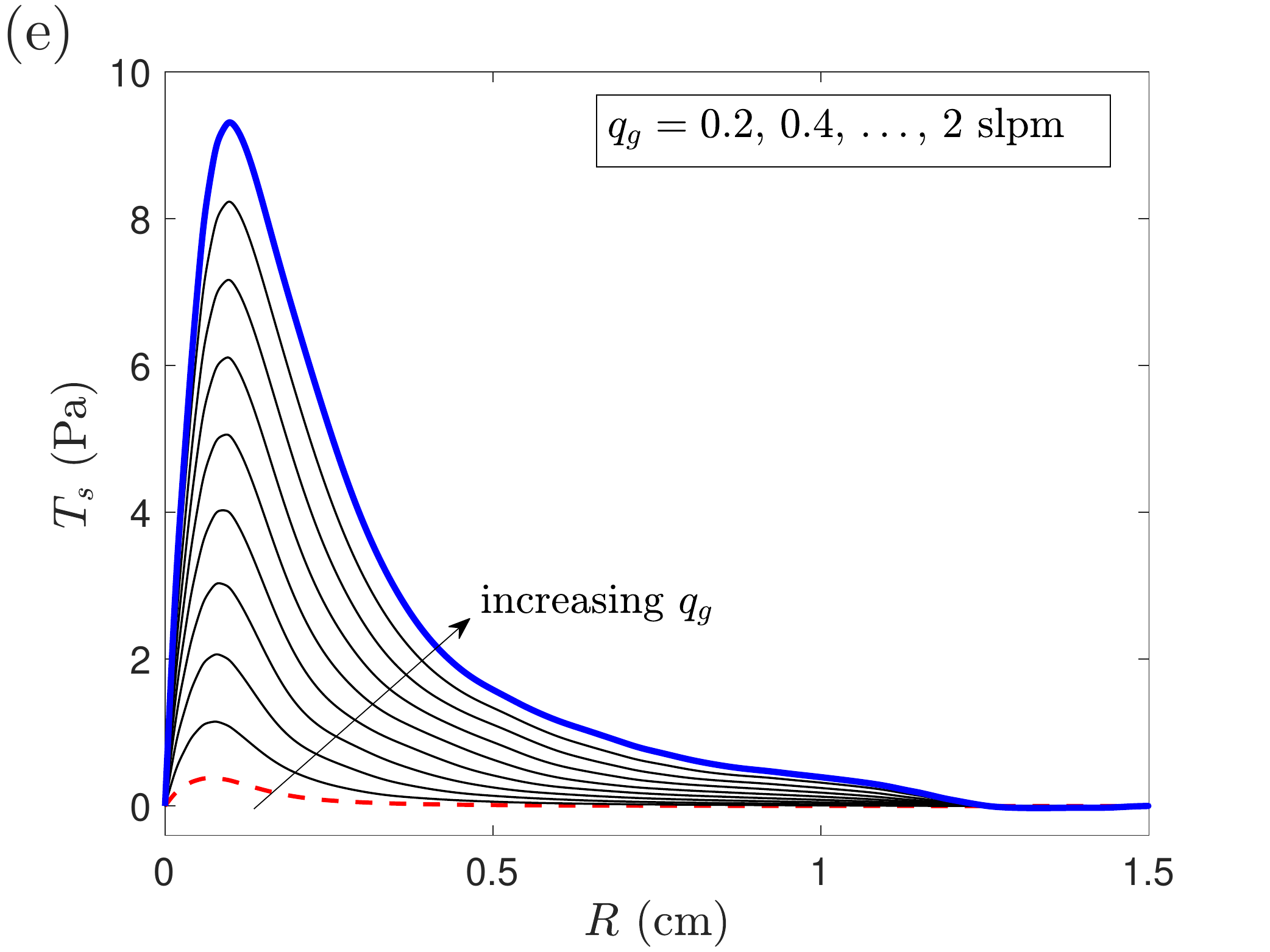}\\
\includegraphics[scale=0.28]{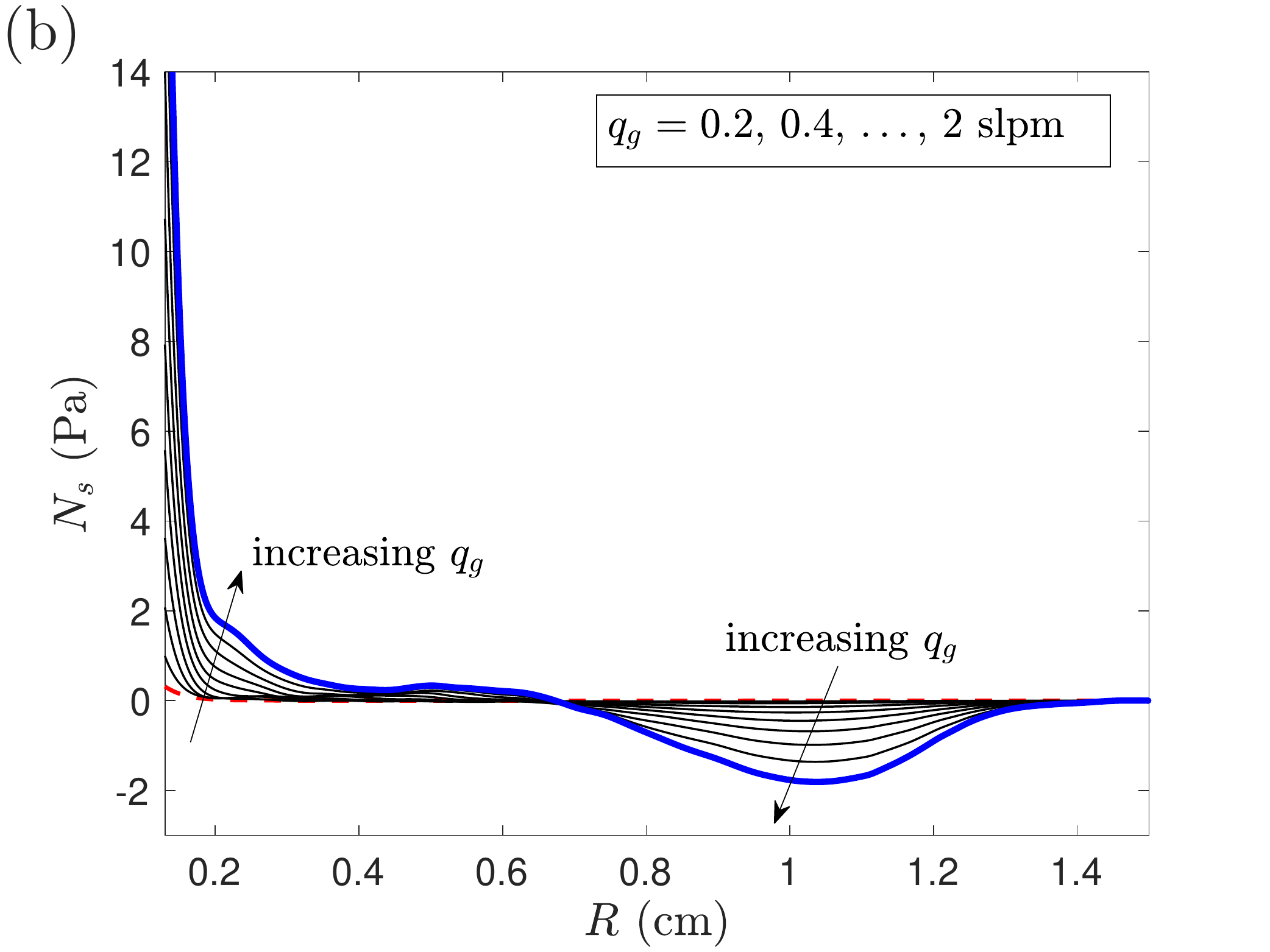}
\hspace{0.01\textwidth}
\includegraphics[scale=0.28]{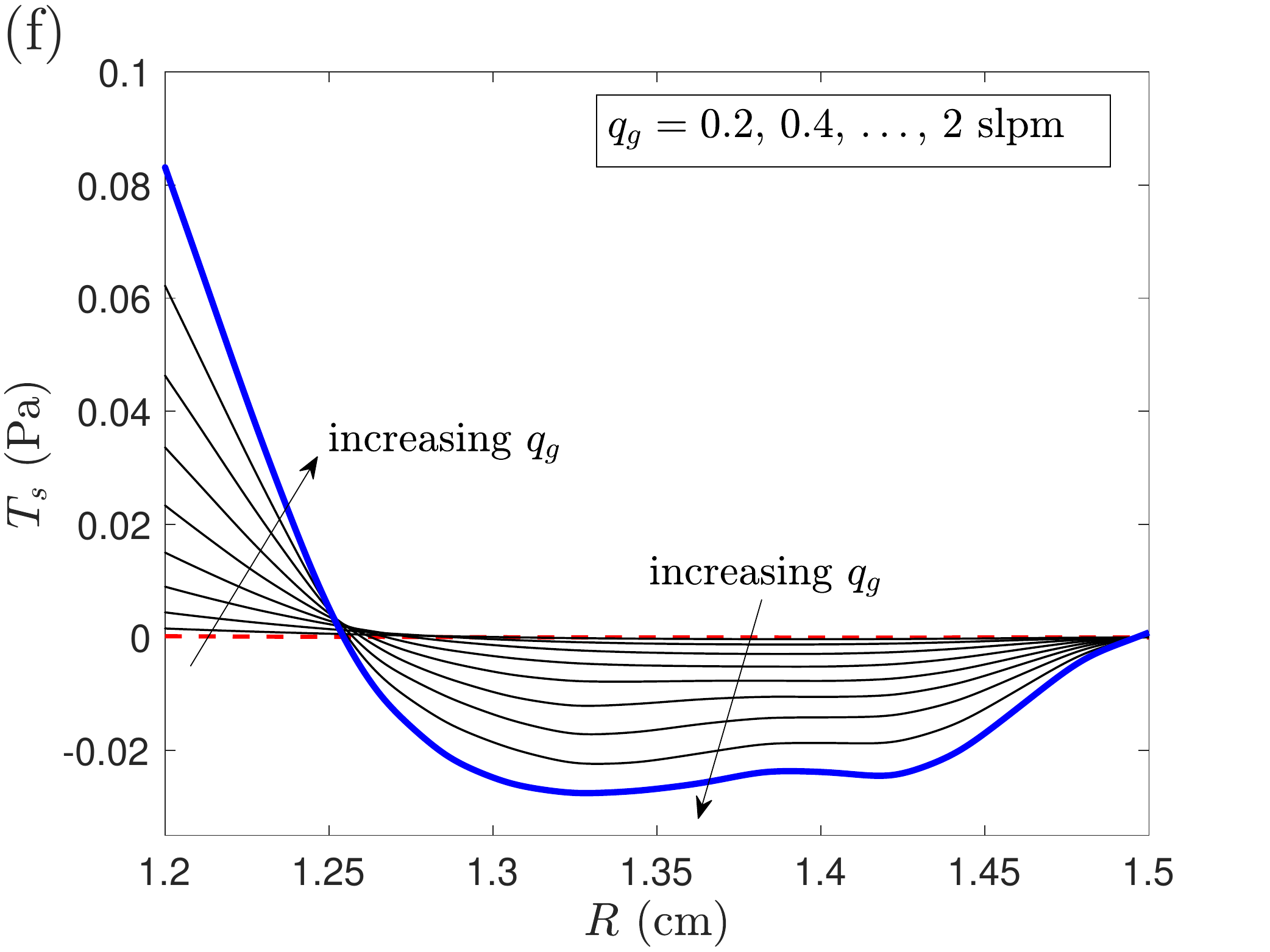}\\
\includegraphics[scale=0.28]{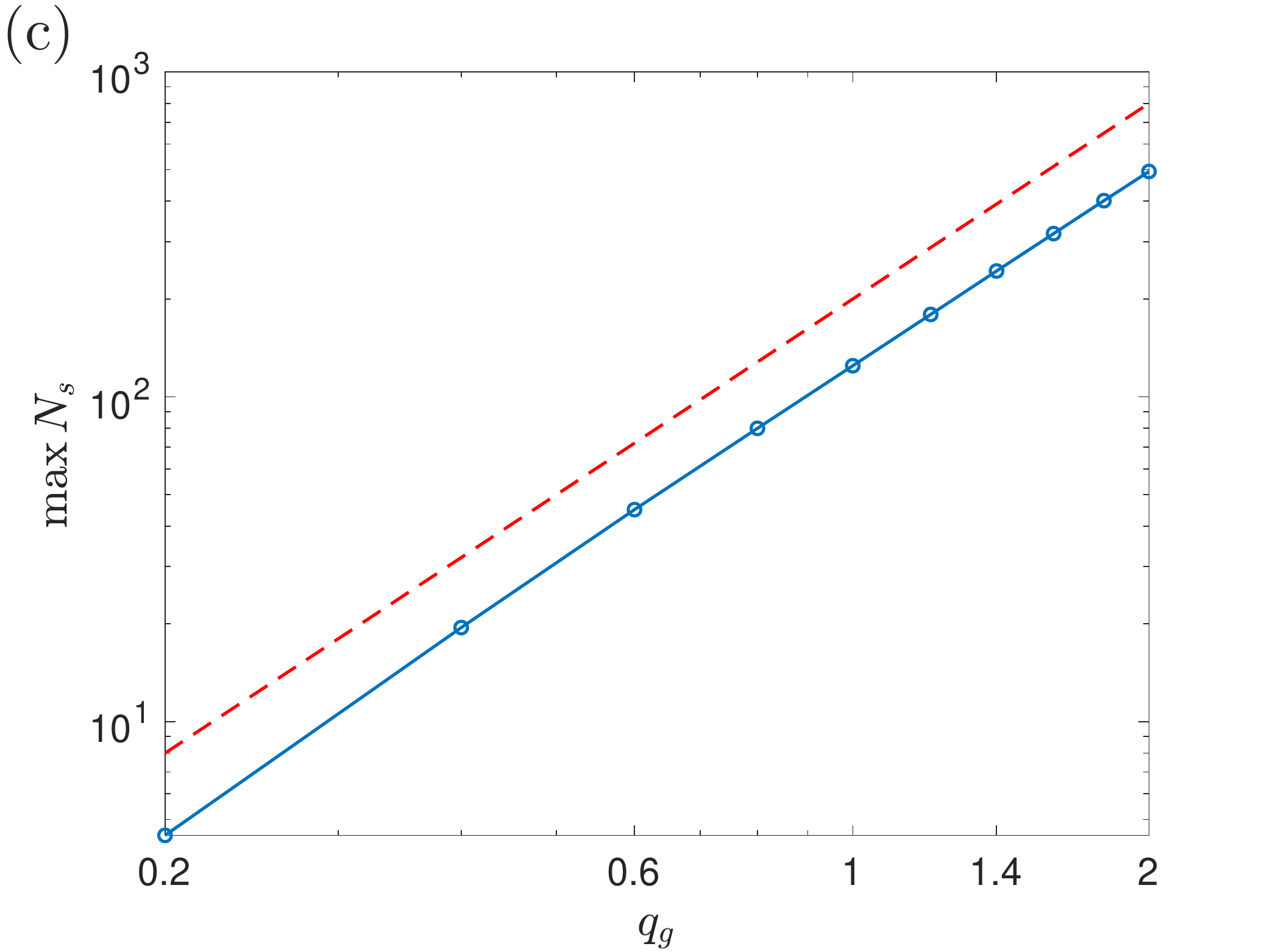}
\hspace{0.01\textwidth}
\includegraphics[scale=0.28]{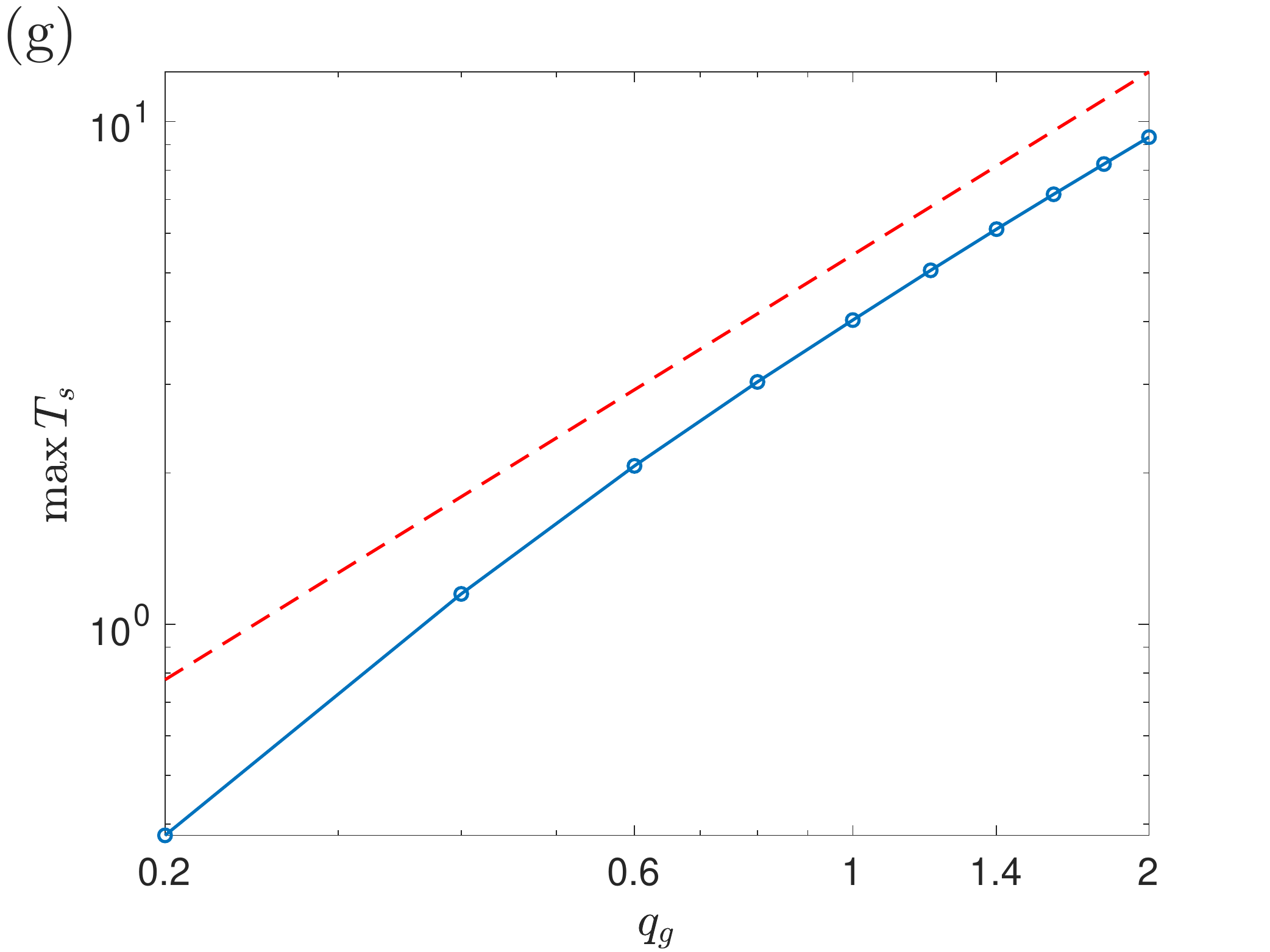}\\
\includegraphics[scale=0.28]{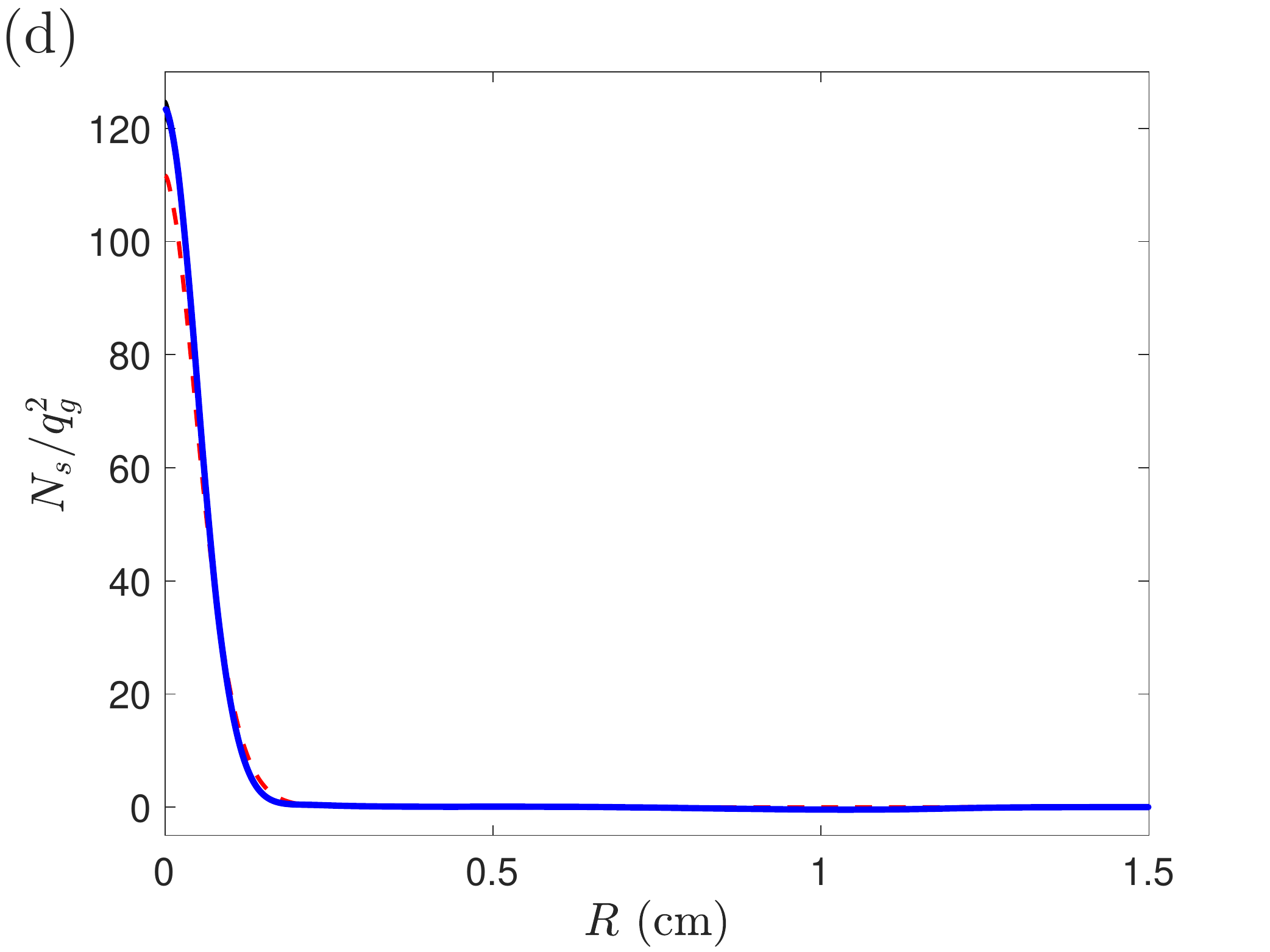}
\hspace{0.01\textwidth}
\includegraphics[scale=0.28]{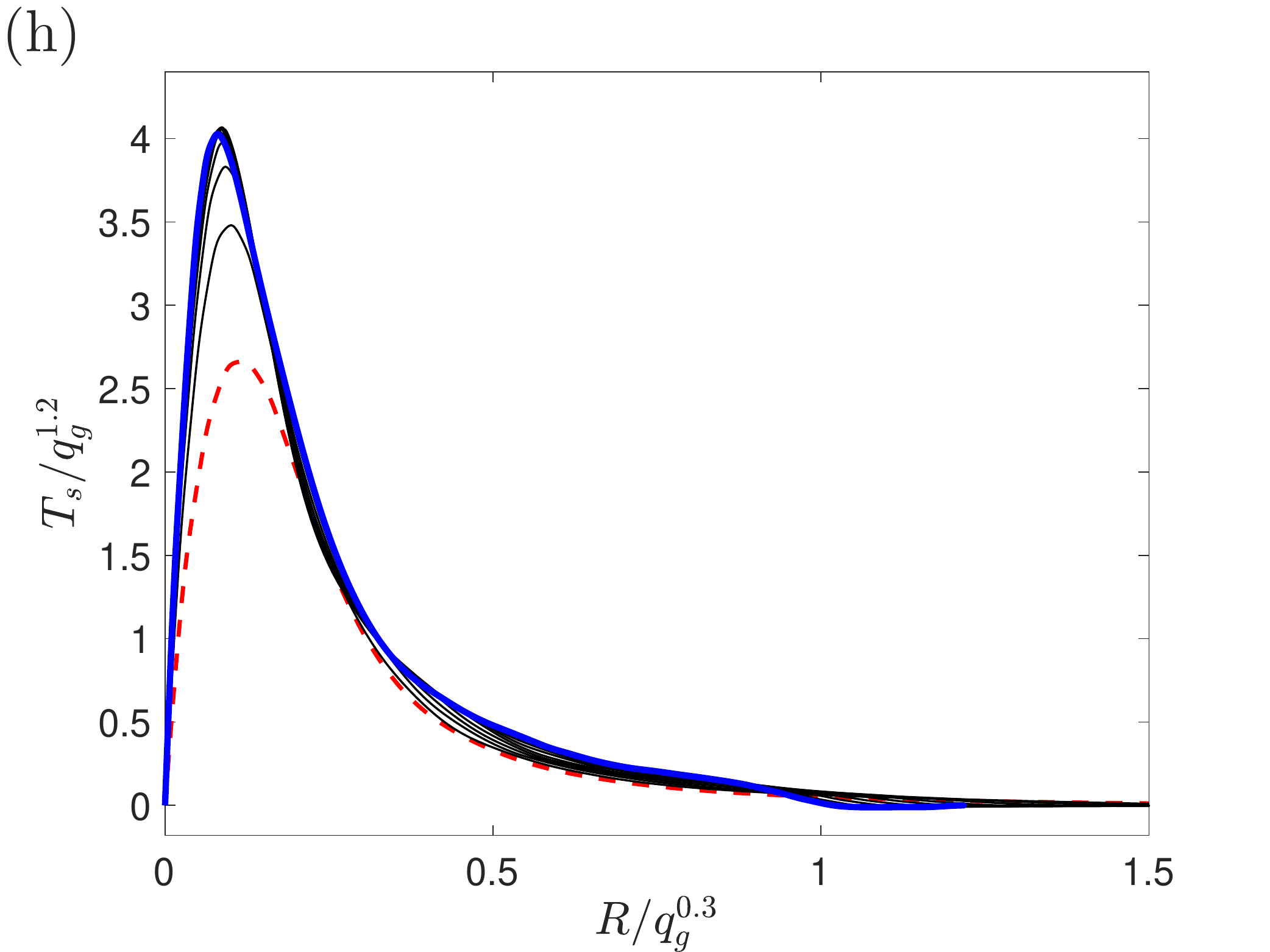}
\caption{Stresses exerted by the gas onto the gas--liquid interface modelled as a flat solid wall. Panels (a)--(d) correspond to normal stresses and (e)--(h) correspond to tangential stresses. (a) and (e) show the stresses for the gas flow rate $q_g$ changing from $0.2$ to $2\,\mathrm{slpm}$ (red dashed and thick blue solid lines, respectively) with the increment of $0.2\,\mathrm{slpm}$. (b) and (e) show zooms of the stresses in the regions away from the centre. (c) and (g) show the maxima of the stresses (blue solid lines) versus $q_g$ on the log-log scale. The red dashed lines have slopes $2$ and $1.2$ for the normal and tangential stresses, respectively. (d) and (h) show the rescales stresses $N_s/q_g^2$ and $T_s/q_g^{1.2}$, respectively. Note that for the tangential stresses in panel (h) the horizontal axis is scaled as $R/q_g^{0.3}$.}
\label{fig:stresses}
\end{figure}

\begin{figure}
\centering
\includegraphics[scale=0.28]{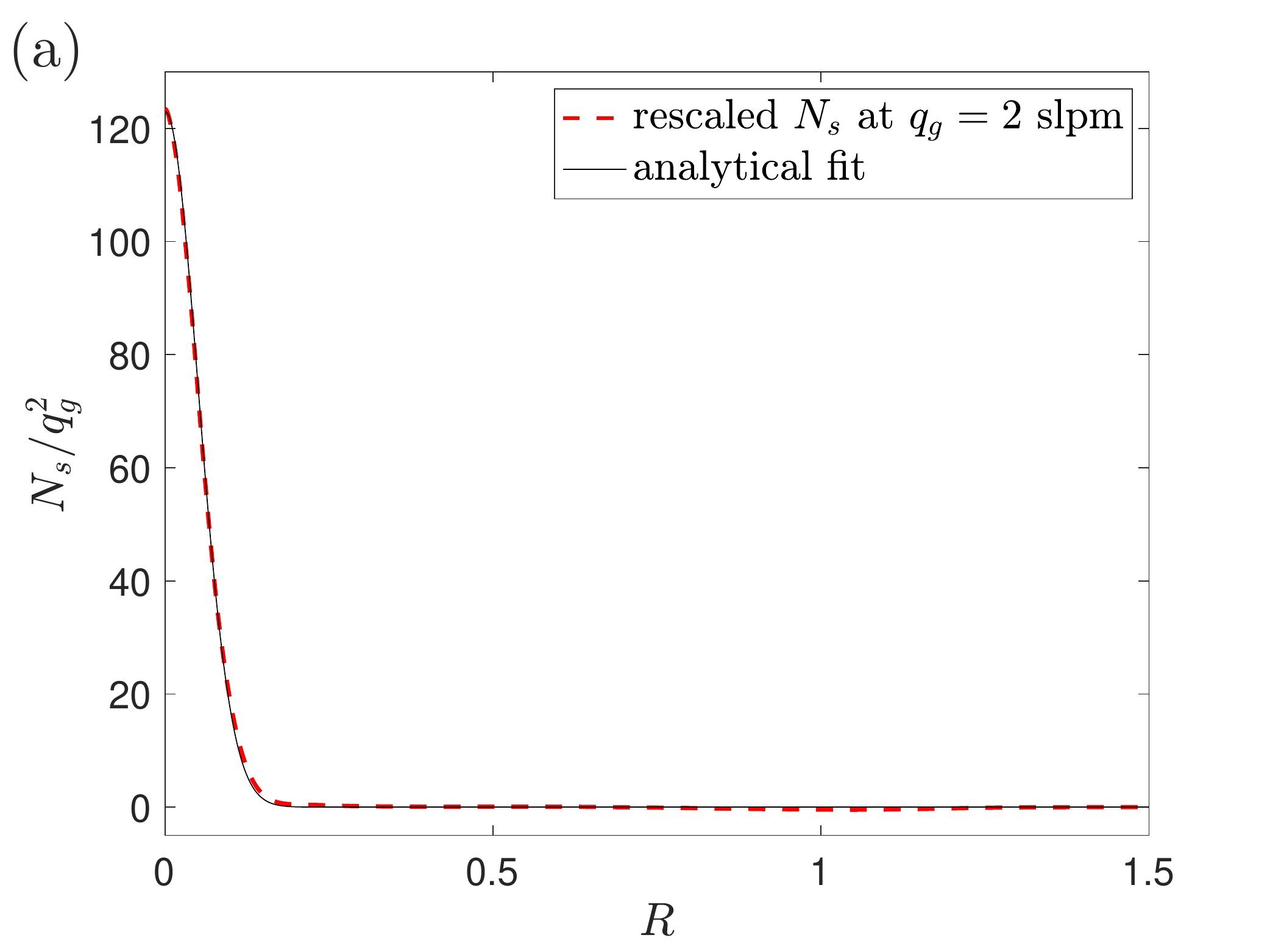}
\hspace{0.01\textwidth}
\includegraphics[scale=0.28]{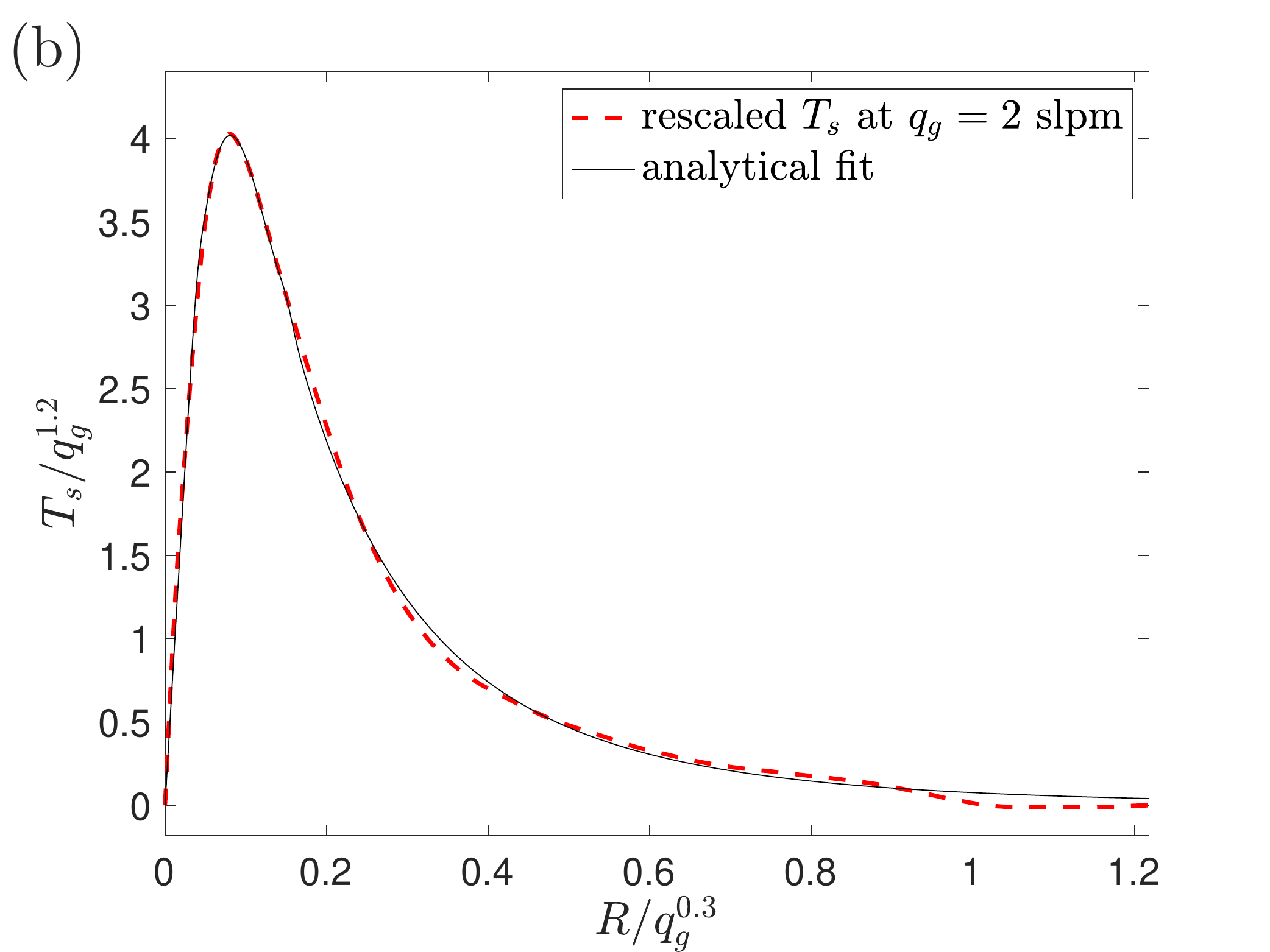}\\
\caption{Analytical fitting of stresses exerted by the gas onto the gas--liquid interface modelled as a flat solid wall. Panel (a) corresponds to the normal stress. The red dashed line shows the rescaled stress $N_s/q_g^2$ for $q_g=2$. The solid line corresponds to the analytical fit $f_1$ given by equation (\ref{eq:normal_fit}). Panel (b) corresponds to the tangential stress. The red dashed line shows the rescaled stress $T_s/q_g^{1.2}$ for $q_g=2$. The solid line corresponds to the analytical fit $f_2$ given by equation (\ref{eq:tangential_fit}).}
\label{fig:stresses_fitting}
\end{figure}

Next, we will analyse in detail how the normal and tangential stresses exerted by the gas onto the interface behave when the gas flow rate varies for the case when the interface is modelled as a flat solid wall. The results are presented in figure~\ref{fig:stresses}. Panels (a) and (e) show the normal and tangential stresses for the gas flow rate $q_g$ changing from $0.2$ (red dashed lines) to $2\,\mathrm{slpm}$ (thick blue solid lines) with the increment of $0.2\,\mathrm{slpm}$. As expected, the stresses grow as the gas flow rate increases. The normal stresses have their maximum values in the centre (at $R=0$) and then rapidly decay as the radial distance increases. However, as is apparent from the zoom in panel (b), the decay is not monotonic, and there is a region where the normal stress becomes negative and then increases. The tangential stresses vanish at the centre and have their maximum values at a distance slightly away from the centre, at approximately $R=0.1\,\mathrm{cm}$. Then they slowly decay as $R$ increases up to approximately $R = 1.2\,\mathrm{cm}$. After this distance, the tangential stresses become small and negative, as can be seen in the zoom in panel (f). This may be associated with the presence of a slow recirculation zone in the corner, as seen in figure~\ref{fig:gas_only1}. Panels (c) and (g) show the maxima of the normal and tangential stresses (blue solid lines), respectively, versus the gas flow rate, $q_g$, on the log-log scale and suggest a power law behaviour, i.e. $\max N_s$ scales approximately as $q_g^2$ and $\max T_s$ scales approximately as $q_g^{\alpha}$, where $\alpha\approx 1.2$. Indeed, the red dashed lines in panels (c) and (g) have slopes $2$ and $1.2$, respectively. It can be observed that the scaling for the normal stresses works well for all the values of the flow rate, whereas for the tangential stresses the scaling works better for larger values of the flow rate. We plotted the rescaled normal and tangential stresses $N_s/q_g^2$ and $T_s/q_g^{1.2}$ in panels (d) and (h), respectively. For the normal stresses, we can observe that the curves seem to collapse onto the same universal curve (except for the smallest gas flow rate for which there is a slight deviation, see the red curve). For the tangential stresses, in order to build towards a universal scaling, the horizontal axis also needs to be rescaled as $R/q_g^{\beta}$, where $\beta\approx 0.3$. The scaling works well only for relatively large gas flow rates.  The scaling for the normal stress follows from the fact that the stagnation point pressure (i.e. the pressure or normal stress where the gas jet impinges on the wall at $R=0$) is associated with the dynamic pressure at the centre line of the gas jet, which follows from Bernoulli's theorem (assuming incompressible flow), see e.g. \cite{cheslak1969cavities,clancy2006aerodynamics}. The scaling then follows from the fact that the dynamic pressure is proportional to the square of the jet velocity, which in turn is proportional to the flow rate $q_g$. 
The reason for the apparent scaling for the tangential stresses is not immediately obvious from the governing equations and is left as  a topic for future investigation. 

We conclude that the scalings for the gas stresses may be utilised for larger values of the gas flow rate 
(particularly for thin films where the interface is not significantly deformed). 
For smaller gas flow rates, these scalings do not work as well, and the stresses need to be recomputed for each value of $q_g$ 
(this particularly applies to tangential stresses) in order to obtain not only qualitative but also good quantitative descriptions of the liquid deformation 
under the gas jet and the generated flow in the liquid. Nevertheless, as we will show below, good agreement with DNS and experiments can still be obtained 
when these scalings are used even for lower gas flow rates. To do so, it is useful to fit analytical expressions to the apparent universal shapes in figures~\ref{fig:stresses}(d) and \ref{fig:stresses}(h). Proposed fits are demonstrated in figure~\ref{fig:stresses_fitting}. In panel~(a), the rescaled normal stress for $q_g=2\,\mathrm{slpm}$ is plotted by the red dashed line, and the solid line corresponds to the analytical 
fit given by the following Gaussian function \cite[a form inspired by e.g.][]{Lunz_Howel_2018}:
\begin{equation}
f_1(R)=123\exp\bigl(-(R/0.071)^2\bigr).
\label{eq:normal_fit}
\end{equation} 
In panel (b), the rescaled tangential stress for $q_g=2\,\mathrm{slpm}$ is plotted over $R/q_g^{0.3}$ by the red dashed line, and the solid line corresponds to an analytical fit. It may be appropriate to suggest various fits. We used the following assumption:
\begin{equation}\label{eq:f2_fit}
f_2(\widetilde{R})\approx\begin{cases}
f_{2a}(\widetilde{R})\equiv\widetilde{R}/0.015, & \text{if } 0\leq\widetilde{R}\lesssim 0.02,\\
f_{2b}(\widetilde{R})\equiv p_3\widetilde{R}^3+p_2\widetilde{R}^2+p_1\widetilde{R}+p_0, & \text{if } 0.02\lesssim\widetilde{R}\lesssim 0.14, \\
f_{2c}(\widetilde{R})\equiv a{(\widetilde{R}+b)^{-c}}, & \text{if } \widetilde{R}\gtrsim 0.14,
\end{cases}
\end{equation} 
where $\widetilde{R}$ represents $R/q_g^{0.3}$, and $p_0=-0.7$, $p_1=138.75$, $p_2=-1279.9$, $p_3=3478.3$, and also $a=0.7$, $b=0.6$, $c=4.8$. These values were obtained using the curve fitting tool \texttt{cftool} in Matlab. In order to obtain smooth transitions at $\widetilde{R}\approx0.02$ and $\widetilde{R}\approx0.14$, we, in fact, use the following function:
\begin{equation}
f_2(\widetilde{R})=f_{2a}(\widetilde{R})+\bigl(f_{2b}(\widetilde{R})-f_{2a}(\widetilde{R})\bigr)H_d(\widetilde{R}-0.02)+\bigl(f_{2c}(\widetilde{R})-f_{2b}(\widetilde{R})\bigr)H_d(\widetilde{R}-0.14),
\label{eq:tangential_fit}
\end{equation}
where $H_d(x)=0.5\bigl(1+\tanh(x/d)\bigr)$ is a smoothed-out Heaviside function
with the steepness parameter $d=0.008$.

%%%
\begin{figure}
\centering
\hspace{0.57cm}\includegraphics[width=0.43\textwidth, height=4.64cm]{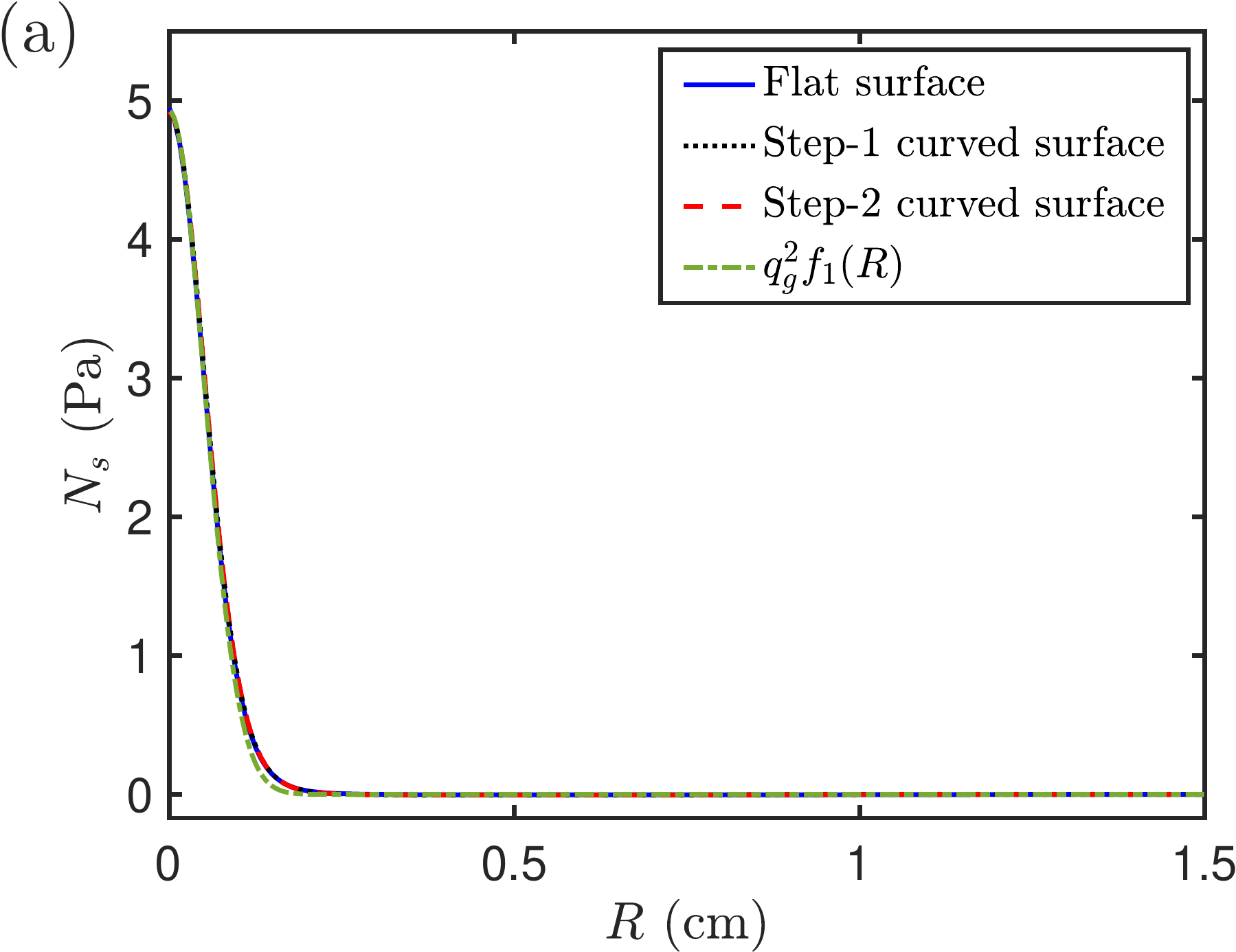}
\hspace{0.009\textwidth}
\includegraphics[width=0.445\textwidth]{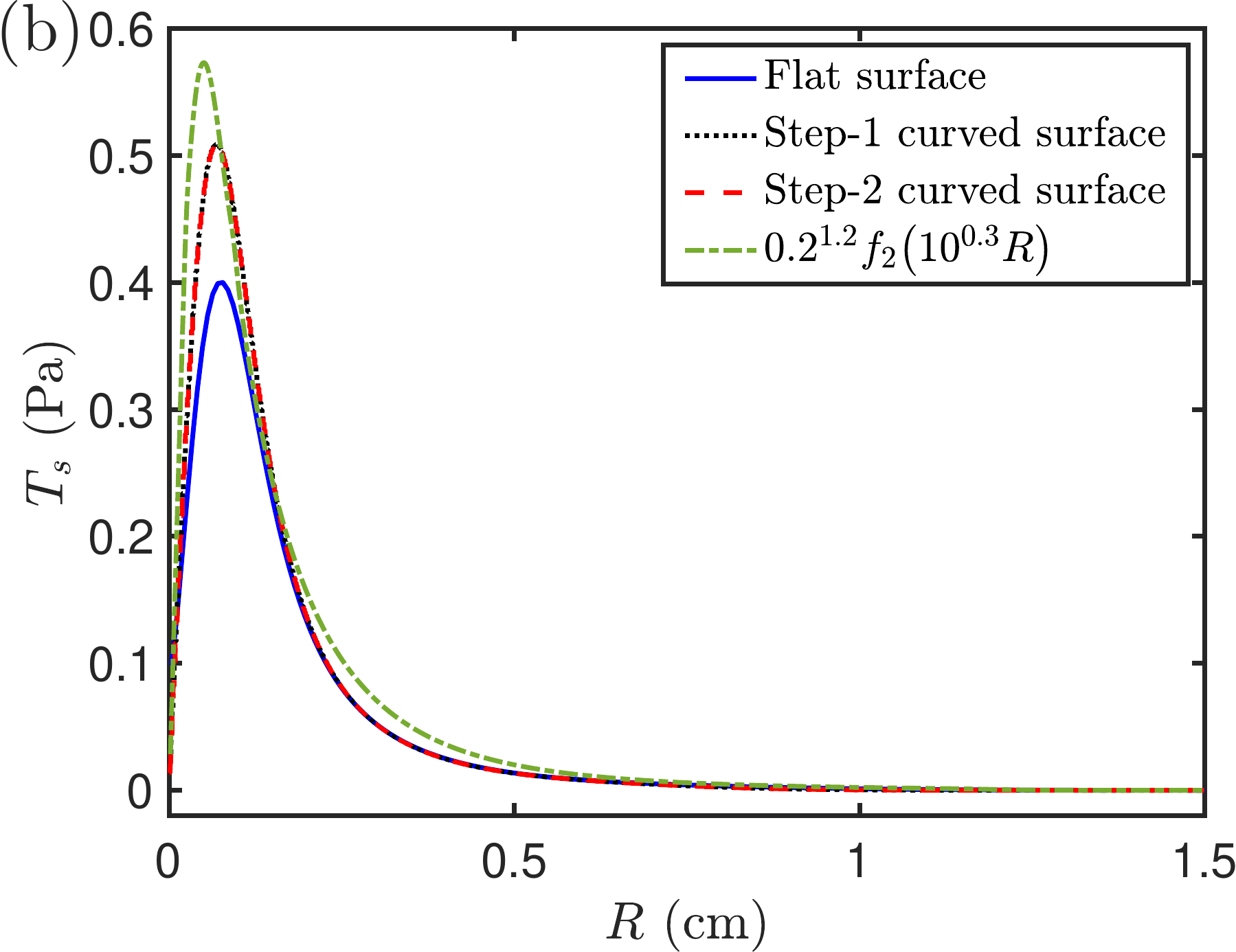}\,\,\,\,\,\,\\[0.3cm]
\includegraphics[width=0.45\textwidth, height=4.7cm]{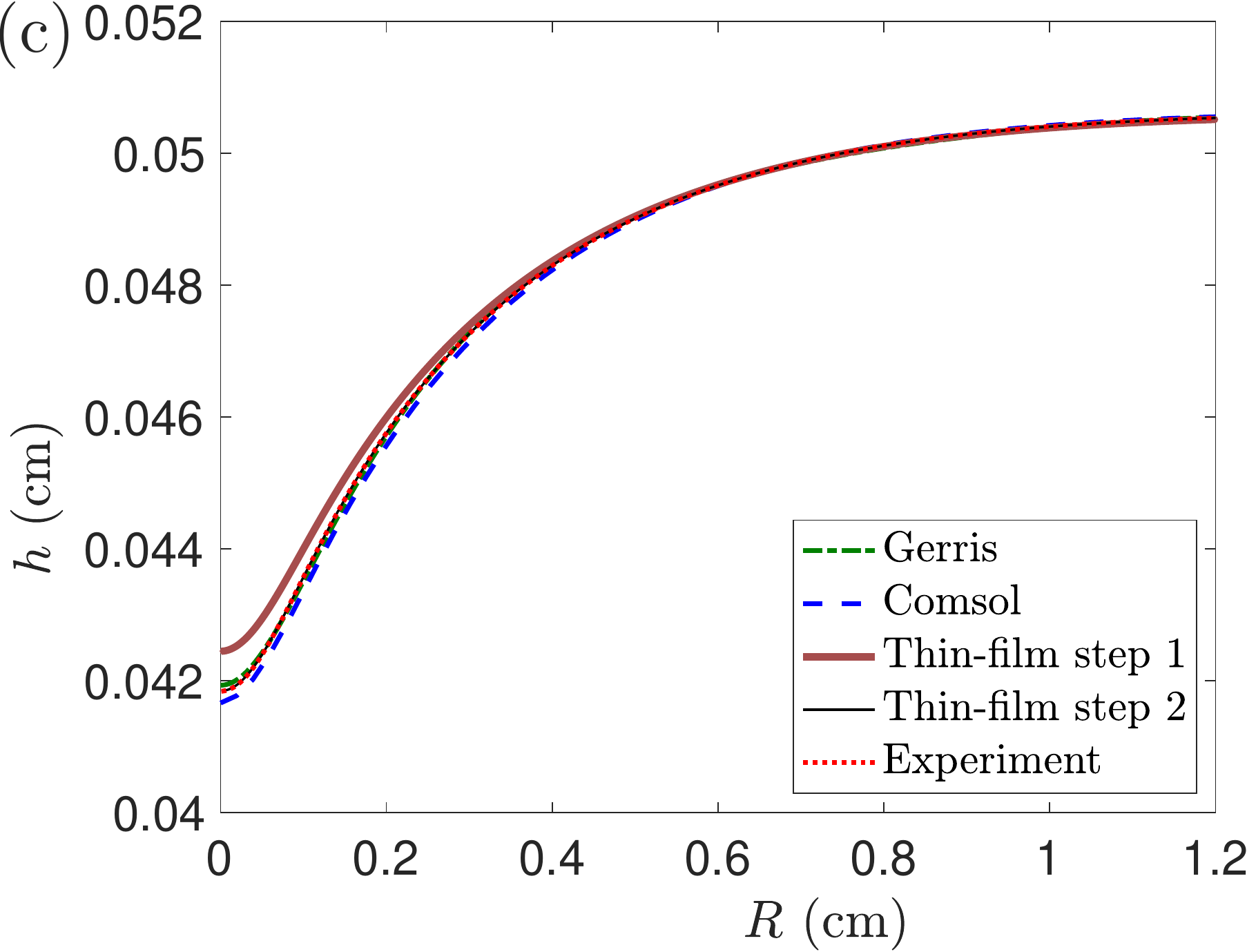}\,\,\,\,
\includegraphics[width=0.44\textwidth,height=4.6cm]{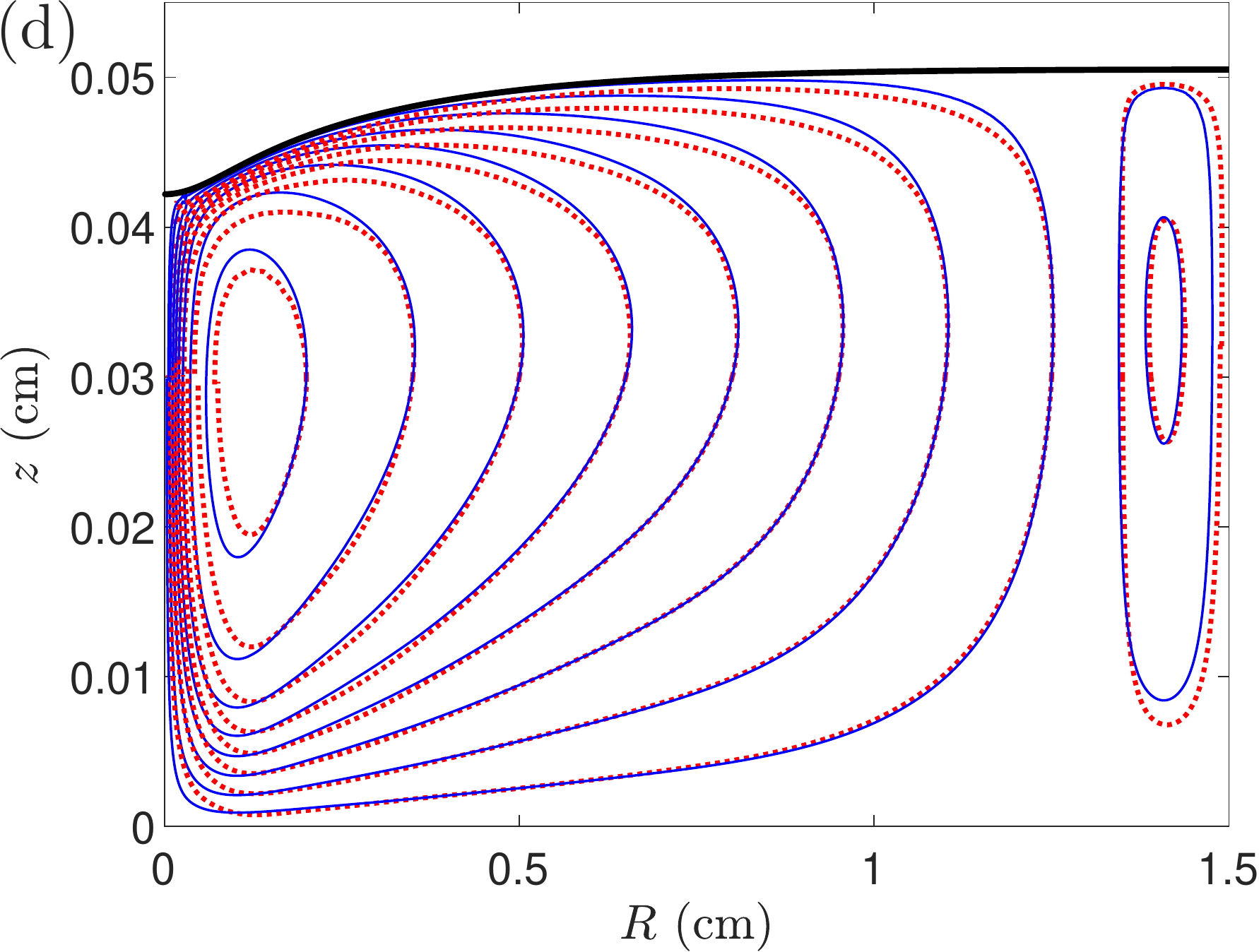}
\caption{Numerical results for a gas jet of flow rate $q_g=0.2\,\mathrm{slpm}$ impinging onto a liquid film of thickness $h_0=0.5\,\mathrm{mm}$. Panels (a) and (b) show the normal and tangential stresses exerted by the gas on the interface and computed using the iterative procedure discussed in \S~\ref{sect:gas_only}. The first three iterations are shown: for a flat interface and for two subsequent deformed interfaces, as indicated in the legends.  The stresses obtained using the scaling laws and analytical fits suggested in \S~\ref{sect:gas_only} are also shown (green dash-dotted lines). Panel (c) shows the resulting interface deformations at steady state obtained using $\mathpzc{Gerris}$ and COMSOL for the fully coupled gas--liquid model, as well as the thin-film equation (\ref{eq:th_eq_axisymm}) with the iterative procedure, as indicated in the legend. An experimental result is also shown. Panel (d) shows the streamlines in the liquid film obtained using the thin-film equation and COMSOL (the blue solid and red dotted lines, respectively).
}
\label{fig:0.5mm_0.2slpm}
\end{figure}
%%%

%%%
\begin{figure}
\centering
\includegraphics[width=0.45\textwidth]{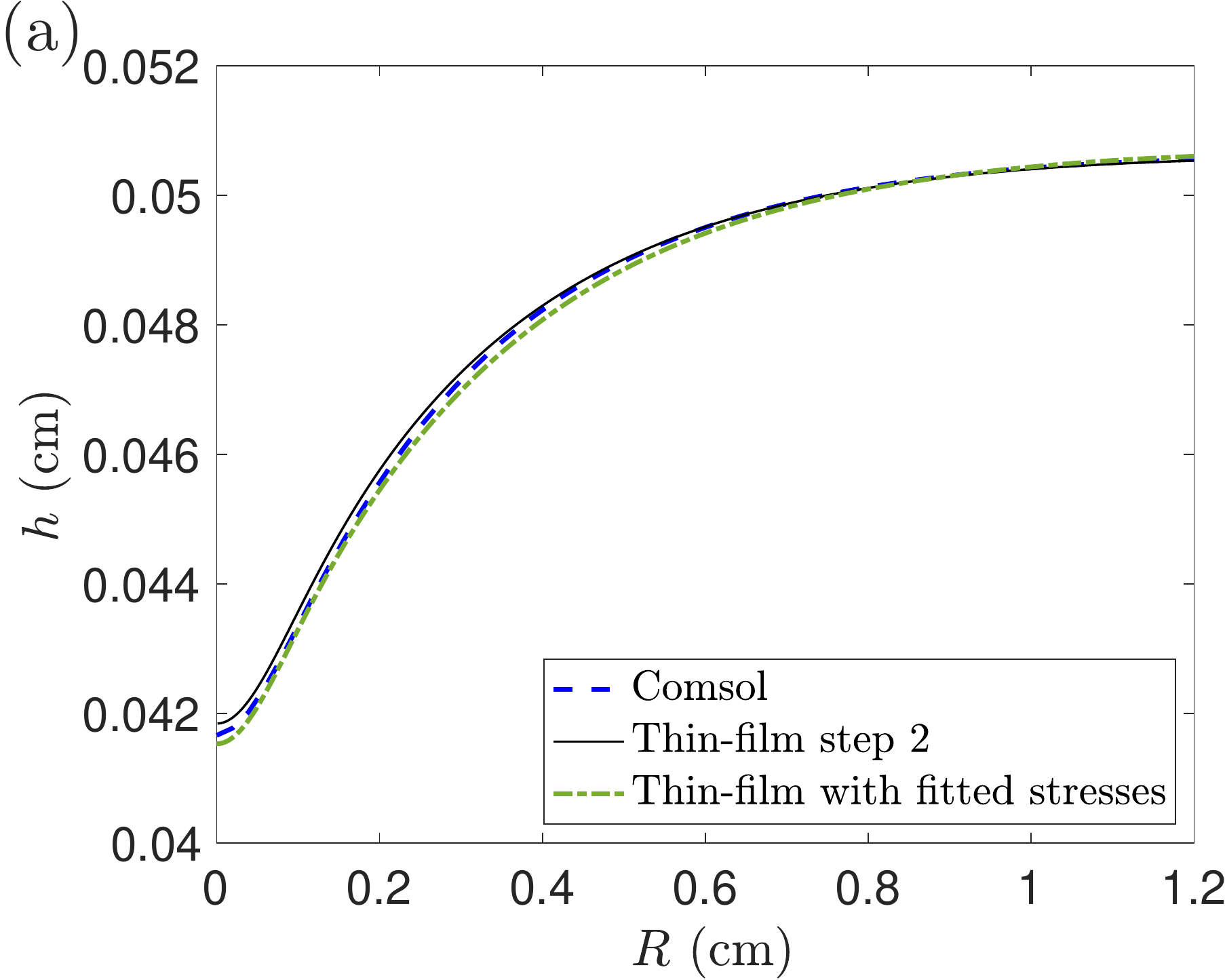}\,\,\,\,
\includegraphics[width=0.45\textwidth]{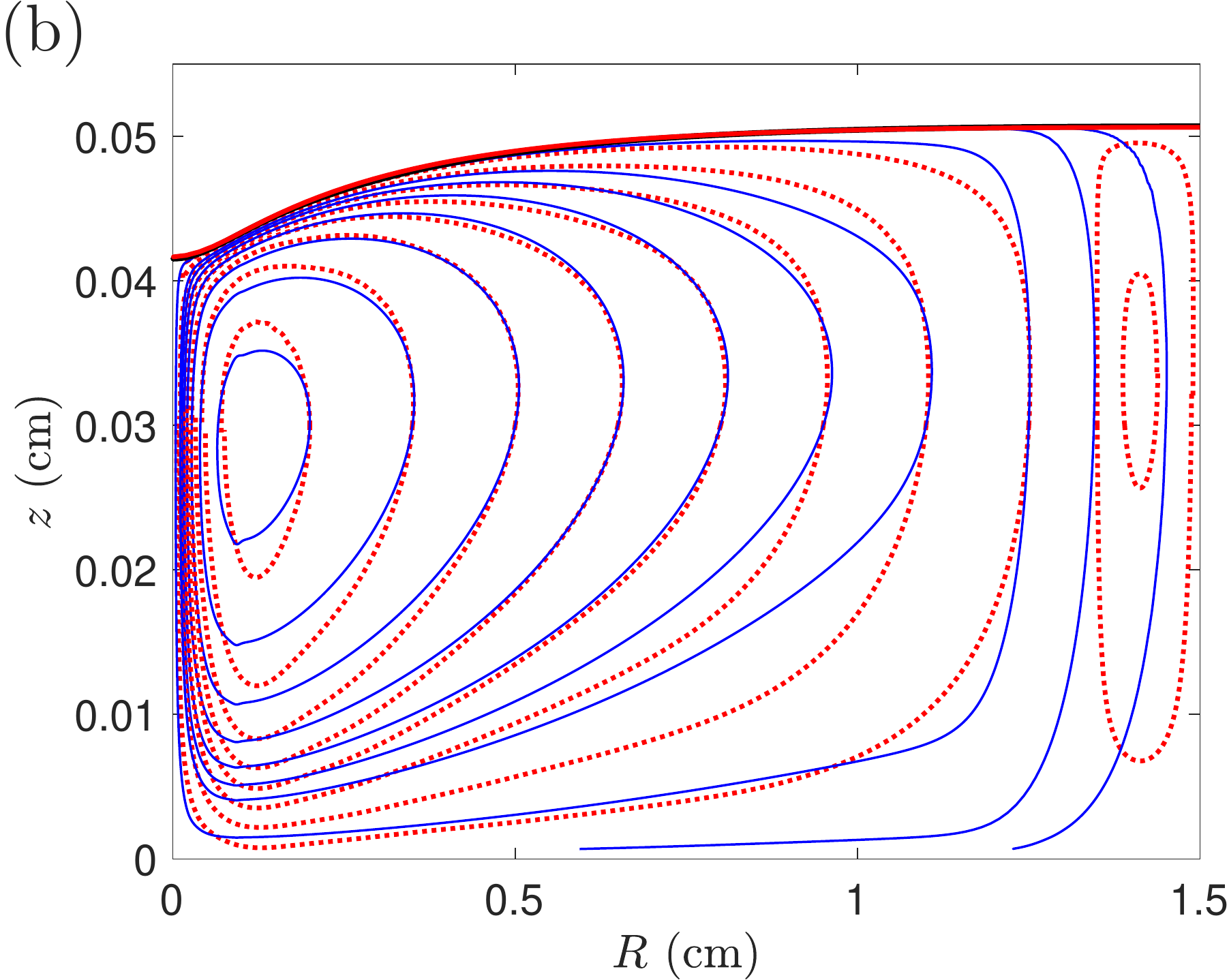}
\caption{Numerical results for a gas jet of flow rate $q_g=0.2\,\mathrm{slpm}$ impinging onto a liquid film of thickness $h_0=0.5\,\mathrm{mm}$. Panel (a) shows the resulting interface deformations at steady state obtained using COMSOL for the fully coupled gas--liquid model (blue dashed line), as well as the thin-film equation (\ref{eq:th_eq_axisymm}) with the iterative procedure utilising gas stresses computed in COMSOL (black solid line) and the thin-film equation (\ref{eq:th_eq_axisymm}) with the analytically fitted gas stresses (green dash-dotted line). Panel (b) shows the streamlines in the liquid film obtained using COMSOL and the thin-film equation with the fitted gas stress (the red dotted and blue solid lines, respectively).
}
\label{fig:0.5mm_0.2slpm_unirersal}
\end{figure}
%%%

\subsection{Comparison of the thin-film model with DNS and experiments}

In this section, we present results for a film of thickness $0.5\,\mathrm{mm}$. We start with the gas flow rate of $0.2\,\mathrm{slpm}$ at which the film deforms but does not rupture. The results are given in figure~\ref{fig:0.5mm_0.2slpm}. Panels (a) and (b) show the normal and tangential stresses, respectively, exerted on the interface. These are computed using the iterative procedure described above. The blue solid lines correspond to the flat interface, the black dotted lines correspond to the curved interface, obtained by solving the thin-film equation (\ref{eq:th_eq_axisymm}), after the first iteration, and the red dashed lines are obtained for the curved interface after the second iteration. It can be observed that two iterations are sufficient in this case and convergence is achieved. There is only a minor difference in the results for the normal stresses. However, there is a noticeable difference between the tangential stresses computed for flat and curved interfaces. The green dash-dotted lines in panels (a) and (b) are obtained using the suggested scaling laws and analytical fits given by equations (\ref{eq:normal_fit}) and (\ref{eq:tangential_fit}), respectively. Interestingly, the analytical fit for the tangential stress is in better agreement with the results for the curved interfaces than with the results for the flat interface, although the fit itself was obtained under the assumption of a flat interface.

The resulting interface deformations for this gas flow rate are shown in panel (c) after the computational time $t=5\,\mathrm{s}$, although to converge to a steady state approximately $2$--$3$ seconds was found to be sufficient. In panel (c), we compare the thin-film results computed with the gas stresses obtained using the different stages of the iterative procedure (the brown thick solid and black thin solid lines) with the $\mathpzc{Gerris}$ and COMSOL results for the fully coupled gas--liquid model (the green dash-dotted and blued dashed lines, respectively). An experimental result is also shown (the red dotted line). There is a slight difference in the region near $R=0$ between the $\mathpzc{Gerris}$ and COMSOL results for the fully coupled model and the thin-film result when the gas stress were computed by assuming that the interface was flat. This difference is nearly eliminated after the iterative procedure, and we conclude that the thin-film equation performs well even for a film thickness well beyond the expected range of validity of the equation. Panel (d) shows the streamlines in the liquid film obtained using the thin-film model (blue solid lines) and COMSOL (red dotted lines). We can observe that there is a relatively large eddy close to the axis of symmetry at $R=0$ and there is a smaller and slower eddy near the side wall of the beaker. This second eddy appears due to small and slow eddy in the gas in the corner between the gas--liquid interface and the side wall, see figure~\ref{fig:gas_only1}. We note that the $\mathpzc{Gerris}$ results for the 
streamlines are in qualitative agreement with this.

Figure~\ref{fig:0.5mm_0.2slpm_unirersal} presents in addition the results for the thin-film equation with the gas normal and tangential stresses obtained using the scaling laws and analytical fits (equations (\ref{eq:normal_fit}) and (\ref{eq:tangential_fit})) discussed above. Panel (a) compares the resulting interface profile with the previously presented COMSOL result and the thin-film result found using the iterative procedure. In can be seen that the interface profile for the thin-film equation with the analytically fitted stresses is in good agreement with the previously computed results. However, the dimple is slightly deeper. This is expected as the scaling/fitting procedure slightly overestimates the gas tangential stress for $q_g=0.2\,\text{slpm}$ (see figure~\ref{fig:0.5mm_0.2slpm}(b)). Panel~(b) shows the streamlines in the liquid film obtained using the thin-film model with the fitted stresses (blue solid lines) and COMSOL (red dotted lines). We can observe that, although the eddy close to the axis of symmetry at $R=0$ is qualitatively well described, the agreement is not as good as with the thin-film equation with the iterative procedure. In particular, the slower eddy near the side wall of the beaker is not predicted by such a thin-film computation. This is again expected, as the utilised fitting procedure ignores the region near the side wall of the beaker where the gas tangential stress becomes small and negative (see figure~\ref{fig:stresses}(f)). 
There is no immediately obvious scaling for the gas shear stress in this region. 
This would become useful for processes such as mass transport and mixing, thus meriting further attention in the future in an application-specific context.

%%%
\begin{figure}
\centering
\includegraphics[scale=0.4]{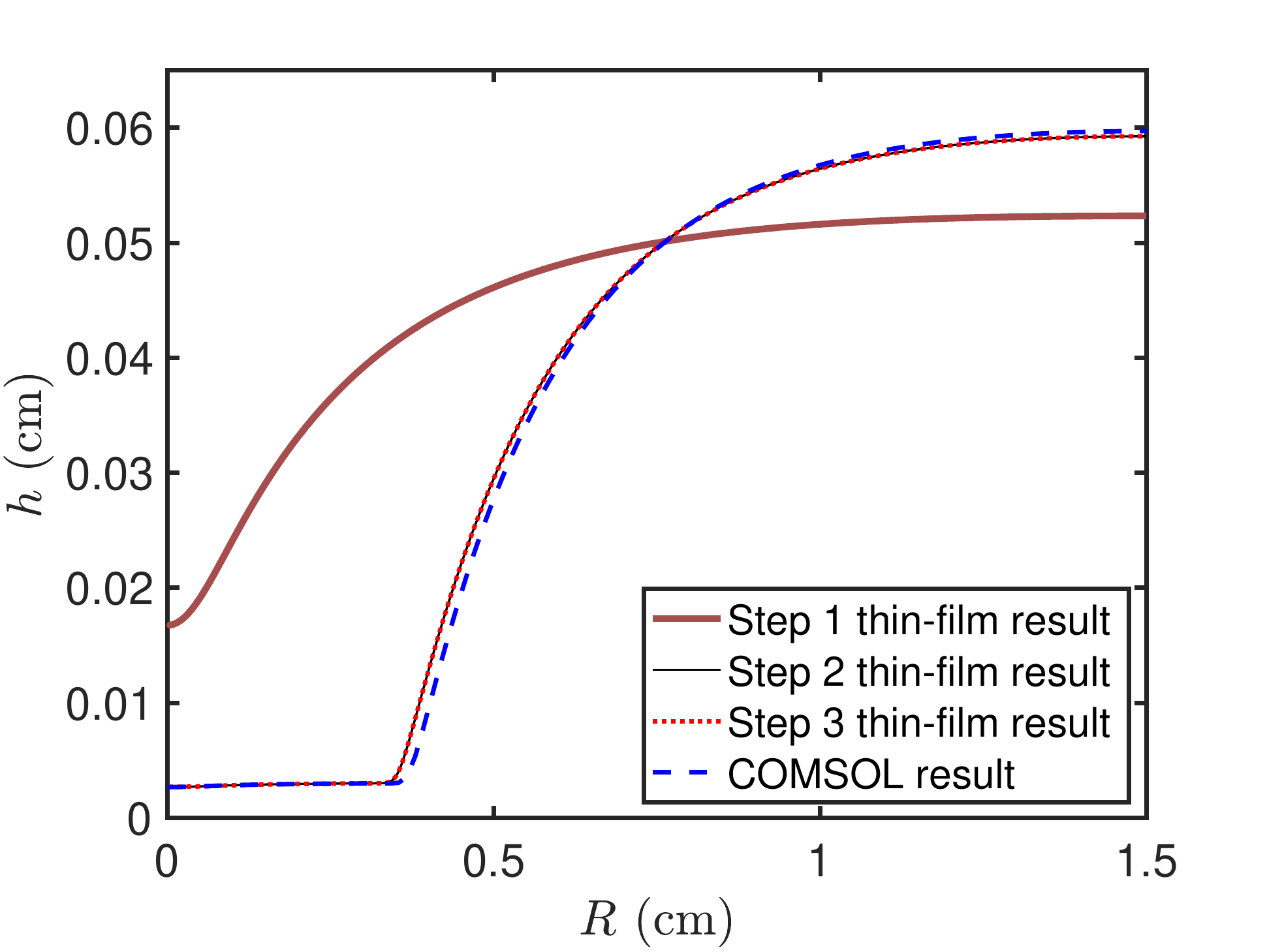}
\caption{Numerical results for a gas jet of flow rate $q_g=0.37\,\mathrm{slpm}$ impinging onto a liquid film of thickness $h_0=0.5\,\mathrm{mm}$. 
Shown are the resulting interface deformations at steady state obtained using the thin-film equation (\ref{eq:th_eq_axisymm}) with the iterative procedure and COMSOL for the fully coupled gas--liquid model, as indicated in the legend.  The numerical formulations include the disjoining pressure (\ref{eq:disj_p1}) giving the equilibrium contact angle $\theta_{eq}=15^\circ$ and the precursor thickness $h_{eq}=0.03\,\mathrm{mm}$.}
\label{fig:0.5mm_0.37slpm}
\end{figure}
%%%

The importance of the iterative procedure in computing gas stresses becomes more apparent when the gas flow rate is close to the value that leads to dewetting of the liquid from the bottom of the beaker. An example is shown in figure~\ref{fig:0.5mm_0.37slpm} for gas flow rate $q_g=0.37\,\mathrm{slpm}$. To account for possible dewetting, the thin-film and COMSOL numerical simulations here include the disjoining pressure (\ref{eq:disj_p1}) with $A=2.9609\times 10^{-11}$ and $B=8.2659\times 10^{-25}$ giving the contact angle $\theta_{eq}=15^\circ$ and the precursor thickness $h_{eq}=0.03\,\mathrm{mm}$. It can be seen that in the COMSOL simulation for the fully coupled gas--liquid model a dry spot for radius of approximately $0.36\,\mathrm{cm}$ appears in the centre (blue dashed line). However, for the first step of the iterative procedure for the thin-film model we find that no dry spot appears. This is then suitably accounted for by the next step of the iterative procedure. Indeed, when we recompute the stresses for the resulting deformed interface and use them in the thin-film equation again, we observe that dewetting is initiated and a dry spot appears of radius that is 
in good agreement with  
the COMSOL result (see the black thin line). The next iteration on the gas stresses and the thin-film equation leads to the result that is indistinguishable from the one in the previous iteration.

\begin{figure}
\centering
\includegraphics[scale=0.4]{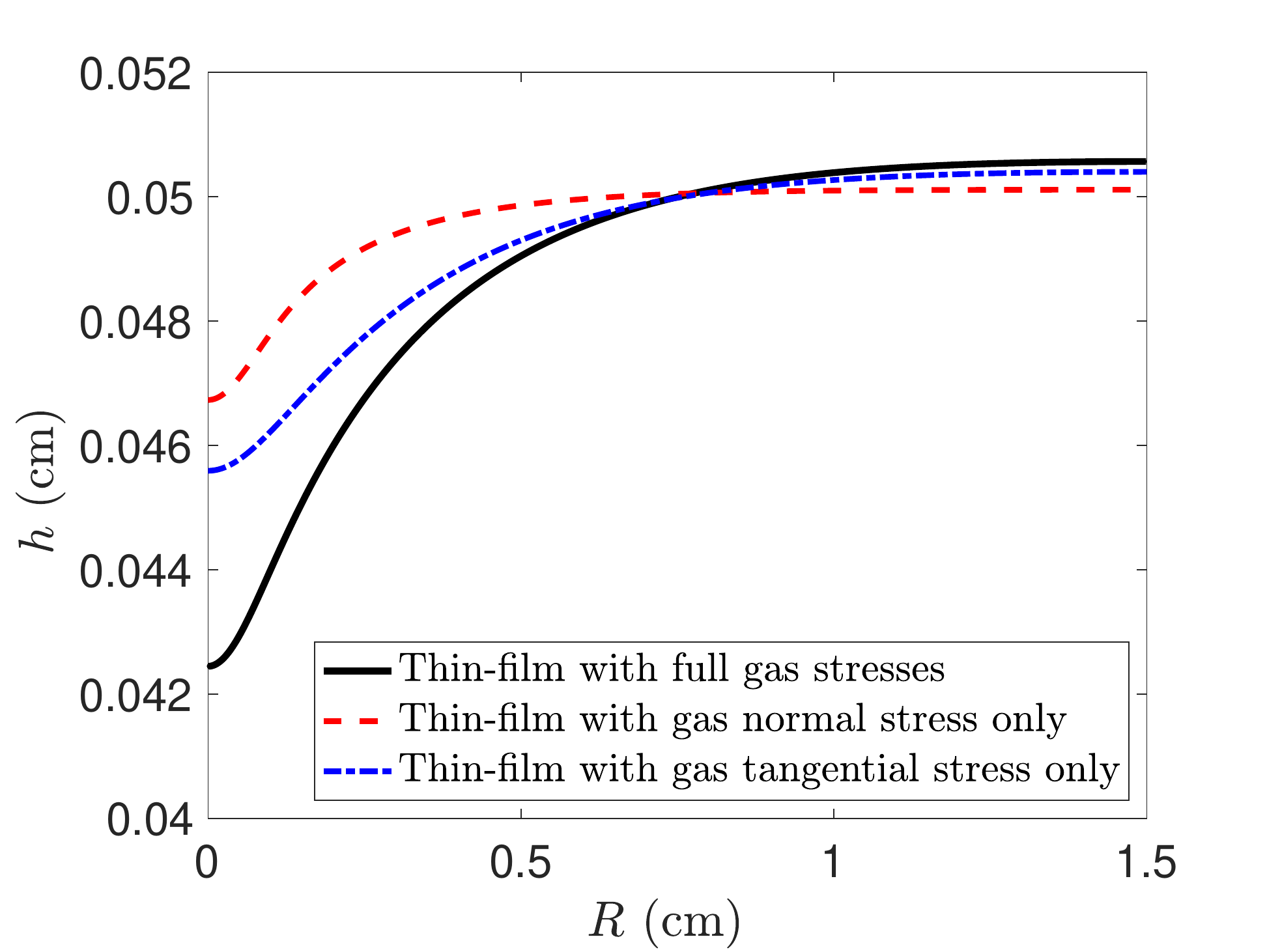}
\caption{Numerical results for a gas jet of flow rate $q_g=0.2\,\mathrm{slpm}$ impinging onto a liquid film of thickness $h_0=0.5\,\mathrm{mm}$. Shown are the resulting steady-state interface deformations obtained with the thin-film equation where both the gas normal stress and the tangential stress are included (black solid line), where only the gas normal stress is included (red dashed line) and where only the gas tangential stress is included (blue dash-dotted line).}
\label{fig:0.5mm_0.2slpm_different_stresses}
\end{figure}

We note that in many previous studies, the effect of the gas on the liquid deformation was modelled only through an assumed pressure distribution (i.e. normal stress) exerted by the gas (typically, of a Gaussian shape), see e.g.~\cite{Kriegsmann_etal_1998,Lunz_Howel_2018}, or only through an assumed  gas tangential stress on the interface, see e.g.~ \cite{Sullivan_etal_2008,Davis_etal_2010}. However, our analysis shows that for the considered setup neglecting either normal or tangential stress leads to significant errors in the results. This is particularly apparent in figure~\ref{fig:0.5mm_0.2slpm_different_stresses}, where the black solid line corresponds to the computation where both gas normal and tangential stresses are included, the red dashed line corresponds to the computation where only the gas normal stress is included and the blue dash-dotted line corresponds to the computation where only the gas tangential stress is included. It can be seen that excluding the gas tangential/normal stress leads to the error of approximately 58\%/40\%, respectively, for the amplitude of the interface deformation. In addition, it should be mentioned that excluding the  gas tangential stress for our setup implies that the liquid velocity is zero at a steady state, and, thus, without gas tangential stresses it is not possible to describe flow patterns, such as eddies, in the liquid film.

\begin{figure}
\centering
\includegraphics[width=0.45\textwidth]{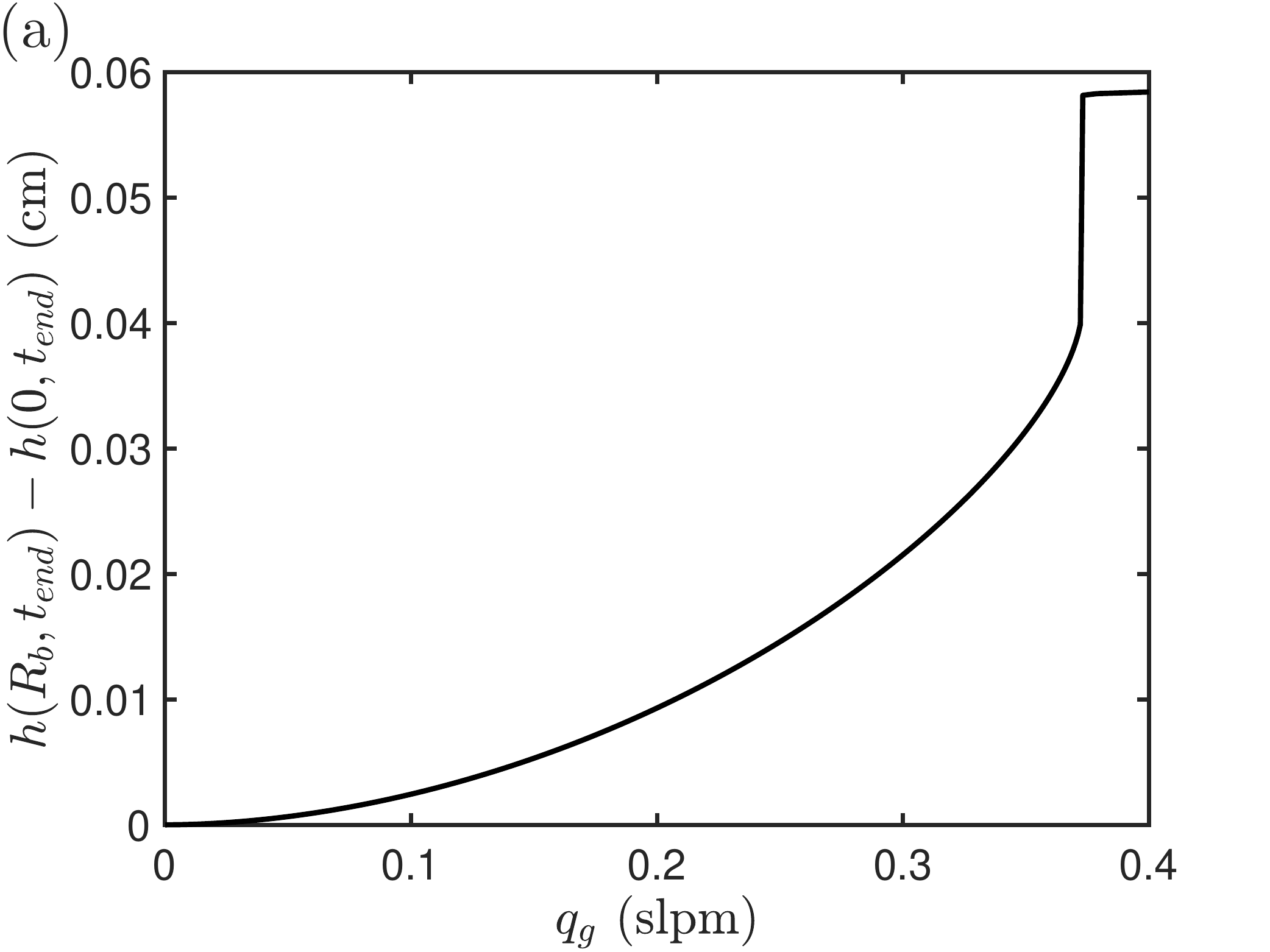}
\includegraphics[width=0.45\textwidth]{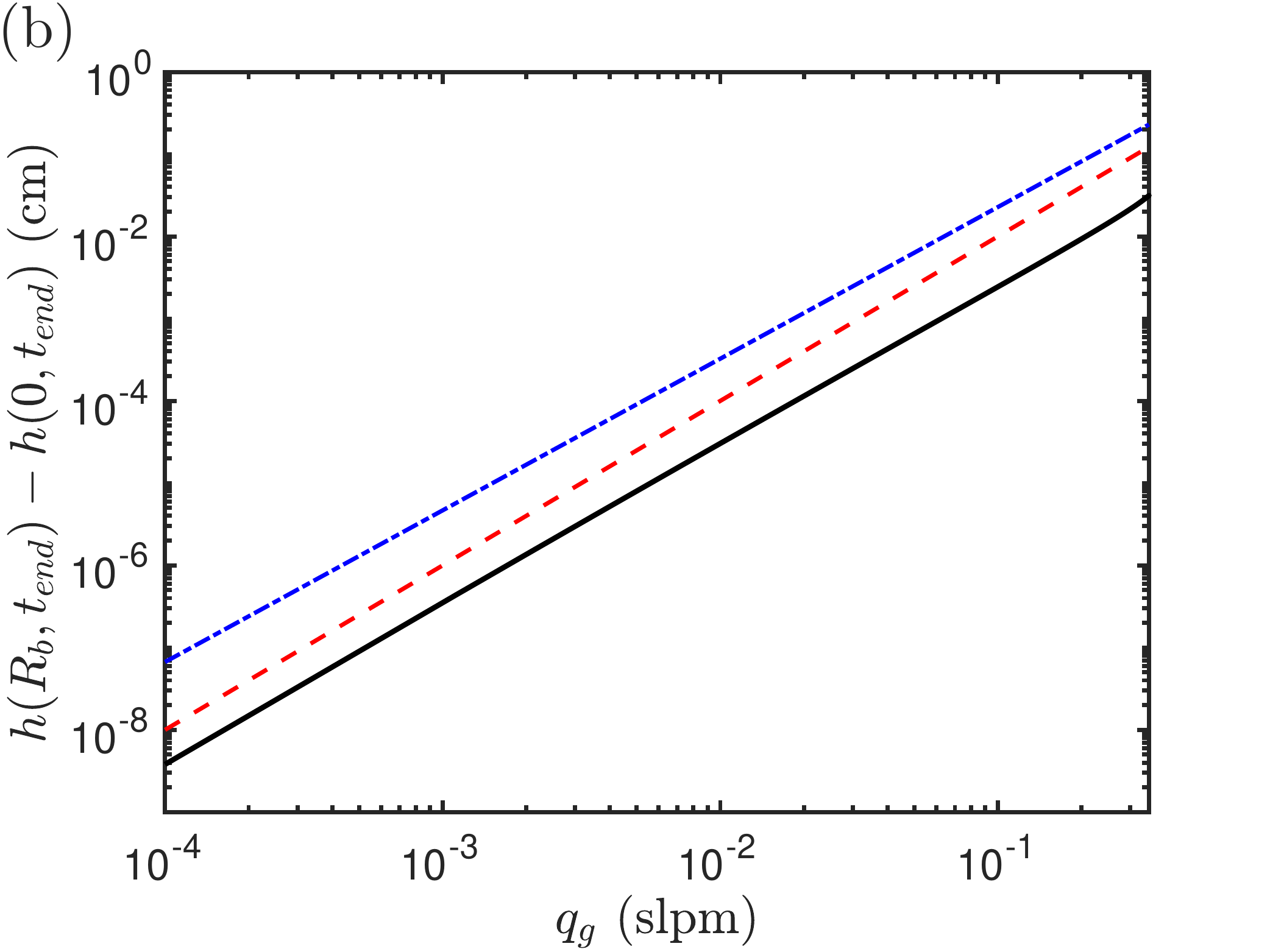}
\caption{Amplitude of the interface deformation of a liquid film of thickness $0.5\,\mathrm{mm}$, 
computed by taking the difference between the liquid thickness at the beaker wall, $h(R_b,t_{end})$, and the liquid thickness at the beaker centre, $h(0,t_{end})$, 
depending on the gas flow rate on normal and log-log scales in panels (a) and (b), respectively (see the black solid lines). 
To compute the interface deformation, the thin-film equation with the analytically-fitted gas stresses was used and it was integrated in time up to $t_{end}=5\,\mathrm{s}$ to ensure that steady-state solutions were reached. In panel (b), the red dashed line corresponds to $q_g^2$ and the blue dash-dotted line corresponds to $q_g^{2.1}\log(q_g)$.}
\label{fig:0.5mm_amplitude}
\end{figure}

Finally, we analyse in more detail how the amplitude of the interface deformation depends on the gas flow rate. We use the thin-film equation with the gas normal and tangential stresses given by the scaling laws and analytical fits (equations (\ref{eq:normal_fit}) and (\ref{eq:tangential_fit})) discussed above. 
Numerical results showing the amplitude of the interface deformation of a liquid film of thickness $0.5\,\mathrm{mm}$ in dependence on the gas flow rate are presented in figure~\ref{fig:0.5mm_amplitude}. The thin-film equation was integrated in time for $5\,$s to ensure that a steady state was reached, and the amplitude was computed for the resulting steady state. Panel (a) corresponds to a normal scale, and it is apparent that a dry spot appears at approximately $q_g=0.37\,\mathrm{slpm}$. This agrees with the DNS results, see figure~\ref{fig:0.5mm_0.37slpm}. Panel (b) represents the result on a log-log scale (see the solid line). There is an evidence of a power-law dependence -- the red dashed line corresponds to $q_g^2$. In addition, we also plot the law $q_g^{2.1}\log(q_g)$ (blue dash-dotted line), and its significance is explained in the analysis below.

Assuming that the gas jet induces a steady-state deformation $h=h(R)$, we obtain that it satisfies the following dimensionless equation:
\begin{equation}
Bo\, h'+[N_s]'-\overline{\Pi}'(h)h'-\biggl(\frac{1}{R}(Rh')'\biggr)'=\frac{3}{2}\frac{T_s}{h},
\label{eq:steady_state}
\end{equation}
where the primes denote differentiation with respect to $R$. Taking into account the scalings for the stresses $N_s$ and $T_s$ introduced above, we can rewrite
\begin{equation}
N_s=q_g^2 f_1(R),\qquad T_s=q_g^\alpha f_2(R/q_g^\beta),
\end{equation}
where $f_1$, $f_2$ (recall expressions~\eqref{eq:normal_fit} and \eqref{eq:f2_fit}) and $q_g$ have been appropriately non-di\-men\-si\-o\-na\-li\-sed, and $\alpha\approx 1.2$, $\beta\approx 0.3$.

To analyse how the amplitude of the interface deformation scales with the gas flow rate, we consider the limit $q_g\ll 1$. In this regime, $h=1+\eta$ with $\eta \ll 1$, and  the disjoining pressure effects can be neglected.  It can be readily seen that if the dominant balance is between the surface tension and gas normal stress terms, then $\eta\sim q_g^2$ (and then the gravity term is also in balance). On the other hand, assuming that the dominant balance is between the surface tension and gas tangential stress terms (it can be shown that the gravity term is of a higher order {\it a posteriori}) and linearising (\ref{eq:steady_state}), we find
\begin{equation}
\biggl(\frac{1}{R}(R\eta')'\biggr)'=-\frac{3}{2}q_g^\alpha f_2(R/q_g^\beta),
\label{eq:steady_state_lin}
\end{equation}
with $\eta'=0$ at $R=0$ and at $R=R_b$ and with $\int_0^{R_b}R\eta\,\mathrm{d}R=0$.  
Rescaling $\eta = q_g^\alpha\tilde\eta$ and denoting $\epsilon = q_g^\beta$, we obtain
\begin{equation}
\biggl(\frac{1}{R}(R\tilde\eta')'\biggr)'=-\frac{3}{2}f_2(R/\epsilon).
\label{eq:steady_state_lin}
\end{equation}
Integrating this equation once and multiplying by $R$ yields 
\begin{equation}
(R\tilde\eta')'=-\frac{3}{2} \epsilon\, R\, F(R/\epsilon) +C_1 R,
\label{eq:steady_state_lin1}
\end{equation}
where $C_1$ is a constant of integration and $F(x)=\int_0^xf_2(s)\,\mathrm{d}s$. Integrating again gives
\begin{equation}
R\tilde\eta'=-\frac{3}{2} \epsilon^3G(R/\epsilon) +\frac{C_1}{2} R^2,
\label{eq:steady_state_lin2}
\end{equation}
where $G(x)=\int_0^x sF(s)\,\mathrm{d}s$. Note that the constant of integration is zero due to the condition $\tilde \eta'=0$ at $R=0$. Also, the condition $\tilde \eta'=0$ at $R=R_b$ implies
\begin{equation}
C_1=\frac{3\epsilon^3}{R_b^2}G(R_b/\epsilon).
\end{equation}
Dividing (\ref{eq:steady_state_lin2}) by $R$ and integrating again, we find 
\begin{equation}
\tilde\eta=-\frac{3}{2} \epsilon^3H(R/\epsilon) +\frac{3\epsilon^3G(R_b/\epsilon)}{4R_b^2} R^2+C_2,
\label{eq:steady_state_lin3}
\end{equation}
where $H(x)=\int_0^x [G(s)/s]\mathrm{d}s$ and $C_2$ is a constant of integration (which can be determined by requiring that  $\int_0^{R_b}R\tilde\eta\,\mathrm{d}R=0$). 

Assuming that $f_2(x)$ is positive and decays sufficiently fast as $x\rightarrow\infty$, it can be concluded that 
${F(x)\sim a_0}$, $G(x)\sim (a_0/2) x^2-a_1$ and $H(x)\sim (a_0/4) x^2-a_1 \log x +a_2$ as $x\rightarrow\infty$ for some constants $a_0>0$, $a_1>0$ and $a_2\in\mathbb{R}$. It can also be shown that $\tilde\eta$ is monotonically increasing, and, therefore, the amplitude of the deformation $\tilde \eta$ is 
\begin{equation}
\tilde\eta(R_b)-\tilde\eta(0)=-\frac{3}{2} \epsilon^3H(R_b/\epsilon) +\frac{3}{4}\epsilon^3G(R_b/\epsilon).
\end{equation}
Using the asymptotic behaviours of $G(x)$ and $H(x)$ as $x\rightarrow\infty$, we conclude that
\begin{equation}
\tilde\eta(R_b)-\tilde\eta(0)=O(\epsilon^{3}\log\epsilon).
\end{equation}
Hence, the amplitude of the interface deformation scales as
\begin{equation}
\eta(R_b)-\eta(0)=O(q_g^{\alpha+3\beta}\log q_g),
\end{equation}
which for $\alpha\approx 1.2$ and $\beta\approx 0.3$ becomes of $O(q_g^{2.1}\log q_g)$. 
This is asymptotically smaller than $O(q_g^2)$, thus, strictly speaking, the gas normal stress must be dominant. 
However, in practice, these orders are close to each other (and, in fact, $q_g^{2.1}\log q_g$ becomes smaller than $q_g^2$ only for $q_g < 2.91\times 10^{-16}$, i.e. 
for extremely small values of $q_g$). Hence, for practical purposes, gas normal and tangential stresses are equally important, which is in agreement with the results presented in figure~\ref{fig:0.5mm_0.2slpm_different_stresses}.

\subsection{Dewetting}

Before discussing dewetting induced by gas jets, it is useful to analyse the linear stability of flat-film equilibrium solutions in the absence of the gas jet as well as to construct bifurcation diagrams of various non-equilibrium solutions. It will be shown later that many of the observed behaviours in gas-jet-induced dewetting can be explained by such an analysis. 

\subsubsection{Linear stability analysis and equilibrium solutions}
\label{sect:analysis}

We will utilise the thin-film equation (\ref{eq:th_eq_axisymm}) for our analysis. We consider a liquid film in the absence of a gas jet and analyse steady-state solutions of the thin-film equation, which takes the form
\begin{equation}
h_t+\frac{1}{R}\bigg[\frac{R h^3}{3}\biggl(\frac{1}{R}(Rh_R)_R-Bo\, h+\overline{\Pi}(h)\biggr)_R\bigg]_R=0.
\label{eq:th_eq_axisymm_no_gas}
\end{equation}

Due to the destabilising effect of the long-range attractive forces, a sufficiently thin film must become linearly unstable. In (\ref{eq:th_eq_axisymm_no_gas}) distances have been non-dimensionalised with the undisturbed film thickness, $h_0$, making the undisturbed dimensionless film thickness is $1$, so varying the dimensional film thickness is equivalent to varying the dimensionless parameters (and the dimensionless domain size). Thus, we consider the dimensionless equation (\ref{eq:th_eq_axisymm_no_gas}) and perform the linear stability analysis of the solution $h\equiv 1$ to find out the linear instability conditions in terms of the dimensionless parameters. Note that the linear stability analysis for a similar equation but with the disjoining pressure containing only a destabilising term was performed by \cite{WitelskiBernoff2000_PhysD}. Substituting $h(R,t)=1+h_1(R,t)$ into  (\ref{eq:th_eq_axisymm_no_gas})  and assuming that $|h_1(R,t)|\ll 1$, we obtain the following linearised equation:
\begin{equation}
h_{1t}=\mathcal{L}[h_1]\equiv-\frac{1}{R}\bigg[\frac{R}{3}\biggl(\frac{1}{R}(Rh_{1R})_R+\bigl(\overline{\Pi}'(1)-Bo\bigr) h_{1}\biggr)_{\!R}\bigg]_R.
\end{equation}
The associated boundary conditions are
\begin{equation}
h_{1R}=0,\qquad h_{1RRR}=0\qquad\text{at}\qquad R=0
\end{equation}
and
\begin{equation}
h_{1R}=0,\qquad Rh_{1RRR}+h_{1RR}=0\qquad\text{at}\qquad R=\overline{R}_b.
\end{equation}
It can be concluded from \cite{WitelskiBernoff2000_PhysD} that the eigenvalues of the operator $\mathcal{L}$ are given by
\begin{equation}
\lambda_n=\frac{1}{3}\bigl[-\Lambda_n^4+\bigl(\overline{\Pi}'(1)-Bo\bigr)\Lambda_n^2\bigr],\qquad n=1,\,2,\,\ldots,
\end{equation}
where $\Lambda_n$'s are the eigenvalues of the Helmholtz problem
\begin{equation}
\frac{1}{R}(Rh_{1R})_R+\Lambda^2 h_{1}=0,
\end{equation}
with $h_R=0$ at $R=0$ and at $R=\overline{R}_b$. The eigenfunctions of $\mathcal{L}$ are the corresponding eigenfunctions of the Helmholtz problem. The eigenvalues of the Helmholtz problem are real, and it is enough to consider the positive ones. Assuming that the smallest positive eigenvalue is $\Lambda_1$, the condition for linear instability becomes $\lambda_1>0$, or, equivalently,
\begin{equation}
Bo<\overline{\Pi}'(1)-\Lambda_1^2.
\end{equation}
The eigenvalues of the Helmholtz problem are solutions of the equation $J_1(\Lambda \overline{R}_b)=0$, where $J_1$ is the first-order Bessel function of the first kind. Denoting the smallest positive root of $J_1$ by $x_1\approx 3.832$, we then find that $\Lambda_1=x_1\big/\,\overline{R}_b$. The corresponding eigenfunction is $J_0(\Lambda_1 R)$, where $J_0$ is the zeroth-order Bessel function of the first kind. The linear instability condition becomes
\begin{equation}
Bo<\overline{\Pi}'(1)-\frac{x_1^2}{\overline{R}_b^2}.
\end{equation}
In terms of the dimensional parameters, this condition can be rewritten as
\begin{equation}
\bigg(\frac{\rho_l g}{\gamma}+\frac{x_1^2}{R_b^2}\biggr)h_0^7-\frac{3A}{\gamma}h_0^3+\frac{6B}{\gamma}<0.
\label{eq:linstab_ineq}
\end{equation}

%%%
\begin{figure}
\centering
\includegraphics[scale=0.45]{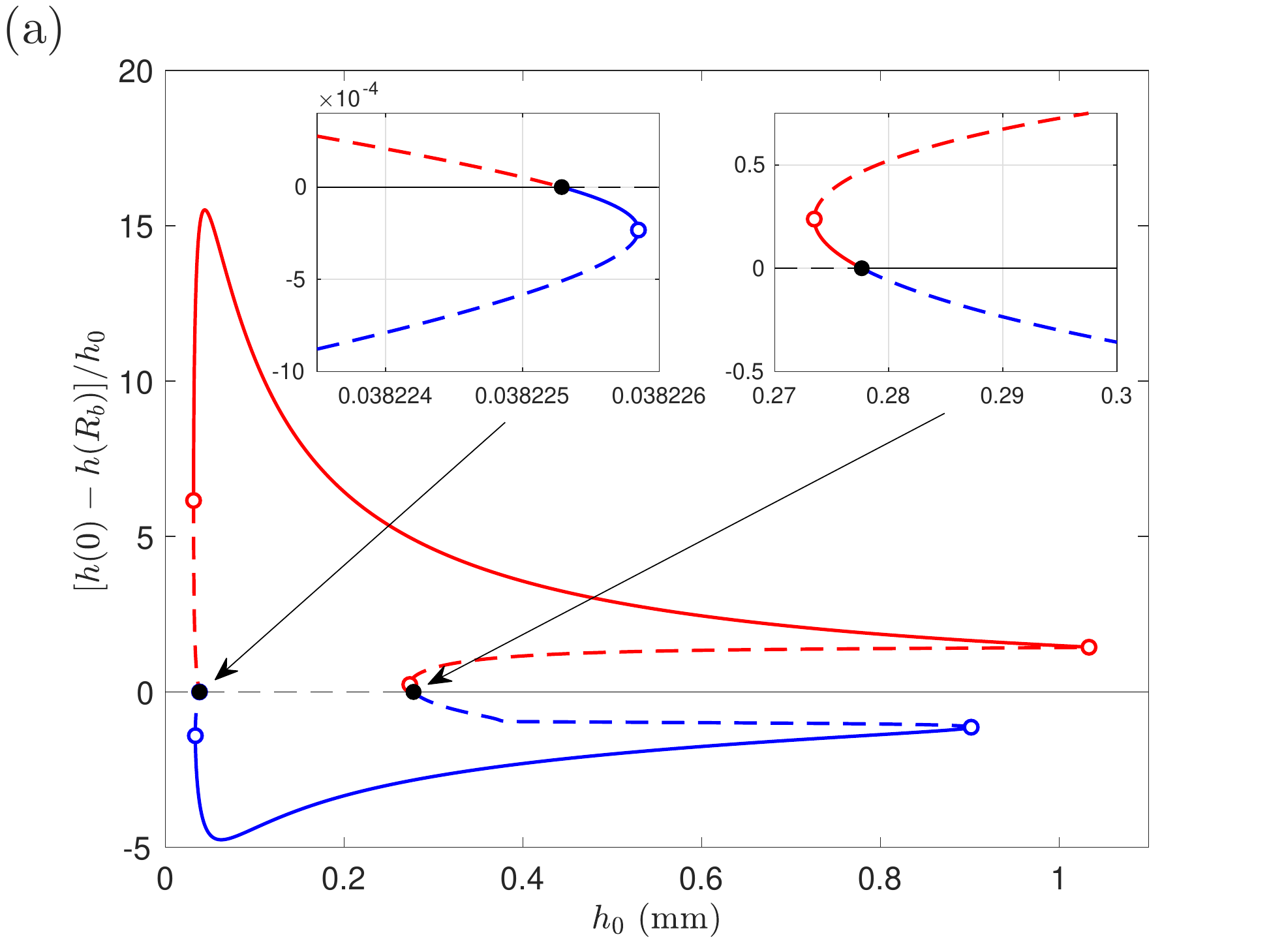}
\centering
\includegraphics[scale=0.45]{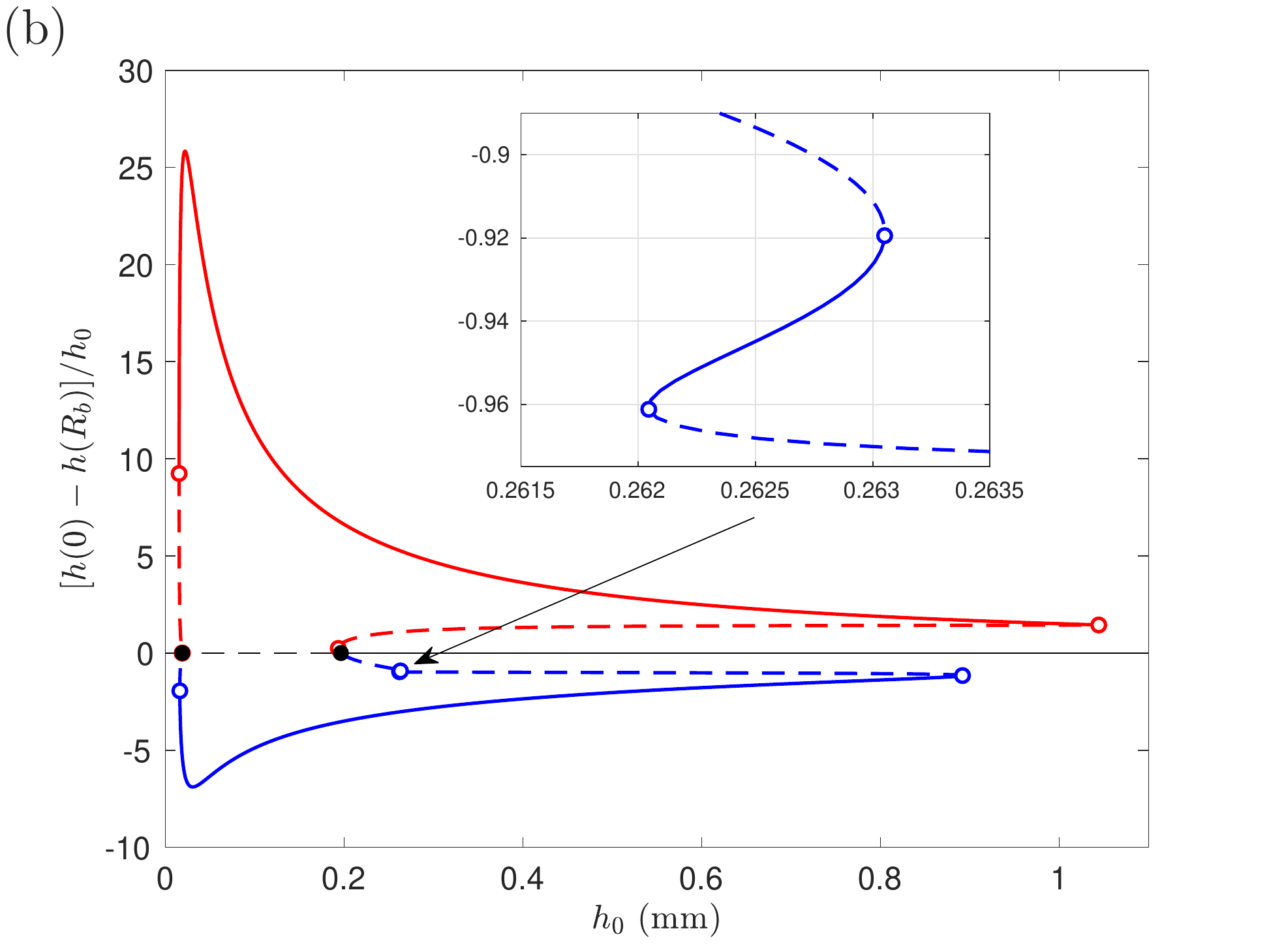}
\caption{Bifurcation diagrams of steady-state solutions of the thin-film equation (\ref{eq:th_eq_axisymm_no_gas}) when the equilibrium contact angle is $\theta_{eq}=30^\circ$ and the precursor thickness is (a)~$h_{eq}=0.03\,\mathrm{mm}$ or (b)~$h_{eq}=0.015\,\mathrm{mm}$ showing the dependence of the dimensionless quantity $[h(0)-h(R_b)]/h_0$ on the undisturbed film thickness $h_0$. The branches of flat solutions are shown in black. The branches of solutions with a minimum at $R=0$ are below the horizontal line at zero (and are shown in blue). The branches of solutions with a maximum at $R=0$ are above the horizontal line at zero (and are shown in red). The parts of the branches corresponding to stable and unstable solutions are shown with solid and dashed lines, respectively. The primary bifurcation points on the branches of flat-film solutions are indicated with black filled circles. Empty circles correspond to turning points on the branches.}
\label{fig:bif_diag_30deg_0p03mm}
\end{figure}
%%%

For a given equilibrium precursor thickness, it can be verified that the inequality (\ref{eq:linstab_ineq}) has solutions if the contact angle is sufficiently large. For example, taking $h_{eq}=0.03\,\mathrm{mm}$, we find that the liquid film may become linearly unstable if $\theta\gtrsim 1.21^\circ$, and taking $h_{eq}=0.015\,\mathrm{mm}$, we find that the liquid film may become linearly unstable if $\theta\gtrsim 0.61^\circ$. If the latter condition is satisfied, we find that there exist a range of film thicknesses, $h_0\in(h_a,h_b)$, for which the flat film solution is linearly unstable. For example, for $h_{eq}=0.03\,\mathrm{mm}$ and $\theta_{eq}=30^\circ$, we find that $h_a\approx 0.03822516\,\mathrm{mm}$ and $h_b\approx 0.27959481\,\mathrm{mm}$, and for $h_{eq}=0.015\,\mathrm{mm}$ and $\theta_{eq}=30^\circ$, we find that $h_a\approx 0.01911091\,\mathrm{mm}$ and $h_b\approx 0.19778522\,\mathrm{mm}$. (Overall, $h_a$ and $h_b$ become smaller as $h_{eq}$ decreases for a given value of $\theta_{eq}$.) According to the bifurcation theory, for film thicknesses close to the values $h_a$ and $h_b$ at which linear stability of the uniform-thickness solution changes, there exist non-uniform solutions that bifurcate from the uniform solutions. (There of course exist other branches of solutions that bifurcate from the points where more modes become unstable. Such solutions are, however, less relevant to the present study and we do not discuss them here.) The nature of the bifurcations can be analysed by utilising a weakly nonlinear analysis to obtain the evolution equation for the amplitude of the unstable mode $J_0(\Lambda_1 R)$, see e.g.\ \cite{WitelskiBernoff2000_PhysD} for a similar analysis. It turns out that the bifurcations are transcritical, and this is demonstrated e.g.\ in figure~\ref{fig:bif_diag_30deg_0p03mm}(a) for $h_{eq}=0.03\,\mathrm{mm}$ and $\theta_{eq}=30^\circ$, 
which shows the bifurcation diagram of the non-uniform solutions when $h_0$ is used as the bifurcation parameter. We use the dimensionless quantity $[h(0)-h(R_b)]/h_0$ 
as a measure of the interfacial shape distortion of the solutions. The uniform solutions then correspond to the horizontal line with the measure equal to zero (see the black line). The solid lines show stable solutions and the dashed lines show unstable solutions. The black filled circles show the primary bifurcation points. We find that these points are connected with each other by branches of non-uniform solutions, and, in fact, we find that from each of the primary bifurcation points there emerge two types of solutions. Namely, we obtain solution which have a maximum at $R=0$ (the corresponding branch is above the horizontal line at zero and is shown in red). For $h_0=h_a$, we find that this branch initially goes to the left and is initially unstable up to the first turning point (the turning points are shown with empty circles). For $h_0=h_b$, we find that this branch also initially goes to the left and is initially stable. We also obtain solution which have a minimum at $R=0$ (the corresponding branch is below the horizontal line at zero and is shown with blue colour). Solutions of this type are more relevant to the study of the deformation of liquid films under gas jets, and we will, therefore, discuss them in more detail. For $h_0=h_a$, we find that this branch initially goes to the right and is initially stable for a very small range of values of $h_0$ (up to the turning point at $h_0=h_{t1}\approx 0.03822585\,\mathrm{mm}$). After the turning point at $h_0=h_{t1}$ the branch becomes unstable up to the second turning point at $h_0=h_{t2}\approx 0.03355743\,\mathrm{mm}$. Then, the solutions become stable up to the next turning point at $h_0=h_{t3}\approx 0.9016\,\mathrm{mm}$. The next part of the bifurcation curve is unstable and goes to the left terminating at the primary bifurcation point, $h_0=h_b$. The insets in the figure zoom into the regions around the primary bifurcation points and confirm the transcritical nature of the bifurcations. 

%%%
\begin{figure}
\centering
\includegraphics[scale=0.4]{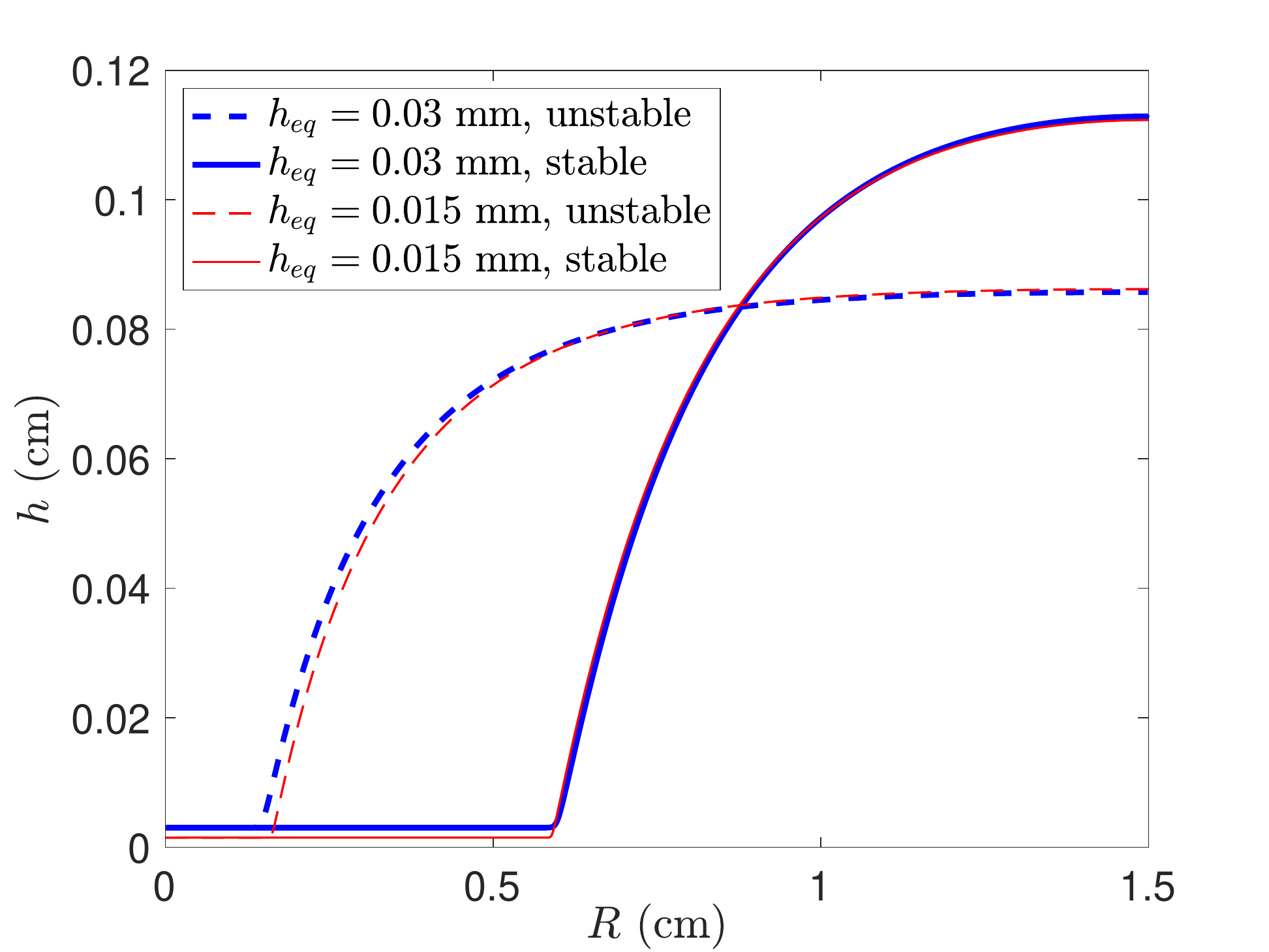}
\caption{Steady-state solutions with a minimum at $R=0$ of the thin-film equation (\ref{eq:th_eq_axisymm_no_gas}) when the average film thickness is $h_0=0.8\,\mathrm{mm}$, the equilibrium contact angle is $\theta_{eq}=30^\circ$ and the precursor thickness is $h_{eq}=0.03\,\mathrm{mm}$ (thick blue lines) or $h_{eq}=0.015\,\mathrm{mm}$ (thin magenta lines). The solid and dashed lines correspond to stable and unstable solutions, respectively.}
 \label{h_30deg_0p8}
\end{figure}
%%%

We conclude that for $h_0>h_{t3}$ stable non-uniform solutions having a minimum in the centre do not exist. Thus, if we consider a film of an average thickness $h_0>h_{t3}$ that was initially deformed by a gas jet with the gas flow switched off after some time, we expect the dry spot to heal with time with the liquid film eventually levelling up and returning to a uniform-thickness state. Note that the healing process of liquid films has been analysed in detail by \cite{Dijksman_etal_2015}, \cite{Bostwick_etal_2017}, \cite{zheng_etal_2018}, and in \cite{zheng_etal_2018} in particular it has been shown that there 
exists self-similarity in the healing process and the solutions that govern the healing process are self-similar solutions of the second kind.

For $h_0\in (h_b,h_{t3})$, in addition to stable uniform-thickness solutions, there exist stable non-uniform-thickness solutions which have a form of dewetted liquid films with a dry spot in the centre. (As mentioned above, we ignore solutions with a maximum at $R=0$ as these are not relevant to our study.) These solutions are separated from the uniform-thickness solutions by an unstable part of the branch of non-uniform solutions. The solutions of this unstable part of the branch also have a form of dewetted liquid films with a dry spot in the centre for $h_0\gtrsim0.38\,\mathrm{mm}$ but with the smaller radii of the dry spots compared to the dry spots for the corresponding stable solutions. For $h_0\lesssim0.38\,\mathrm{mm}$ the solutions of the unstable part have a localised minimum in the centre, with the thickness greater than the precursor film thickness, so that a dry spot does not appear. The stable and unstable solutions for $h_0=0.8\,\mathrm{mm}$ are shown in figure~\ref{h_30deg_0p8} with the thick (blue) solid and dashed lines, respectively. As regards a liquid film deformed by a gas jet for such film thicknesses, we expect that if the gas flow rate is not strong enough, the liquid film may initially dewet in the centre but then return to the uniform state after the gas jet is switched off. However, if the gas jet is strong enough to push the liquid film profile beyond the unstable non-uniform solution, we expect that the liquid film remains dewetted in the centre even after the gas flow is switched off. 

For $h_0\in (h_{t1},h_b)$, the uniform thickness solution is linearly unstable and we therefore expect that for any gas flow rate the liquid film will dewet and remain dewetted even after the gas flow is switched off. We are not interested in the behaviour of very thin liquid films of thicknesses close to or smaller than the precursor-film thickness and thus we do not discuss the expected behaviour for $h_0<h_{t1}$.

In figure~\ref{fig:bif_diag_30deg_0p03mm}(b), we  show the bifurcation diagram of the non-uniform solutions for the same equilibrium contact angle ($\theta_{eq}=30^\circ$) as in figure~\ref{fig:bif_diag_30deg_0p03mm}(a) but for a twice as thin 
equilibrium precursor thickness,  $h_{eq}=0.015\,\mathrm{mm}$. We notice that the bifurcation diagram agrees well with the one for $h_{eq}=0.03\,\mathrm{mm}$, 
particularly for larger values of $h_0$. One qualitative difference is the appearance of two additional turning points on the branch of non-uniform solutions with the minimum in the centre, at $h_0=h_{t4}\approx 0.262$ and at $h_0=h_{t5}\approx 0.263$. This can be clearly seen in the inset zooming into the region around these turning points. This implies that there is a small range of the film thicknesses, $h_0\in (h_{t4},h_{t5})$, where there exist additional stable non-uniform-thickness solutions that have a localised minimum in the centre (without a dry spot).  To confirm qualitative and quantitative similarity of the results for $h_{eq}=0.015\,\mathrm{mm}$ with the results for $h_{eq}=0.03\,\mathrm{mm}$ for larger values of $h_0$, we show in figure~\ref{h_30deg_0p8} the stable and unstable solutions for $h_0=0.8\,\mathrm{mm}$ when $h_{eq}=0.015\,\mathrm{mm}$ with the thin (red) solid and dashed lines, respectively, in addition to the corresponding solutions for $h_{eq}=0.03\,\mathrm{mm}$. We indeed observe that the results for $h_{eq}=0.03\,\mathrm{mm}$ and $h_{eq}=0.015\,\mathrm{mm}$ agree very well.

\subsubsection{Dewetting induced by gas jets}

\begin{figure}
\centering
\includegraphics[scale=0.33]{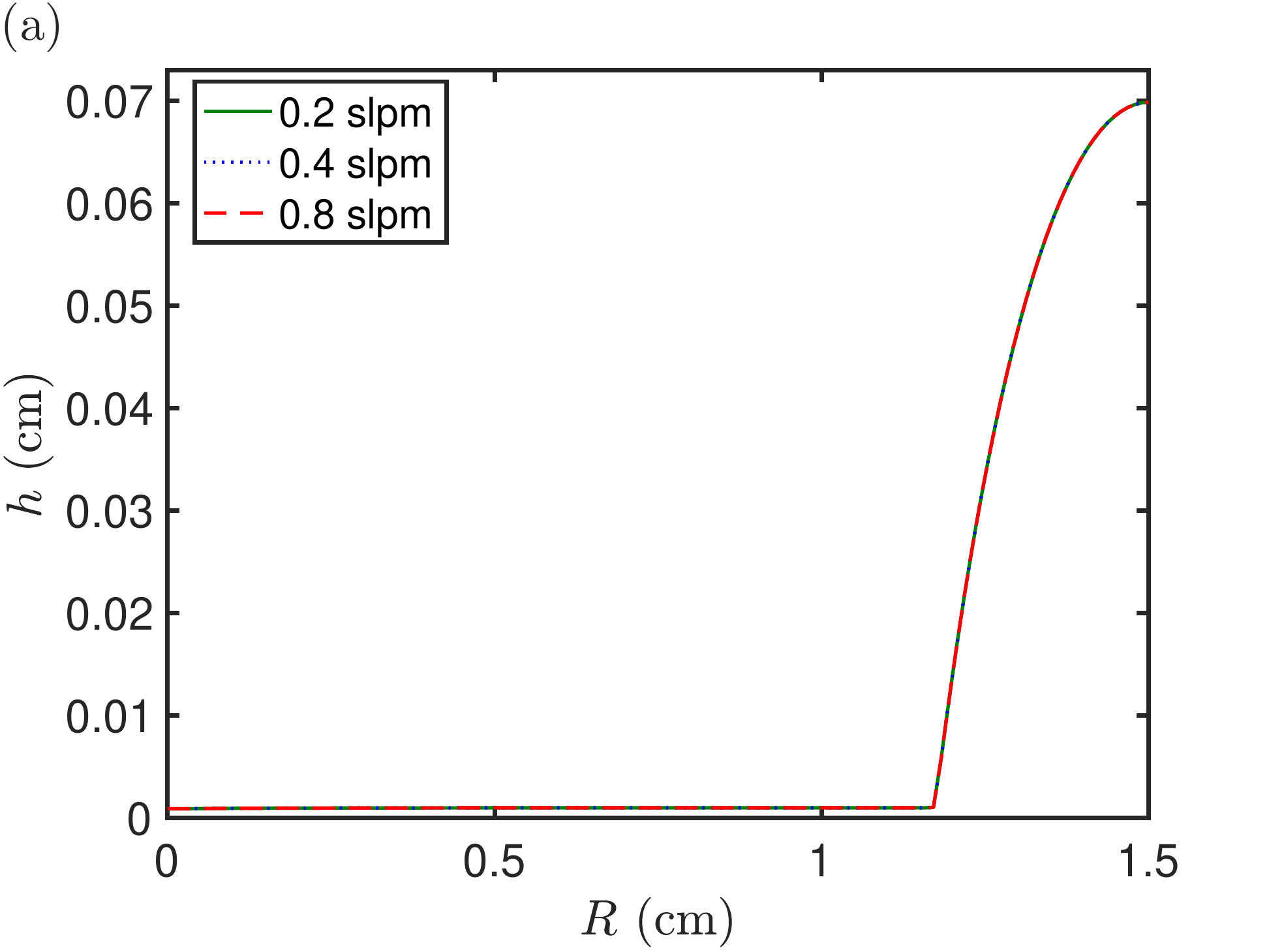}
\includegraphics[scale=0.25]{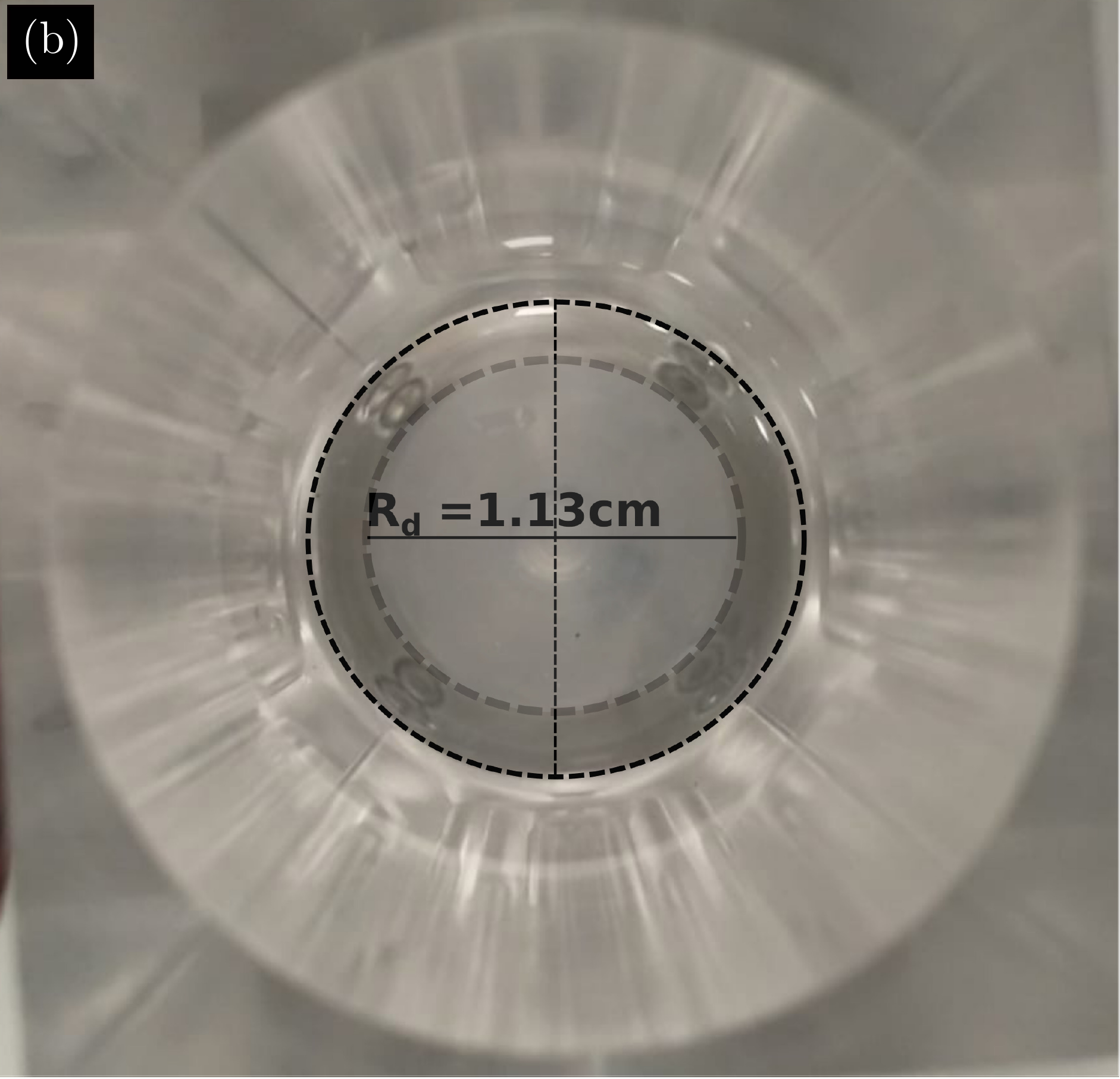}
\caption{(a) COMSOL numerical solutions for dewetting of a water film of initial thickness $0.2\,\mathrm{mm}$ induced by the gas jets of flow rates $q_g=0.2,$ $0.4$ and $0.8\,\mathrm{slpm}$ (green solid, blued dotted and red dashed lines, respectively) obtained using $\theta_{eq}=30^\circ$ and $h_{eq}=0.01\,\mathrm{mm}$. The profiles correspond to time $t=1\,\mathrm{s}$, at which steady states are reached. (b) A top view for the final profile for an experiment of dewetting of a water film of thickness $0.2\,\mathrm{mm}$ induced by a gas jet of flow rate $0.4\,\mathrm{slpm}$.}
\label{f3}
\vspace{0.2cm}
\centering
\includegraphics[scale=0.33]{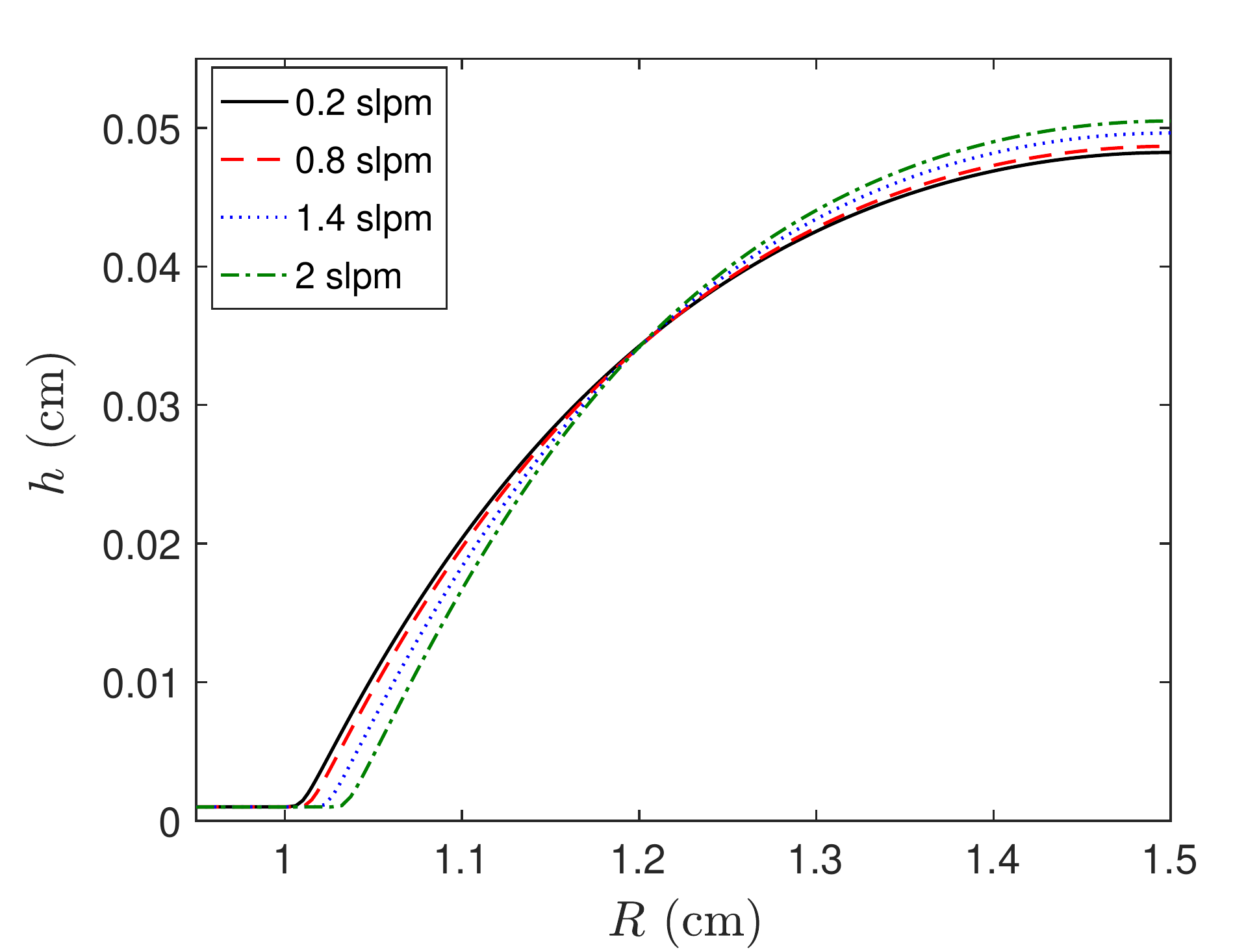}
\caption{COMSOL numerical solutions for dewetting of a water film of initial thickness $0.2\,\mathrm{mm}$ induced by the gas jets of flow rates $q_g=0.2,$ $0.8$, $1,4$ and $2\,\mathrm{slpm}$, as indicated in the legend, obtained using $\theta_{eq}=15^\circ$ and $h_{eq}=0.01\,\mathrm{mm}$. The profiles correspond to time $t=1\,\mathrm{s}$, at which steady states are reached.}
\label{f4}
\end{figure}

We first consider a liquid film of thickness $h_0=0.2\,\mathrm{mm}$. For such a thin film, we used $h_{eq}=0.01\,\mathrm{mm}$ for the COMSOL and thin-film computations. We use the equilibrium contact angle $\theta_{eq}= 30^\circ$, which agrees with the experimental value, see Appendix~\ref{appendix:contact_angle}. Using the analysis from the previous section, it can be shown that  for such values of $\theta_{eq}$ and $h_{eq}$, a water film of thickness $h_0=0.2\,\mathrm{mm}$ is linearly stable, but this value is close the value $h_b\approx 0.1615\,\mathrm{mm}$ below which the liquid film becomes linearly unstable. This indicates that a gas jet of a relatively weak flow rate can destabilise the liquid film so that it would dewet leaving a dry spot in the centre of the beaker. This is confirmed by our numerical simulations using all the three different approaches ($\mathpzc{Gerris}$, COMSOL and the thin-film model). Figure~\ref{f3}(a) shows dewetted liquid film profiles after $t=1\,\mathrm{s}$ for the gas flow rates $q_g=0.2,$ $0.4$ and $0.8\,\mathrm{slpm}$ using DNS in COMSOL. It can be observed that the profiles converge to the same steady state independent of the gas flow rate. We note that the simulations with the thin-film model produce the profiles which are in excellent agreement with the COMSOL results. The $\mathpzc{Gerris}$ simulations also agree very well with these results (with a very small difference in the amplitude for the final film profile due to the fact that for the $\mathpzc{Gerris}$ simulations there is no precursor film and thus the volume of the liquid in the final wetted region is slightly bigger). We, therefore, do not show the thin-film and $\mathpzc{Gerris}$ 
results in figure~\ref{f3}(a) (detailed comparisons are discussed below for thicker water films). 
Note that for the time evolution computations with the thin-film equation in this section we used gas stresses computed under the assumption of a flat interface for the decoupled gas problem.
%%%
\begin{figure}
\centering
\includegraphics[scale=0.36]{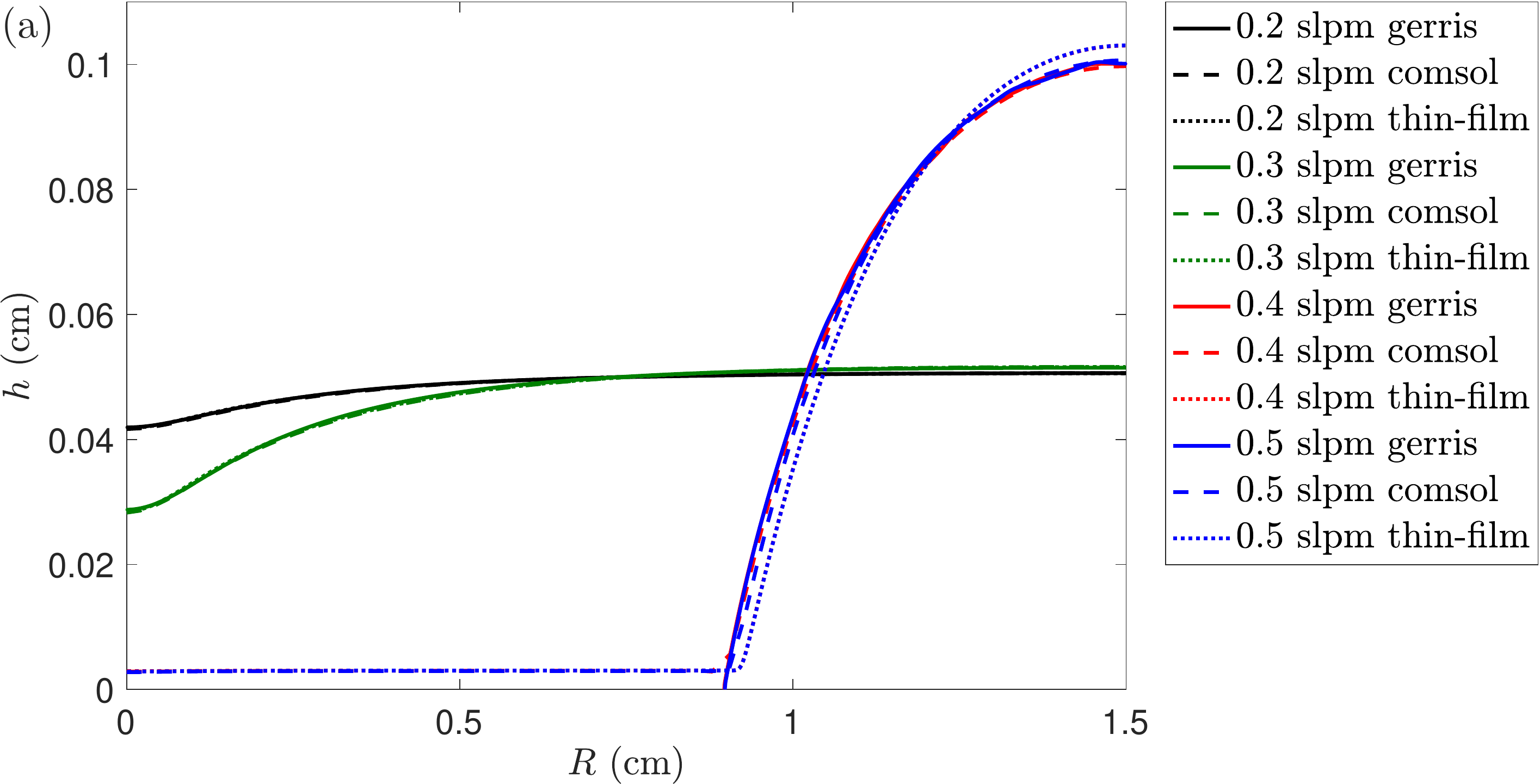}\\[0.35cm]
\includegraphics[scale=0.36]{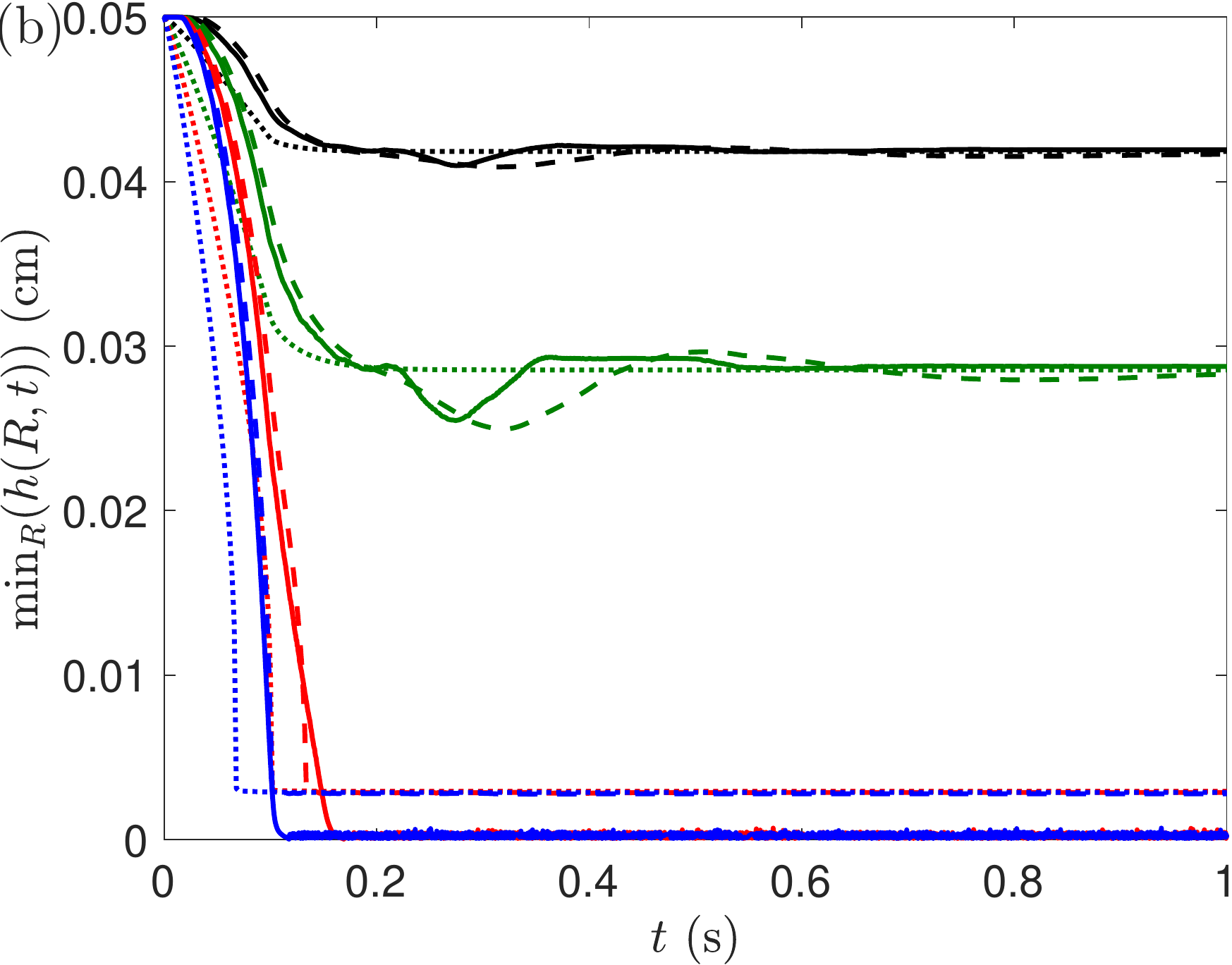}\,\,\,\,\,
\includegraphics[scale=0.36]{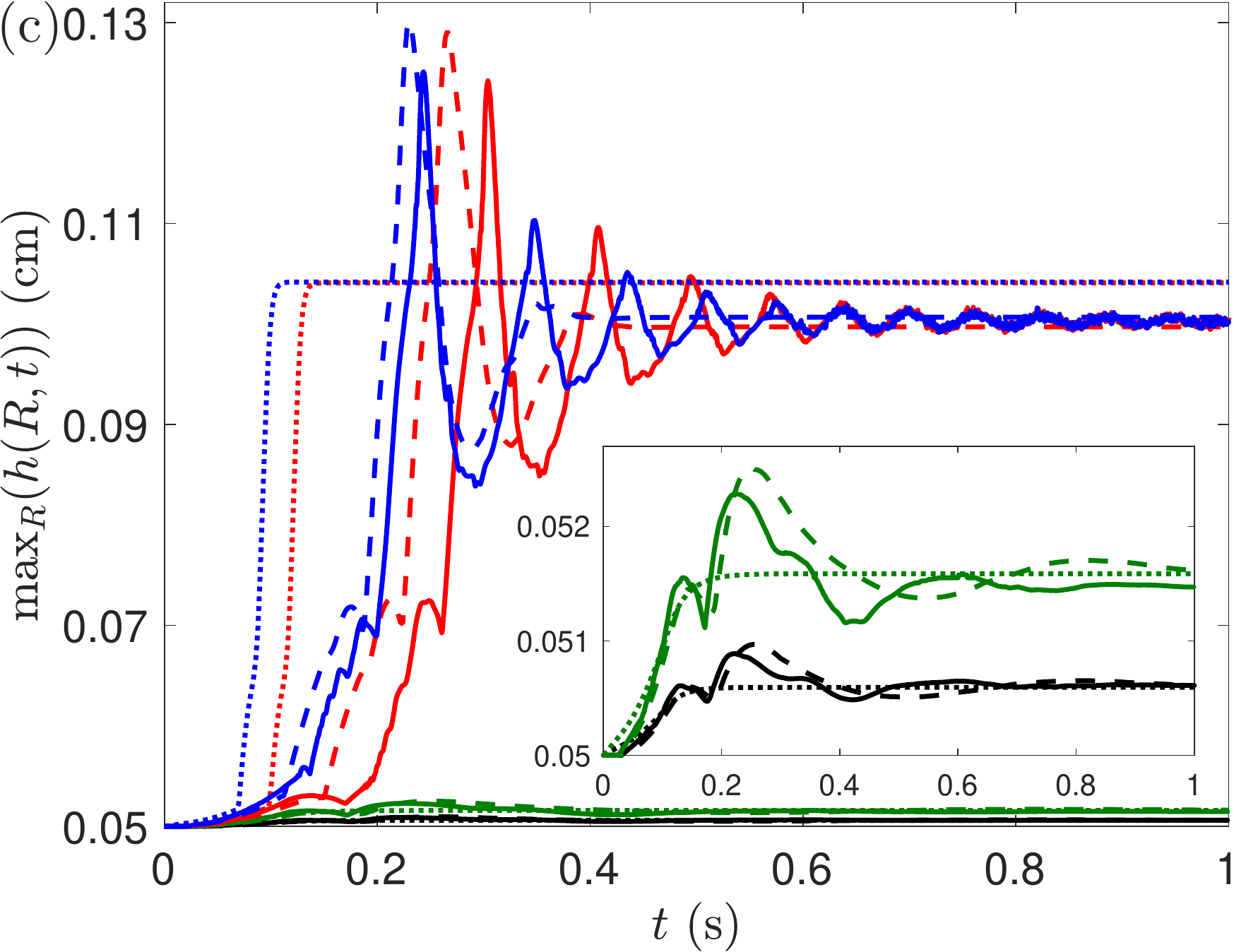}
\caption{Numerical solutions for the evolution of a water film of thickness $0.5\,\mathrm{mm}$ under the gas jets of flow rates $q_g=0.2$, $0.3$, $0.4$ and $0.5\,\mathrm{slpm}$ (black, green, red and blue lines, respectively) obtained using $\theta_{eq}=30^\circ$ and $h_{eq}=0.03\,\mathrm{mm}$. The results obtained using $\mathpzc{Gerris}$, COMSOL and the thin-film equation are shown with the solid, dashed and dotted lines, respectively. Panel (a) shows the profiles at time $t=1\,\mathrm{s}$. Panels (b) and (c) show the evolutions of the minimum and maximum values of the profiles, respectively.}
\label{f11}
\end{figure}
%%%
An experimental result for a dewetted water film of 
thickness $0.2\,\mathrm{mm}$ induced by a gas jet of flow rate $0.4\,\mathrm{slpm}$ is given in figure~\ref{f3}(b) which shows a top view for the final profile. 
The radius of the dry spot turns out to be $R_d\approx 1.13\,\mathrm{mm}$, 
which agrees well with the numerical simulations in which the radius is approximately $R_d\approx 1.12\,\mathrm{mm}$ in the $\mathpzc{Gerris}$ calculation and 
$1.17\,\mathrm{mm}$ in COMSOL. In this case, we can conclude that the gas jet is important for initialising dewetting, but dewetting itself is dominated by the receding contact line motion until the equilibrium contact angle is reached. The gas jet does not affect the final steady-state profile. This can be explained by the fact that for the equilibrium solution the wetted region is such that the gas normal and tangential stresses are negligibly small, see figures~\ref{fig:stresses}(a,d). Numerical results obtained in COMSOL for the final steady-state profiles for the case of a smaller contact angle and the same film thickness are shown in figure~\ref{f4} for the gas flow rates $q_g=0.2$, $0.8$, $1.4$ and $2\,\mathrm{slpm}$. In this case, the radius of the dry spot for the steady-state solutions is approximately $1\,\mathrm{mm}$, and the influence of the gas jet, although weak, becomes noticeable. As expected, the stronger the gas flow the larger the radius of the dry spot becomes. By looking at the gas normal and tangential stresses in figures~\ref{fig:stresses}(a,d), we can notice again that normal stresses are negligible in the wetted area, but now there are small but non-negligible tangential stresses when $R\approx 1\,\mathrm{mm}$, which push the liquid away from the centre.

\begin{figure}
\centering
\includegraphics[scale=0.36]{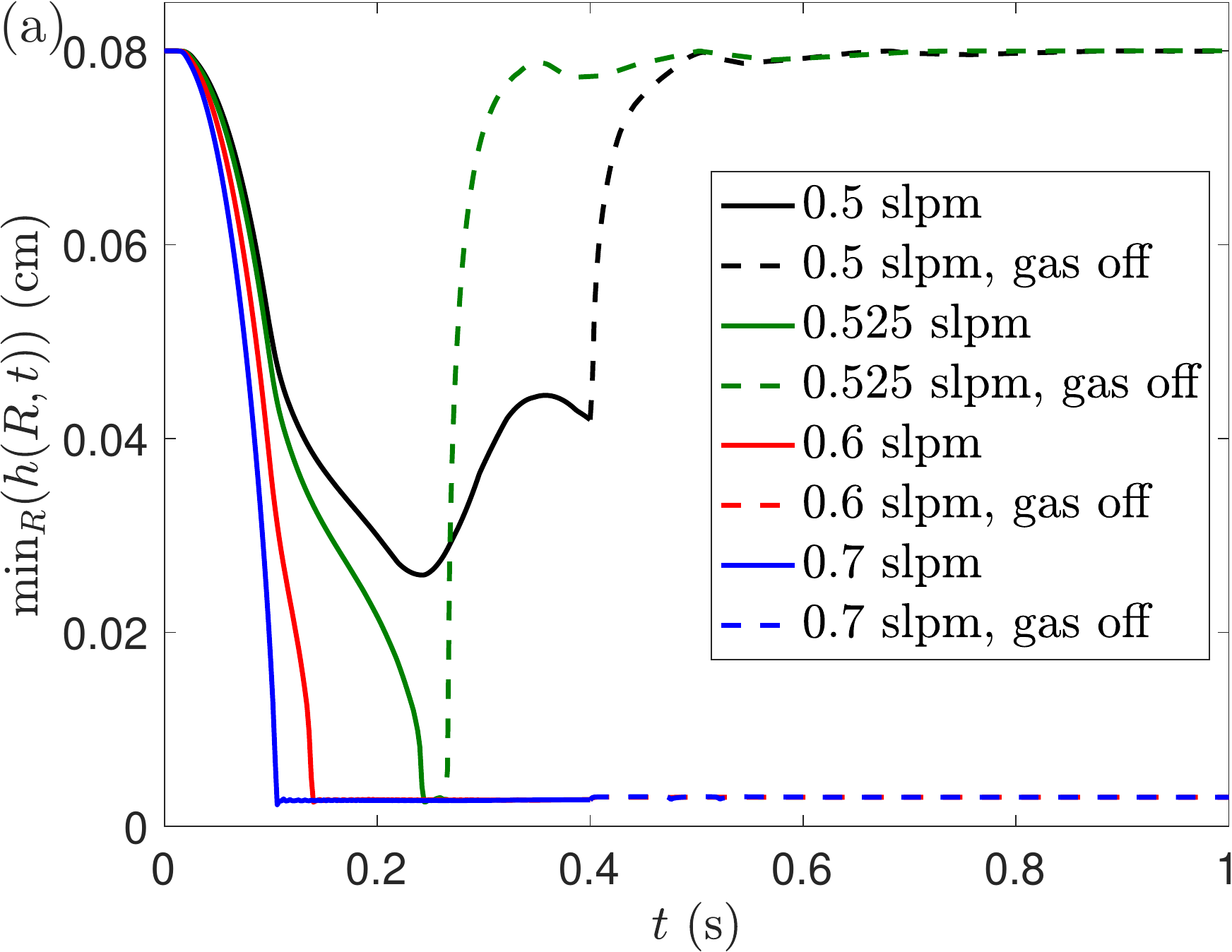}\,\,\,\,\,
\includegraphics[scale=0.36]{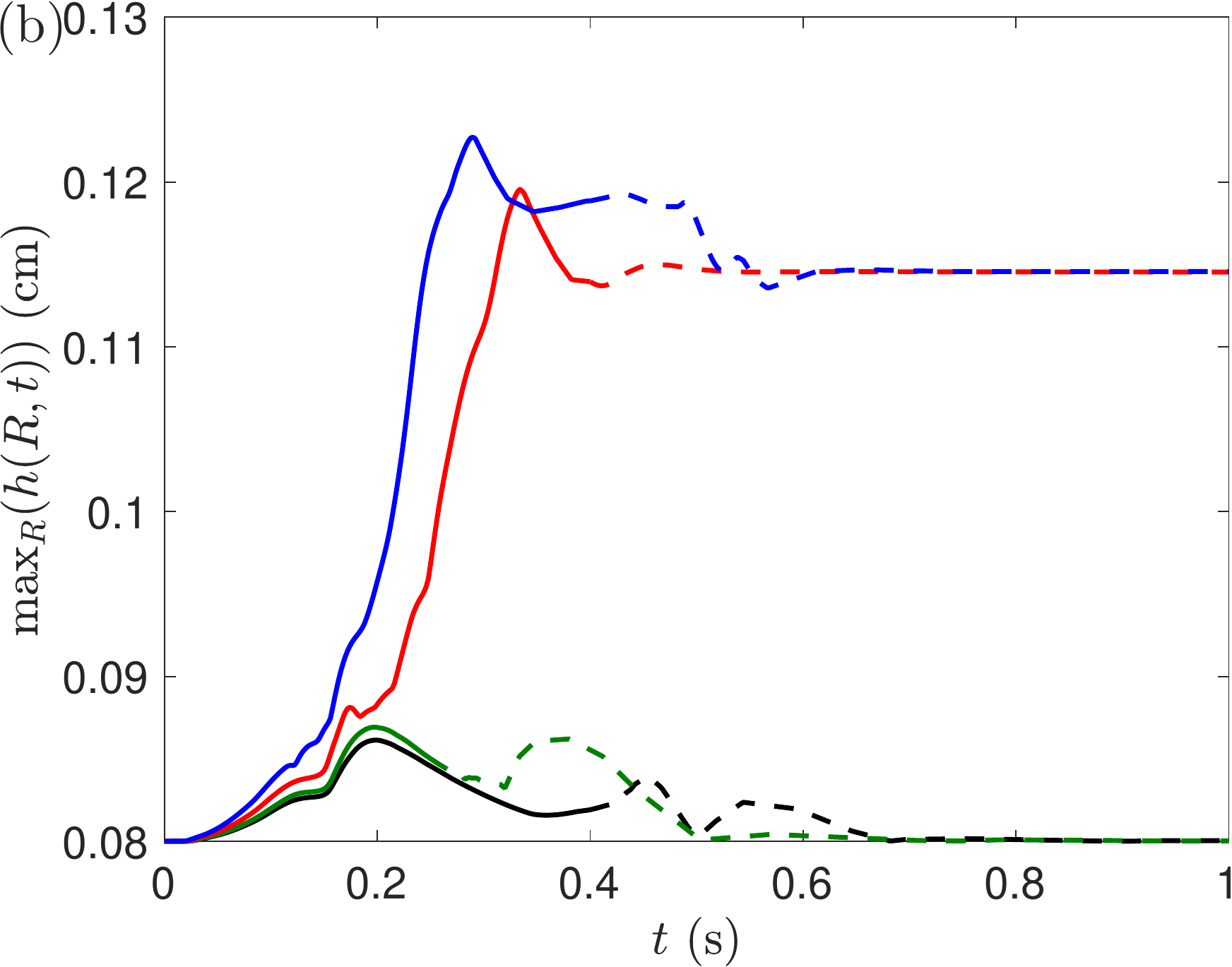}
\caption{COMSOL numerical solutions for the evolution of a water film of thickness $0.8\,\mathrm{mm}$ under the gas jets of flow rates $q_g=0.5$, $0.525$, $0.6$ and $0.7\,\mathrm{slpm}$ (black, green, red and blue lines, respectively) obtained using $\theta_{eq}=30^\circ$ and $h_{eq}=0.03\,\mathrm{mm}$. Panels (a) and (b) show the evolutions of the minimum and maximum values of the profiles, respectively. The gas flow was switched off at $t=0.25\,\mathrm{s}$. 
}
\label{f22}
\vspace{0.5cm}

\includegraphics[scale=0.36]{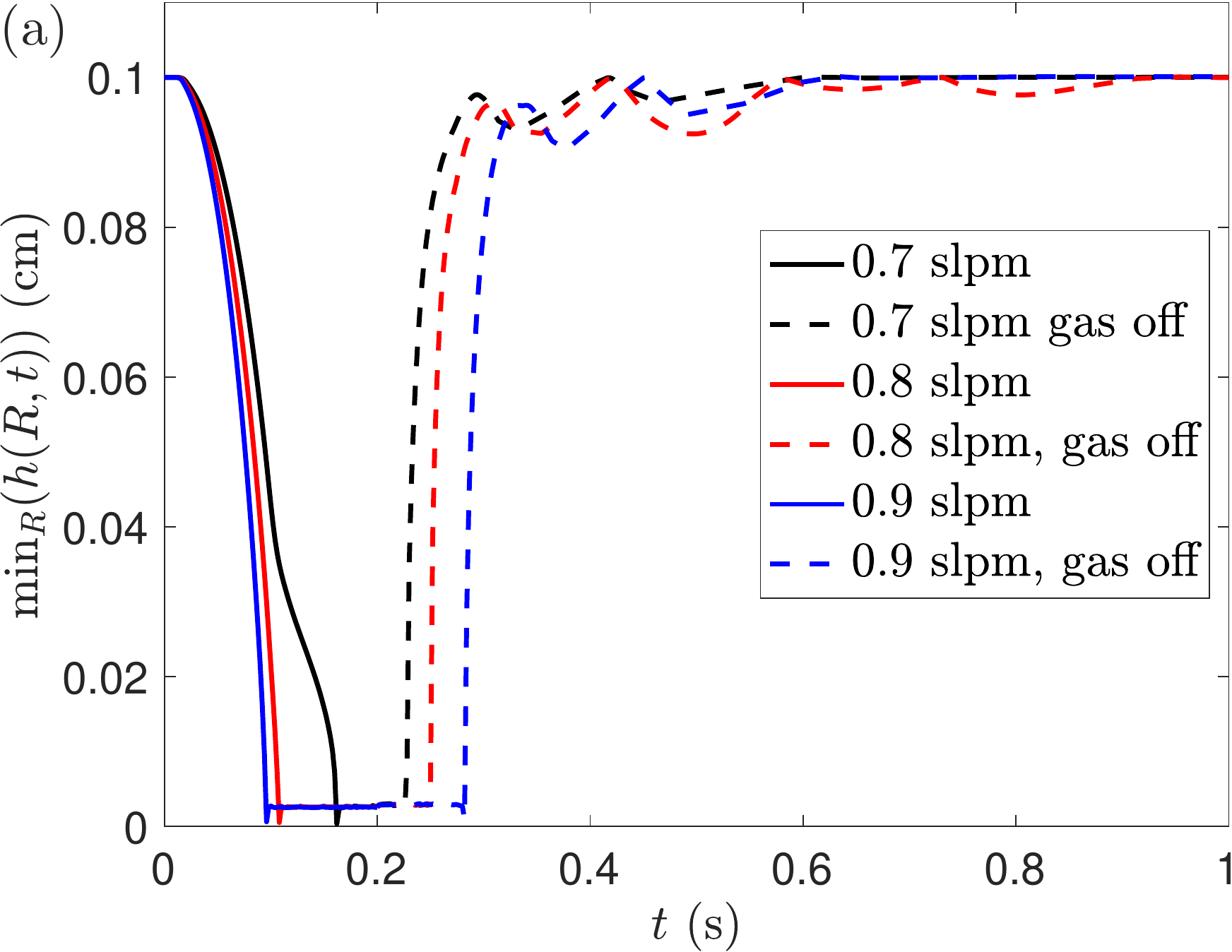}\,\,\,\,\,
\includegraphics[scale=0.36]{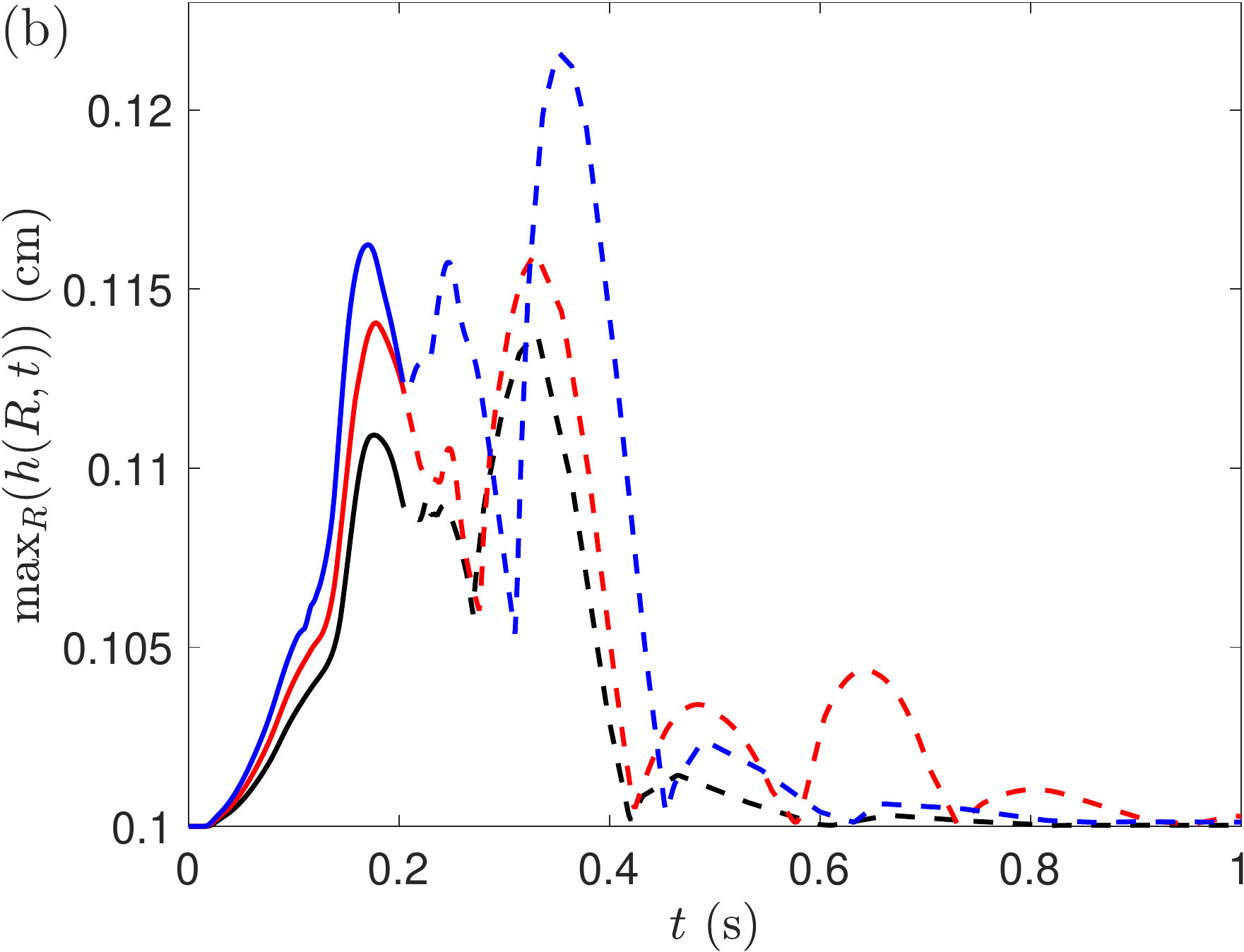}
\caption{COMSOL numerical solutions for the evolution of a water film of thickness $1\,\mathrm{mm}$ under the gas jets of flow rates $q_g=0.7$, $0.8$ and $0.9\,\mathrm{slpm}$ (black, red and blue lines, respectively) obtained using $\theta_{eq}=30^\circ$ and $h_{eq}=0.03\,\mathrm{mm}$. Panels (a) and (b) show the evolutions of the minimum and maximum values of the profiles, respectively. The gas flow was switched off at $t=0.2\,\mathrm{s}$.
}
\label{f33}
\end{figure}

Now we consider a thicker water film with $h_0=0.5\,\mathrm{mm}$ and we assume that $\theta_{eq}=30^\circ$ and $h_{eq}=0.03\,\mathrm{mm}$. Figure~\ref{f11}(a) shows liquid film profiles after $t=1\,\mathrm{s}$ for the gas flow rates $q_g=0.2,$ $0.3$, $0.4$ and $0.5\,\mathrm{slpm}$ (see the black, green, red and blue lines, respectively). Note that at $t=1\,\mathrm{s}$ steady state profiles are reached. This can be confirmed in figures~\ref{f11}(b,c) showing the evolutions of the minimum and maximum values of the profiles, respectively.  We show the results obtained using $\mathpzc{Gerris}$, COMSOL and the thin-film equation, see the solid, dashed and dotted lines, respectively. All the models agree very well, particularly as far as the final profiles are concerned. The evolutions of the minimum and maximum values obtained using $\mathpzc{Gerris}$ and COMSOL show qualitative agreement and indicate oscillatory approach toward steady states indicating the presence of interfacial oscillations/waves. The results obtained using the thin-film equation also show reasonably good agreement with the $\mathpzc{Gerris}$ and COMSOL results, but do not feature oscillations. This may be due to the fact that for such a thickness of the film inertial effects become important and cannot be neglected to accurately describe the evolution of the liquid film. We can observe that dewetting is initiated for a gas flow rate between $0.3$ and $0.4\,\mathrm{slpm}$, and, as before, the final dewetted profiles do not depend much on the strength of the gas jet. This also agrees with our experimental observations.

Next, we consider even thicker water films with $h_0=0.8\,\mathrm{mm}$ and $h_0=1\,\mathrm{mm}$, see figures~\ref{f22} and \ref{f33}, respectively. As before, we assume that $\theta_{eq}=30^\circ$ and $h_{eq}=0.03\,\mathrm{mm}$. Panels (a) and (b) show the evolutions of the minimum and maximum values of the profiles, respectively. For $h_0=0.8\,\mathrm{mm}$ we used the gas flow rates $q_g=0.5$, $0.525$, $0.6$ and $0.5\,\mathrm{slpm}$ (see the black, green, red and blue lines, respectively, in figure~\ref{f22}) and we switched off the gas flow at $t=0.25\,\mathrm{s}$. The solid parts of the curves correspond to the gas flow switched on and the dashed parts correspond to the gas flow switched off. We can observe that dewetting is initiated for a gas flow rate between $0.5$ and $0.525\,\mathrm{slpm}$. However, for $q_g=0.525\,\mathrm{slpm}$ we find out that after the gas flow is switched off, the dry spot in the centre heals and the liquid films returns to the uniform thickness state. This agrees with the theoretical analysis of \S~\ref{sect:analysis}, where we predicted (using the thin-film equation) the coexistence of stable uniform thickness and dewetted solutions in the absence of gas flow. We also predicted that for a relatively weak gas flow the liquid film may dewet in the centre but then heal and return to the uniform thickness state after the gas flow is switched off, as we indeed observe for $q_g=0.525\,\mathrm{slpm}$. For $q_g=0.6$ and $q_g=0.7\,\mathrm{slpm}$, we observe that the liquid film remains dewetted even after the gas flow is switched off, in agreement with the theoretical prediction of \S~\ref{sect:analysis}. This is also confirmed using $\mathpzc{Gerris}$ simulations.

For $h_0=1\,\mathrm{mm}$, the theoretical prediction was that the liquid film should heal after switching off the gas flow, no matter how strong the gas flow was. This is confirmed in figures~\ref{f33}, where we used the gas flow rates $q_g=0.7$, $0.8$ and $0.9\,\mathrm{slpm}$ (see the black, red, and green lines, respectively). We switched off the gas flow at $t=0.2\,\mathrm{s}$. The solid parts of the curves correspond to the gas flow switched on and the dashed parts correspond to the gas flow switched off. In all the cases, the liquid film returns to the uniform thickness state.

\begin{figure}
\centering
\includegraphics[scale=0.38]{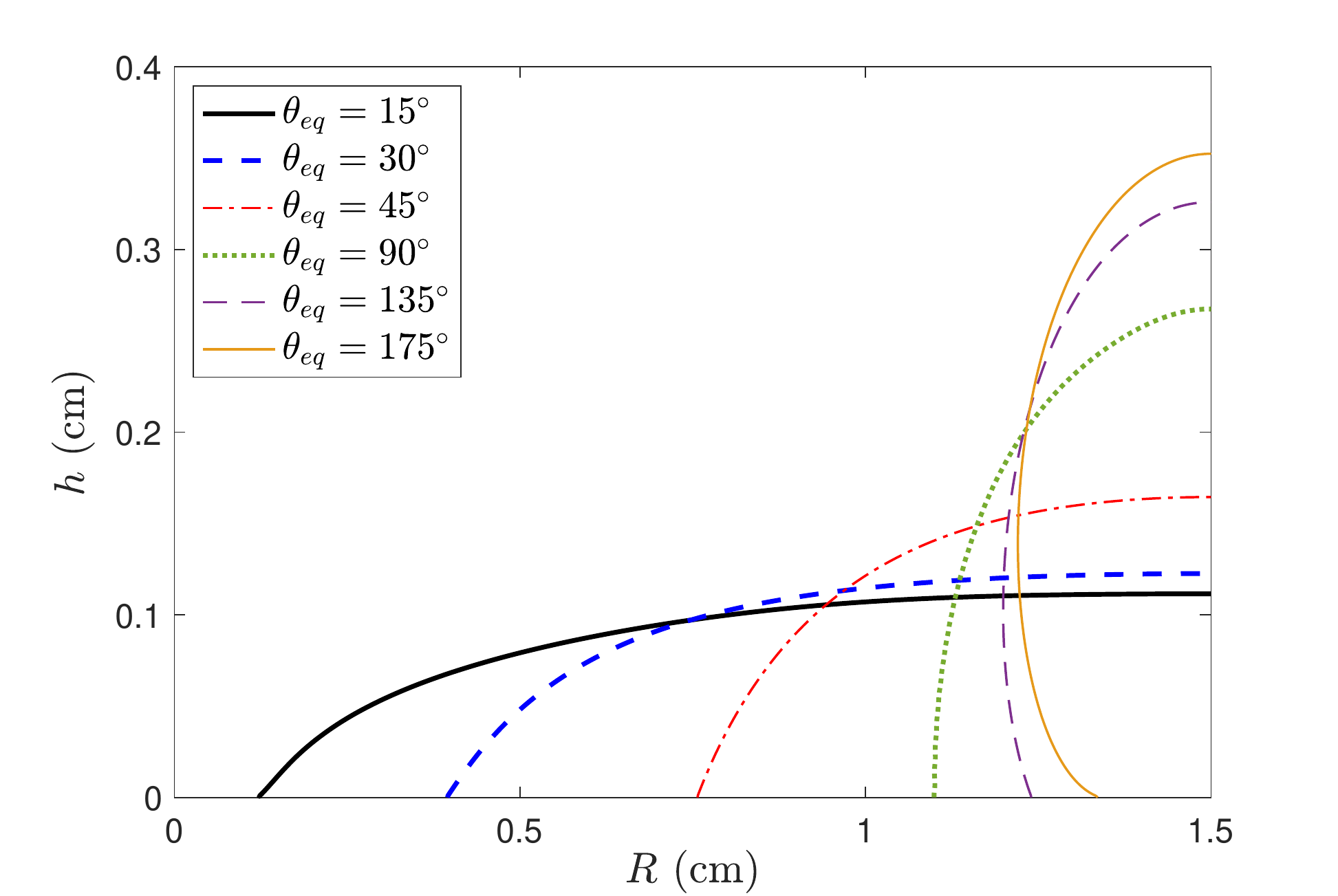}
\caption{$\mathpzc{Gerris}$ numerical solutions for dewetting of a water film of thickness $1\,\mathrm{mm}$ induced by the gas jet of flow rate $q_g=1\,\mathrm{slpm}$ for $\theta_{eq}=15^\circ$, $30^\circ$, $45^\circ$, $60^\circ$, $90^\circ$, $135^\circ$ and $175^\circ$ as indicated in the legend. The profiles correspond to time $t=3\,\mathrm{s}$, at which steady states are reached. 
}
\label{f333}
\end{figure}

Finally, we analyse dewetting of a water film of thickness $h_0=1\,\mathrm{mm}$ for the gas flow rate $q_g=1\,\mathrm{slpm}$ but for different equilibrium contact angles. 
We consider $\theta_{eq}=15^\circ$, $30^\circ$, $45^\circ$, $60^\circ$, $90^\circ$, $135^\circ$ and $175^\circ$ in figure~\ref{f333}, illustrating interface profiles at time $t=3\,\mathrm{s}$, at which steady states are reached. These results were obtained using $\mathpzc{Gerris}$, as for contact angles greater than $90^\circ$ our COMSOL implementation and thin-film model are not suitable. Videos showing the evolution of the water film for $\theta_{eq}=15^\circ$ and $135^\circ$ are included in the supplementary material. 
It can be observed that the approach to a steady state is non-monotonic in both cases, described instead by an oscillatory behaviour. In figure~\ref{f333},  we can observe that for larger contact angles the radius of the dry spot is larger, as expected. For $\theta_{eq}=15^\circ$, we have confirmed the dry spot in the centre heals after the gas flow is switched off (as for $\theta_{eq}=30^\circ$ as we discussed above). However, for the larger equilibrium contact angles the liquid film remains dewetted when the gas flow is switched off.

\section{Conclusions} 

We have examined both experimentally and theoretically the flow arising from a gas jet (air) impinging axisymmetrically on a liquid (water) in a cylindrical beaker. The calculations were carried out using two  direct numerical 
simulation techniques (employing COMSOL and $\mathpzc{Gerris}$, respectively) and a third based on a model employing thin-film approximation
 and making use of a novel iterative process to improve the accuracy of the transfer of stress information from the gas flow to the fluid interface. 
 We have used the wide range of methodologies to explore aspects pertaining to the nonlinear evolution of the flow, 
from deformation to dewetting of the interface. This has resulted in both mathematical understanding of the underlying physics through 
the use of bifurcation analysis of the model equations within their range of validity, as well as the identification of more complex features 
such as domain finite-size effects as well as the generated flow inside the liquid in these more realistic conditions.

The deformation of the liquid surface, determined experimentally when a steady state had been reached, was shown to be in good agreement 
with the DNS results for all water depths, when the gas flow rate was low, and in the case of a thin liquid film with all three models. 
It was also shown that interface shapes and streamline patterns calculated using the thin film model were in agreement 
with DNS for film thicknesses much larger than those for which the thin-film approximation was strictly valid.

Experiments were used to determine interface shapes both in the steady state and during the development of a steady flow after the gas jet was switched on. The contact angle between the liquid and its container was also measured experimentally and this was used as an input into the theoretical models. In the thin-film case, when the gas flow rate was high enough, dewetting of the film from the surface occurred. Using the experimentally measured contact angle of 30$^\circ$ and a precursor thickness of $h_{eq}= 0.1\,\mathrm{mm}$, a parameter determined from the disjoining pressure incorporating long-range and short-range intermolecular forces and used in the COMSOL formulation and the thin-film equation, the conditions for dewetting determined by all three models were found to be in good agreement with experiment.

Dewetting was also investigated using linear stability analysis of various steady-state solutions of the thin-film model for a range of values of initial film thicknesses, contact angles and values of $h_{eq}$. This analysis identified the various regimes in which dewetting could occur in agreement with the DNS and thin-film models. Regimes where the liquid would remain in its dewetted state or heal after the gas jet was turned off were identified.

For thicker films, the agreement between the models was less good for the time dependent flow before the final steady state was achieved, although the agreement for the steady states was still good. DNS results feature decaying interfacial oscillations/waves which were not present in the thin-film model, possibly due to the neglect of the inertial terms in the thin-film approximation. Experiment\textcolor{blue}{s} also showed the oscillations persisting for longer times than those predicted by the models 
 (a video of an experiment for a $5\,\mathrm{mm}$-thick water film and a gas jet of flow rate $1\,\mathrm{slpm}$ showing interfacial oscillations 
is included in the supplementary material). Indeed, preliminary experiments and DNS carried out over a range of gas flow rates for thicker films show that in some cases the oscillations do not decay at all, resulting in `self-sustained oscillations' instead. The conditions for this to occur will be the subject of a future study.

Finally it should be pointed out that in the context of falling liquid films, accurate reduced-order models taking into account inertia have been developed using, for example, the weighted integral-boundary-layer approach \cite[see e.g.][]{Ruyer-Quil2000,Kalliadasis2011,JFM2011,Denner_etal_2016}. Derivation and analysis of such models in the present context is left as a topic for future investigation.

\vspace{0.2cm}
We acknowledge the support of the UK Fluids Network (EPSRC Grant EP/N032861/1) in funding participation in a Special Interest Group meeting and a week-long Short Research Visit of RC to Loughborough University that shaped the direction of this work. CJO would like to acknowledge the Adventure Mini-CDT on Gas-Plasma Interactions with Organic Liquids at Loughborough University for the PhD studentship.\\[-0.1cm]

Declaration of Interests. The authors report no conflict of interest.

\appendix
\section{Substrate--liquid contact angle}\label{appendix:contact_angle}

\begin{figure}
\centering
\includegraphics[scale=0.4]{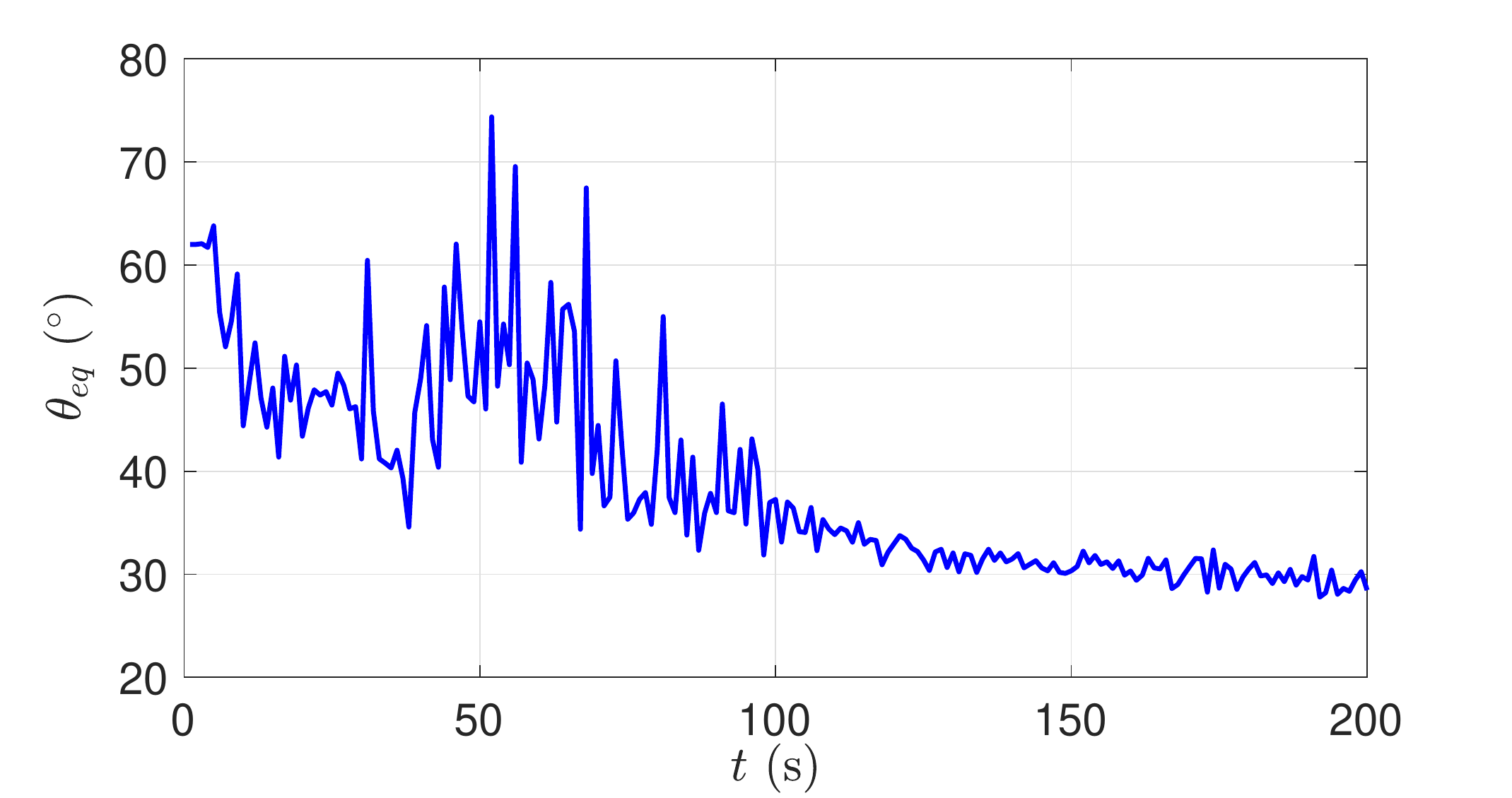}
\caption{An experimental result for the evolution of the contact angle that a drop of water makes with the acrylic polymer used in experiments for gas-jet induced dewetting measured using the Drop Shape Analyser DSA100 by KR\"USS.}
\label{fig:contact_angle}
\end{figure}

To compare our computational results with experiments, we measured the equilibrium contact angle for the bottom of the beaker made of an acrylic polymer. This was done by placing a drop of water onto a plate made of the acrylic polymer and measuring the resulting angle using the Drop Shape Analyser DSA100 by KR\"USS. The experimental results showing the evolution of the contact angle $\theta_{eq}$ over time are given in figure~\ref{fig:contact_angle} and indicate that $\theta_{eq}\approx 30^\circ\,(\pm 2^\circ)$.

%\nocite{*}
 
\bibliographystyle{jfm}

\bibliography{ImpingingJetRefs}

\begin{thebibliography}{50}
\expandafter\ifx\csname natexlab\endcsname\relax\def\natexlab#1{#1}\fi

\bibitem[Abr\`amoff {\em et~al.\/}(2004)Abr\`amoff, Magalh\~aes \& Ram]{ImageJ}
{\sc Abr\`amoff, M., Magalh\~aes, P. \& Ram, S.} 2004 Image processing with
  {I}mage{J}. {\em Biophotonics Int.\/} {\bf 111}, 36--42.

\bibitem[Adib {\em et~al.\/}(2018)Adib, Ehteram \& Tabrizi]{ADIB2018510}
{\sc Adib, M., Ehteram, M.~A. \& Tabrizi, H.~B.} 2018 Numerical and
  experimental study of oscillatory behavior of liquid surface agitated by
  high-speed gas jet. {\em Appl. Math. Model.\/} {\bf 62}, 510--525.

\bibitem[Afkhami {\em et~al.\/}(2018)Afkhami, Buongiorno, Guion, Popinet,
  Saade, Scardovelli \& Zaleski]{Afkhami}
{\sc Afkhami, S., Buongiorno, J., Guion, A., Popinet, S., Saade, Y.,
  Scardovelli, R. \& Zaleski, S.} 2018 Transition in a numerical model of
  contact line dynamics and forced dewetting. {\em J. Comput. Phys.\/} {\bf
  374}, 1061--1093.

\bibitem[Banks \& Chandrasekhara(1963)]{banks1963experimental}
{\sc Banks, R.~B. \& Chandrasekhara, D.~V.} 1963 Experimental investigation of
  the penetration of a high-velocity gas jet through a liquid surface. {\em J.
  Fluid Mech.\/} {\bf 15}, 13--34.

\bibitem[Berendsen {\em et~al.\/}(2013)Berendsen, Zeegers \&
  Darhuber]{BERENDSEN2013505}
{\sc Berendsen, C. W.~J., Zeegers, J. C.~H. \& Darhuber, A.~A.} 2013
  Deformation and dewetting of thin liquid films induced by moving gas jets.
  {\em J. Colloid Interf. Sci.\/} {\bf 407}, 505--515.

\bibitem[Berendsen {\em et~al.\/}(2012)Berendsen, Zeegers, Kruis, Riepen \&
  Darhuber]{Berendsen_etal_2012}
{\sc Berendsen, C. W.~J., Zeegers, J. C.~H., Kruis, G. C. F.~L., Riepen, M. \&
  Darhuber, A.~A.} 2012 Rupture of thin liquid films induced by impinging
  air-jets. {\em Langmuir\/} {\bf 28}, 9977--9985.

\bibitem[Berghmans(1972)]{Berghmans_1972}
{\sc Berghmans, J.} 1972 Stability of a gas-liquid interface and its relation
  to weld pool stability. {\em J. Phys. D\/} {\bf 5}, 1096--1105.

\bibitem[Bostwick {\em et~al.\/}(2017)Bostwick, Dijksman \&
  Shearer]{Bostwick_etal_2017}
{\sc Bostwick, J.~B., Dijksman, J.~A. \& Shearer, M.} 2017 Wetting dynamics of
  a collapsing fluid hole. {\em Phys. Rev. Fluids\/} {\bf 2}, 014006.

\bibitem[Cheslak {\em et~al.\/}(1969)Cheslak, Nicholls \&
  Sichel]{cheslak1969cavities}
{\sc Cheslak, F.~R., Nicholls, J.~A. \& Sichel, M.} 1969 Cavities formed on
  liquid surfaces by impinging gaseous jets. {\em J. Fluid Mech.\/} {\bf 36},
  55--63.

\bibitem[Clancy(2006)]{clancy2006aerodynamics}
{\sc Clancy, J.~L.} 2006 {\em Aerodynamics\/}. Sterling Book House.

\bibitem[Craster \& Matar(2009)]{Craster_Matar_2009}
{\sc Craster, R.~V. \& Matar, O.~K.} 2009 Dynamics and stability of thin liquid
  films. {\em Rev. Mod. Phys.\/} {\bf 81}, 1131--1198.

\bibitem[Davis {\em et~al.\/}(2010)Davis, Gratton \& Davis]{Davis_etal_2010}
{\sc Davis, M.~J., Gratton, M.~B. \& Davis, S.~H.} 2010 Suppressing van der
  {W}aals driven rupture through shear. {\em J. Fluid Mech.\/} {\bf 661},
  522--539.

\bibitem[Denner {\em et~al.\/}(2016)Denner, Pradas, Charogiannis, Markides, van
  Wachem \& Kalliadasis]{Denner_etal_2016}
{\sc Denner, F., Pradas, M., Charogiannis, Al., Markides, C.~N., van Wachem, B.
  G.~M. \& Kalliadasis, S.} 2016 Self-similarity of solitary waves on
  inertia-dominated falling liquid films. {\em Phys. Rev. E\/} {\bf 93},
  033121.

\bibitem[Dijksman {\em et~al.\/}(2015)Dijksman, Mukhopadhyay, Gaebler, Witelski
  \& Behringer]{Dijksman_etal_2015}
{\sc Dijksman, J.~A., Mukhopadhyay, S., Gaebler, C., Witelski, T.~P. \&
  Behringer, R.~P.} 2015 Obtaining self-similar scalings in focusing flows.
  {\em Phys. Rev. E\/} {\bf 92}, 043016.

\bibitem[Foster(2017)]{Foster_2017}
{\sc Foster, J.~E.} 2017 Plasma-based water purification: Challenges and
  prospects for the future. {\em Phys. Plasmas\/} {\bf 24}, 055501.

\bibitem[Galvagno {\em et~al.\/}(2014)Galvagno, Tseluiko, Lopez \&
  Thiele]{Galvagno_etal_2014}
{\sc Galvagno, M., Tseluiko, D., Lopez, H. \& Thiele, U.} 2014 Continuous and
  discontinuous dynamic unbinding transitions in drawn film flow. {\em Phys.
  Rev. Lett.\/} {\bf 112}, 137803.

\bibitem[de~Gennes {\em et~al.\/}(2013)de~Gennes, Brochard-Wyart \&
  Qu\'er\'e]{de2013capillarity}
{\sc de~Gennes, P.-G., Brochard-Wyart, F. \& Qu\'er\'e, D.} 2013 {\em
  Capillarity and wetting phenomena: drops, bubbles, pearls, waves\/}. Springer
  New York.

\bibitem[He \& Belmonte(2010)]{he2010deformation}
{\sc He, A. \& Belmonte, A.} 2010 Deformation of a liquid surface due to an
  impinging gas jet: A conformal mapping approach. {\em Phys. Fluids\/} {\bf
  22}, 042103.

\bibitem[Hughes {\em et~al.\/}(2015)Hughes, Thiele \& Archer]{Hughes_etal_2015}
{\sc Hughes, A.~P., Thiele, U. \& Archer, A.~J.} 2015 Liquid drops on a
  surface: Using density functional theory to calculate the binding potential
  and drop profiles and comparing with results from mesoscopic modelling. {\em
  J. Chem. Phys.\/} {\bf 142}, 074702.

\bibitem[Hwang \& Irons(2012)]{Hwang2012}
{\sc Hwang, H.~Y. \& Irons, G.~A.} 2012 A water model study of impinging gas
  jets on liquid surfaces. {\em Metall. Mater. Trans. B\/} {\bf 43}, 302--315.

\bibitem[Kalliadasis {\em et~al.\/}(2011)Kalliadasis, Ruyer-Quil, Scheid \&
  Velarde]{Kalliadasis2011}
{\sc Kalliadasis, S., Ruyer-Quil, C., Scheid, B. \& Velarde, M.~G.} 2011 {\em
  Falling liquid films\/}. {\em Series on Applied Mathematical Sciences\/} 176.
  Springer, London.

\bibitem[Kriegsmann {\em et~al.\/}(1998)Kriegsmann, Miksis \&
  Vanden-Broeck]{Kriegsmann_etal_1998}
{\sc Kriegsmann, J.~J., Miksis, M.~J. \& Vanden-Broeck, J.-M.} 1998 Pressure
  driven disturbances on a thin viscous film. {\em Phys. Fluids\/} {\bf 10},
  1249--1255.

\bibitem[Lacanette {\em et~al.\/}(2006)Lacanette, Gosset, Vincent, Buchlin \&
  Arquis]{Lacanette_etal_2006}
{\sc Lacanette, D., Gosset, A., Vincent, S., Buchlin, J.-M. \& Arquis, \'E.}
  2006 Macroscopic analysis of gas-jet wiping: numerical simulation and
  experimental approach. {\em Phys. Fluids\/} {\bf 18}, 042103.

\bibitem[Liu {\em et~al.\/}(2015)Liu, Chen, Hu, Xie \& Fu]{Liu_etal_2015}
{\sc Liu, Q., Chen, W., Hu, L., Xie, H. \& Fu, X.} 2015 Experimental
  investigation of cavity stability for a gas-jet penetrating into a liquid
  sheet. {\em Phys. Fluids\/} {\bf 27}, 082106.

\bibitem[Lunz \& Howell(2018)]{Lunz_Howel_2018}
{\sc Lunz, D. \& Howell, P.~D.} 2018 Dynamics of a thin film driven by a moving
  pressure source. {\em Phys. Rev. Fluids\/} {\bf 3}, 114801.

\bibitem[Molloy(1970)]{molloy1970impinging}
{\sc Molloy, N.~A.} 1970 Impinging jet flow in a two-phase system: the basic
  flow pattern. {\em J. Iron Steel Inst.\/} {\bf 208}, 943--950.

\bibitem[Mordasov {\em et~al.\/}(2016)Mordasov, Savenkov \&
  Chechetov]{Mordasov2016}
{\sc Mordasov, M.~M., Savenkov, A.~P. \& Chechetov, K.~E.} 2016 Method for
  analyzing the gas jet impinging on a liquid surface. {\em Technical
  Physics\/} {\bf 61}, 659--668.

\bibitem[Mu{\~n}oz-Esparza {\em et~al.\/}(2012)Mu{\~n}oz-Esparza, Buchlin,
  Myrillas \& Berger]{munoz2012numerical}
{\sc Mu{\~n}oz-Esparza, D., Buchlin, J.~M., Myrillas, K. \& Berger, R.} 2012
  Numerical investigation of impinging gas jets onto deformable liquid layers.
  {\em Appl. Math. Model.\/} {\bf 36}, 2687--2700.

\bibitem[Nguyen \& Evans(2006)]{nguyen2006computational}
{\sc Nguyen, A.~V. \& Evans, G.~M.} 2006 Computational fluid dynamics modelling
  of gas jets impinging onto liquid pools. {\em Appl. Math. Model.\/} {\bf 30},
  1472--1484.

\bibitem[Olmstead \& Raynor(1964)]{olmstead1964depression}
{\sc Olmstead, W.~E. \& Raynor, S.} 1964 Depression of an infinite liquid
  surface by an incompressible gas jet. {\em J. Fluid Mech.\/} {\bf 19},
  561--576.

\bibitem[Pismen(2002)]{PISMEN200211}
{\sc Pismen, L.~M.} 2002 Mesoscopic hydrodynamics of contact line motion. {\em
  Colloids Surf. A\/} {\bf 206}, 11--30.

\bibitem[Popinet(2009)]{popinet2}
{\sc Popinet, S.} 2009 An accurate adaptive solver for surface-tension-driven
  interfacial flows. {\em J. Comput. Phys.\/} {\bf 228}, 5838--5866.

\bibitem[Pryor(2011)]{pryor2011multiphysics}
{\sc Pryor, R.~W.} 2011 {\em Multiphysics modeling using COMSOL\circledR: a
  first principles approach\/}. Jones \& Bartlett Learning.

\bibitem[Rauscher \& Dietrich(2008)]{Rauscher_Dietrich_2008}
{\sc Rauscher, M. \& Dietrich, S.} 2008 Wetting phenomena in nanofluidics. {\em
  Annu. Rev. Mater. Res.\/} {\bf 38}, 143--172.

\bibitem[Ruyer-Quil \& Manneville(2000)]{Ruyer-Quil2000}
{\sc Ruyer-Quil, C. \& Manneville, P.} 2000 Improved modeling of flows down
  inclined planes. {\em Eur. Phys. J. B\/} {\bf 15}, 357--369.

\bibitem[Sibley {\em et~al.\/}(2012)Sibley, Savva \&
  Kalliadasis]{sibley2012slip}
{\sc Sibley, D.~N., Savva, N. \& Kalliadasis, S.} 2012 Slip or not slip? a
  methodical examination of the interface formation model using two-dimensional
  droplet spreading on a horizontal planar substrate as a prototype system.
  {\em Physics of Fluids\/} {\bf 24}, 082105.

\bibitem[Sol\'orzano-L\'opez {\em et~al.\/}(2011)Sol\'orzano-L\'opez, Zenit \&
  Ram\'irez-Arg\'aez]{SOLORZANOLOPEZ20114991}
{\sc Sol\'orzano-L\'opez, J., Zenit, R. \& Ram\'irez-Arg\'aez, M.~A.} 2011
  Mathematical and physical simulation of the interaction between a gas jet and
  a liquid free surface. {\em Appl. Math. Model.\/} {\bf 35}, 4991--5005.

\bibitem[{Sullivan} {\em et~al.\/}(2008){Sullivan}, {Wilson} \&
  {Duffy}]{Sullivan_etal_2008}
{\sc {Sullivan}, J.~M., {Wilson}, S.~K. \& {Duffy}, B.~R.} 2008 A thin rivulet
  of perfectly wetting fluid subject to a longitudinal surface shear stress.
  {\em Q. J. Mech. App. Math.\/} {\bf 61}, 25--61.

\bibitem[Thiele(2007)]{Thiele_2007}
{\sc Thiele, U.} 2007 {\em Thin Films of Soft Matter\/}, chap. Structure
  Formation in Thin Liquid Films, pp. 25--93. Springer, Vienna.

\bibitem[Thornton \& Graff(1976)]{Thornton1976}
{\sc Thornton, J.~A. \& Graff, H.~F.} 1976 An analytical description of the jet
  finishing process for hot-dip metallic coatings on strip. {\em Metall. Trans.
  B\/} {\bf 7}, 607--618.

\bibitem[Tian \& Kushner(2014)]{Tian_2014}
{\sc Tian, W. \& Kushner, M.~J.} 2014 Atmospheric pressure dielectric barrier
  discharges interacting with liquid covered tissue. {\em J. Phys. D\/} {\bf
  47}, 165201.

\bibitem[Tseluiko \& Kalliadasis(2011)]{JFM2011}
{\sc Tseluiko, D. \& Kalliadasis, S.} 2011 Nonlinear waves in counter-current
  gas--liquid film flow. {\em J. Fluid Mech.\/} {\bf 673}, 19--59.

\bibitem[Tuck(1975)]{tuck1975air}
{\sc Tuck, E.~O.} 1975 On air flow over free surfaces of stationary water. {\em
  The ANZIAM Journal\/} {\bf 19}, 66--80.

\bibitem[Turkdogan(1966)]{turkdogan1966fluid}
{\sc Turkdogan, E.~T.} 1966 Fluid dynamics of gas jets impinging on surface of
  liquids. {\em Chem. Eng. Sci.\/} {\bf 21}, 1133--1144.

\bibitem[Turkdogan(1996)]{turkdogan1996fundamentals}
{\sc Turkdogan, E.~T.} 1996 {\em Fundamentals of Steelmaking\/}. Institute of
  Materials.

\bibitem[Vanden-Broeck(1981)]{vanden1981deformation}
{\sc Vanden-Broeck, J.~M.} 1981 Deformation of a liquid surface by an impinging
  gas jet. {\em SIAM J. Appl. Math.\/} {\bf 41}, 306--309.

\bibitem[Vellingiri {\em et~al.\/}(2015)Vellingiri, Tseluiko \&
  Kalliadasis]{JFM2015}
{\sc Vellingiri, R., Tseluiko, D. \& Kalliadasis, S.} 2015 Absolute and
  convective instabilities in counter-current gas--liquid film flows. {\em J.
  Fluid Mech.\/} {\bf 763}, 166--201.

\bibitem[Verlackt {\em et~al.\/}(2018)Verlackt, Van~Boxem \&
  Bogaerts]{C7CP07593F}
{\sc Verlackt, C. C.~W., Van~Boxem, W. \& Bogaerts, A.} 2018 Transport and
  accumulation of plasma generated species in aqueous solution. {\em Phys.
  Chem. Chem. Phys.\/} {\bf 20}, 6845--6859.

\bibitem[Witelski \& Bernoff(2000)]{WitelskiBernoff2000_PhysD}
{\sc Witelski, T.~P. \& Bernoff, A.~J.} 2000 Dynamics of three-dimensional thin
  film rupture. {\em Physica D\/} {\bf 147}, 155--176.

\bibitem[Zheng {\em et~al.\/}(2018)Zheng, Fontelos, Shin, Dallaston, Tseluiko,
  Kalliadasis \& Stone]{zheng_etal_2018}
{\sc Zheng, Z., Fontelos, M.~A., Shin, S., Dallaston, M.~C., Tseluiko, D.,
  Kalliadasis, S. \& Stone, H.~A.} 2018 Healing capillary films. {\em J. Fluid
  Mech.\/} {\bf 838}, 404--434.

\end{thebibliography}
 
\end{document}